\documentclass[a4paper,11pt]{article}
\pdfoutput=1
\usepackage{jheppub}

\usepackage{dsfont}
\usepackage{graphicx}
\usepackage{subfig}

\DeclareMathOperator{\arctanh}{arctanh}

\newcommand\fft[2]{\frac{#1}{#2}}
\newcommand\nn{\nonumber}
\renewcommand{\Re}{\operatorname{Re}}
\renewcommand{\Im}{\operatorname{Im}}
\usepackage{xcolor}
\begin{document}

\title{Repulsive Black Holes and Higher-Derivatives}
\author[a]{Sera Cremonini,}
\emailAdd{cremonini@lehigh.edu}
\affiliation[a]{
Department of Physics, Lehigh University, Bethlehem, PA, 18018, USA
}
\author[b]{Callum R. T. Jones,}
\emailAdd{cjones@physics.ucla.edu}
\affiliation[b]{
Mani L. Bhaumik Institute for Theoretical Physics, Department of Physics and Astronomy, University of California Los Angeles, Los Angeles, CA 90095, USA
}
\author[c]{James T. Liu,}
\emailAdd{jimliu@umich.edu}
\affiliation[c]{Leinweber Center for Theoretical Physics, Randall Laboratory of Physics, University of Michigan, Ann Arbor, MI 48109-1040, USA}
\author[d]{Brian McPeak,}
\emailAdd{brian.mcpeak@df.unipi.it}
\affiliation[d]{Department of Physics, University of Pisa and INFN, \\Largo Pontecorvo 3, I-56127 Pisa, Italy}
\author[a]{Yuezhang Tang,}
\emailAdd{yut318@lehigh.edu}

\abstract{
In two-derivative theories of gravity coupled to matter, charged black holes 
are self-attractive at large distances, with the force vanishing at zero temperature.
However, in the presence of massless scalar fields and four-derivative corrections, zero-temperature black holes no longer need to obey the no-force condition. In this paper, we show how to calculate the long-range force between such black holes. We develop an efficient method for computing the higher-derivative corrections to the scalar charges when the theory has a shift symmetry, and compute the resulting force in a variety of examples.
We find that higher-derivative corrected black holes may be self-attractive or self-repulsive, depending on the value of the Wilson coefficients and the VEVs of scalar moduli. Indeed, we find black hole solutions which are both superextremal and self-attractive. 
Furthermore, we present examples where no choice of higher-derivative coefficients allows for self-repulsive 
black hole states in all directions in charge space. This suggests that, unlike the Weak Gravity Conjecture, which may be satisfied by the black hole spectrum alone, the Repulsive Force Conjecture requires additional constraints on the spectrum of charged particles. }

\maketitle
\flushbottom

%%%%%%%%%%
\section{Introduction}
\label{sec:intro}

Despite being treated often as classical objects, 
black hole solutions have provided remarkable insights into quantum gravity. 
This stems from the fact that they are thermal objects which obey the laws of thermodynamics -- they are equipped with a  temperature and entropy and must decay over time due to the emission of black-body radiation. 
The efforts to understand this process in detail, and to reconcile it with the basic principles of quantum mechanics, have led to a number of sharp paradoxes and hints towards understanding their quantum properties.

A somewhat more modern perspective on these issues is provided by effective field theory (EFT), where unknown ``high-energy" physics is parameterized by higher-derivative corrections to the low-energy action. These corrections are particularly interesting when the two-derivative solutions to the theory are marginal, \textit{i.e.} in cases where a small change to the solution has a major effect on the associated physics. One example of such a situation is in the context of the Weak Gravity Conjecture (WGC) \cite{ArkaniHamed:2006dz, Kats:2006xp}, which has been studied extensively in recent years \cite{Cheung:2014vva, Cottrell:2016bty, Chen:2019qvr, Cheung:2018cwt, Andriolo:2018lvp, Bellazzini:2019xts, Hamada:2018dde, Charles:2019qqt, Goon:2019faz, Jones:2019nev, Cremonini:2019wdk, Loges:2019jzs, Andriolo:2020lul, Aalsma:2020duv, Cremonini:2020smy, Cano:2019oma, Cano:2019ycn, Cano:2021tfs}. According to the WGC, gravitational EFTs which may be UV completed into a consistent quantum theory of gravity must have states with charge greater than mass, i.e. states for which 
``gravity is the weakest force."
Such states -- referred to as superextremal -- may be described by fundamental particles 
or non-perturbative objects such as black holes.
In many two-derivative theories, such as Einstein-Maxwell theory, demanding that black hole singularities be hidden by horizons imposes ``extremality bounds" which 
require the black hole charge to be less than its mass.
However, higher-derivative corrections can change such bounds, and the maximum charge may differ from the mass of the black hole solution -- it can be slightly more or slightly less, depending on linear inequalities involving the EFT coefficients. The WGC can then be invoked to place constraints on such 
combinations of coefficients. Indeed, there are choices of EFT coefficients for which the WGC is trivially satisfied by the existence of superextremal black holes.

Another way to attempt to quantify the strength of gravity  is to examine the long-range forces between charged states in the theory, whether fundamental particles or black holes. 
This line of thought has led to the Repulsive Force Conjecture (RFC), which, in its simplest incarnation, was already postulated in \cite{ArkaniHamed:2006dz} and reemphasized more recently in \cite{Palti:2017elp}. At its core is the idea that long-range repulsive gauge forces between identical charged particles should be at least as strong as all long-range attractive forces.
In other words, EFTs consistent with quantum gravity must have a state which is ``self-repulsive." This is what we will call the ``weak RFC." When multiple charges are present, it is natural to enforce this requirement at each direction in charge space. Refinements of the  RFC have been discussed in \cite{Heidenreich:2019zkl}, where a stronger form was taken as a working definition  
-- 
in every direction in charge space there should be a \emph{strongly self-repulsive multiparticle state}. 
This is a multiparticle state where each constituent state repels every other state, including itself. 
We will refer to this as the ``strong RFC" and comment on it further in the Discussion.
While clearly the WGC and the RFC are related to each other, the two conjectures are in fact distinct
-- as already emphasized in \cite{Heidenreich:2019zkl} --
and it is natural to ask to what extent they can be combined or generalized into a stronger statement about constraining EFTs from quantum gravity.

Generically, two copies of a black hole placed at rest asymptotically far apart will attract each other, unless the black hole is extremal, in which case the net force will vanish \cite{Heidenreich:2020upe}.\footnote{Throughout this paper, \textit{force} always refers to the static or velocity independent part of the interaction. Even BPS black holes are generically subject to non-vanishing, long-range, velocity dependent forces \cite{Gibbons:1986cp,Ferrell:1987gf}, but these are not relevant to the considered formulation of the RFC.} Remarkably, this conclusion holds in any two-derivative theory of gravity, even in cases involving moduli -- massless scalar fields with no potential -- that mediate new long-range forces.
Indeed, the role of such scalars is to precisely offset the 
gravitational and electrostatic forces between the black holes, which otherwise would not cancel.
In the simple case of a black hole of mass $M$ and electric charge $Q$ in theories with a single light scalar, the force is given by
\begin{align}
\label{introforce}
    f(r) = - G_N \frac{M^2}{r^2} + e^2 \frac{Q^2}{r^2} - \frac{1}{4}\frac{\mu^2}{r^2} +\mathcal{O}\left(\frac{1}{r^3}\right)+\mathcal{O}\left(\frac{v^2}{r^2}\right) \, , 
\end{align}
where $\mu$ is the \textit{scalar charge}, to be defined precisely below.

The aim of this paper is to study the effect of higher-derivative corrections to this force law 
for extremal black holes, as well as on 
the relation between the WGC and the RFC. 
%%%
Indeed, higher-derivative operators, which arise naturally at high energies in string theory, are a 
valuable probe to distinguish subtle aspects of the swampland conjectures and 
their range of validity.
%%%
We focus on the weak version of the RFC, the statement that the EFT spectrum should contain a self-repulsive state.
It is important to emphasize that we are still only considering the long-range (i.e.\ $1 / r^2$) contribution to the static force. 
The higher-derivative corrections we examine enter by shifting the extremal mass and extremal scalar charge, for a black hole with fixed charges $Q$ and $P$. 
More specifically, for the simple case of (\ref{introforce}), the corrected force $\Delta f$ will take the form
\begin{equation}
\label{introforceshift}
   \lim_{r\rightarrow \infty} r^2 \Delta f(r) = - 2 G_N \, M^{(2)} \, \Delta M
   -\frac{1}{2}\mu^{(2)} \, \Delta \mu  \, ,
\end{equation}
where $\Delta M,\Delta \mu$ refer to the shifts to 
the leading order (two-derivative) mass $M^{(2)}$ and scalar charge $\mu^{(2)}$.

For higher-derivative theories with no scalars, superextremal black holes must be self-repulsive. Recall that superextremal black holes are those which have charge greater than mass ($\Delta M<0$), meaning that the gauge force will be greater than the gravitational attraction, with the net result being a repulsive force.
However, this is no longer necessarily the case when scalars are turned on, because their contribution to (\ref{introforceshift}) can have either sign. 
Thus, a superextremal black hole is no longer necessarily self-repulsive.
Note that this provides a concrete example of the kind of ``marginal" situation we mentioned earlier -- the higher-derivative corrections can have a significant impact on the long-range force between 
extremal black holes, precisely because at the two-derivative level all interactions are exactly balanced and the corresponding force vanishes.

The mass shift $\Delta M$ in theories with higher-derivative corrections has been studied in a number of cases so far,  thanks to its relevance to the WGC \cite{Kats:2006xp, Cheung:2014vva, Cheung:2016wjt, Cottrell:2016bty, Chen:2019qvr, Cheung:2018cwt, Bellazzini:2019xts, Andriolo:2018lvp, Hamada:2018dde, Charles:2019qqt, Jones:2019nev, Goon:2019faz,Cremonini:2019wdk,Loges:2019jzs,Andriolo:2020lul, Aalsma:2020duv,Cremonini:2020smy, Bobev:2021oku}.
In this paper we show how to compute the scalar shift in such theories
%instances 
and develop an efficient method for 
extracting the scalar charges using a derivative formula that applies when the theory has 
a shift symmetry. We also compute the forces for a variety of examples, which will allow us to make several interesting generalizations and conclusions along the way. The main points we will make are:
\begin{itemize}
\item There are theories with scalar fields and higher derivatives
with 
superextremal black holes which are self-attractive.
\item There are examples in such theories where the mass shift at extremality has a definite sign, but the net force changes sign depending on the direction in charge space. 
\item There are examples where no choice of the EFT coefficients allows for self-repulsive black holes in all directions in charge space.
\end{itemize}

We conclude that there is no \textit{mild} 
or black hole version of the RFC, which allows that the repulsive state required by the RFC may be a black hole. More generally, for these higher-derivative theories, the WGC and RFC are clearly fundamentally inequivalent. The WGC may be trivially satisfied by black holes in theories with higher-derivatives, but the RFC cannot.
If the RFC is true, then, it must imply stronger constraints on EFTs than those imposed by the WGC. 
In our example, it would imply the existence of self-repulsive low-energy \emph{particle} states.

The paper is structured as follows. 
In section II we explain how the force between two black holes is computed, including the contribution due to scalar interactions. We first review the calculation of the scalar charge in \cite{Heidenreich:2019zkl}, and then show how to derive the alternative ``derivative formula" in certain cases. In section III we compute the scalar force for a simple model with gravity, electromagnetism and a scalar field-- the so-called $a$-model. We also present a slight generalization of this model where our scalar charge formula of section II cannot be employed. In section IV we analyze the 4d dimensional reduction of 5d $\mathcal{N} = 2$ supergravity and discuss the resulting theory, which has four charges. We compute the force for four different solutions which only have two charges, and find a number of examples where superextremal black holes are not self-repulsive. Finally, we conclude with a discussion on the interpretation of our results and their relevance to both the weak and strong forms of the RFC \cite{Heidenreich:2015nta}. A number of additional details are provided in appendices A through D.

\section{Long-range forces}
\label{sec:forces}

Consider a four-dimensional theory described by the two-derivative Lagrangian
\begin{equation}
    16\pi e^{-1}\mathcal{L}^{(2)} =  R-\fft1{2}\sum_i \mathcal{G}_{ij}(\phi) \, 
    \partial_\mu\phi^i\partial^\mu\phi^j
    -  \frac{1}{4}\sum_{I,J}  \left( \mathcal Z_{IJ} (\phi) F_{\mu\nu}^I  F^{\mu\nu\,J} 
    +\fft12 \mathcal N_{IJ}(\phi) \, \epsilon^{\mu\nu\rho\sigma} F_{\mu\nu}^I  F_{\rho\sigma}^J \right).
\end{equation}
This describes gravity coupled to a number of gauge fields $A^I$ (with $F^I=dA^I$) and scalars $\phi^i$, with non-trivial scalar couplings parametrized by the functions $\mathcal{G}_{ij}(\phi)$, $\mathcal{Z}_{IJ}(\phi)$ and $\mathcal{N}_{IJ}(\phi)$. 
We are working with Gaussian units, i.e. $G_\text{N}=1$ and $\epsilon_0=1/4\pi$.

The force between two black holes of mass $M$ carrying charges $\{Q_I, P^I\}$ and scalar charges $\mu_i$ is given by 
\begin{equation}
    \lim_{r\rightarrow\infty}r^2f(r) = -  M^2
-\fft14  \sum_{i,j} \mathcal{G}_{ij} \mu^i\mu^j + \sum_{I,J} \left( \mathcal{Z}^{IJ} (Q_I+\mathcal N_{IK}P^K)( Q_J+\mathcal N_{JL}P^L) + \mathcal{Z}_{IJ} P^I P^J \right),
\label{eq:total_force}
\end{equation}
where $\mathcal Z^{IJ}=(\mathcal Z_{IJ})^{-1}$ and all scalar matrices are evaluated with $\phi^i_\infty$.  When $f>0$ the black holes are repulsive. Here we follow the normalization conventions of \cite{Gibbons:1996af}.  In particular, the electric (Noether) and magnetic (topological) charges are normalized according to 
\begin{equation}
    Q_I=\fft1{8\pi}\int_{S^2_\infty}(\mathcal Z_{IJ}*F^J -\mathcal N_{IJ}F^J),\qquad P^I=\fft1{8\pi}\int_{S^2_\infty} F^I.
\label{eq:PQdef}
\end{equation}
This corresponds to 
\begin{align}
A_t\sim\fft{2Q}{r} 
,  \qquad A_\phi\sim-2P\cos\theta\, , 
\end{align}
while the scalar charges are defined from the asymptotic expansion of the scalars,
\begin{align}
\label{scalarsasympt}
    \phi^i(r) = \phi^i_{\infty} + \frac{\mu^i}{r} +\cdots \;.
\end{align}
This normalization is consistent with a first law of black hole mechanics of the form
\begin{equation}
    \label{firstlaw}
    dM = TdS+\Phi^I_e dQ_I+ \Phi_{m I} dP^I - \frac{1}{4} \mathcal{G}_{ij} \mu^i d\phi^i_\infty,
\end{equation}
with electric and magnetic scalar potentials defined as 
\begin{equation}
    \label{electricmagneticpotentials}
    \Phi_e^I \equiv \frac{1}{2}\left[A^I_t(\infty) - A^I_t(r_+)\right], \hspace{10mm} \Phi_{m I} \equiv \frac{1}{2}\left[\tilde{A}_{It}(\infty) - \tilde{A}_{It}(r_+)\right],
\end{equation}
where $\tilde{A}$ is the electric-magnetic dual gauge potential. The scalar charges may also be computed ``thermodynamically" using the prescription  \cite{Gibbons:1996af}
\begin{align}
\label{originalprescription}
    \mu^i = -4\mathcal G^{ij}(\phi_\infty) \left( \frac{\partial M}{\partial \phi^j_\infty} \right)_{S, Q_I, P^I} \, , 
\end{align}
where $M$ is the mass of the finite temperature black hole solution, $\mathcal{G}^{ij}$ is the inverse of $\mathcal{G}_{ij}$, and the underlying assumption is that its dependence on the
moduli at infinity $\phi^i_\infty$ is known. 

The scalar charge prescription, (\ref{originalprescription}), requires knowledge of the black hole mass as a function of scalar VEVs at fixed entropy.  However, by using thermodynamic identities and the first law of black hole mechanics, we can re-express this as a variation of the mass at constant temperature.  The reason this is useful is that we are mainly interested in the scalar charges and the overall force at extermality, corresponding to zero temperature, and often the black hole geometry (and therefore the black hole mass) is only known at zero temperature.  To proceed, we write
\begin{eqnarray}
    \left(\frac{\partial M}{\partial \phi^i_\infty}\right)_{T,Q_I,P^I} 
    &=& \left(\frac{\partial M}{\partial \phi^i_\infty}\right)_{S,Q_I,P^I} + \; \left(\frac{\partial M}{\partial S}\right)_{Q_I,P^I,\phi^i_\infty} \left(\frac{\partial S}{\partial \phi^i_\infty}\right)_{T,Q_I,P^I} \nonumber \\ 
    &=& \left(\frac{\partial M}{\partial \phi^i_\infty}\right)_{S,Q_I,P^I} + \; 
    T \, \left(\frac{\partial S}{\partial \phi^i_\infty}\right)_{T,Q_I,P^I} \, .
\end{eqnarray}
Taking the $T\rightarrow 0$ limit of this relation, and making the further assumption%
\footnote{This assumption is natural in all cases 
in which one can apply the attractor mechanism, which tells us that 
the entropy at extremality is independent of the asymptotic values of the moduli. The attractor mechanism is known to apply to all extremal black holes with a near horizon $\text{AdS}_2$ factor, both supersymmetric and non-supersymmetric \cite{Sen:2005wa}. However, note that to make our argument it suffices that the variation of the entropy is finite.}
\begin{equation}
    \lim_{T\rightarrow0}  \; T \left(\frac{\partial S}{\partial \phi^i_\infty}\right)_{T,Q_I,P^I} = 0 \, ,
\end{equation}
we obtain 
\begin{equation}
   \lim_{T\rightarrow0} \; \left(\frac{\partial M}{\partial \phi^i_\infty}\right)_{T,Q_I,P^I} = \quad  \lim_{T\rightarrow0} \;
  \left(\frac{\partial M}{\partial \phi^i_\infty}\right)_{S,Q_I,P^I} \, .
\end{equation}
This allows us to extract the scalar charges at extremality entirely from the mass of the extremal solution,
\begin{equation}
\label{extscalarthermopres}
  \mu^i\Big|_{T=0} = - 4\mathcal G^{ij}(\phi_\infty) \left(\frac{\partial M}{\partial \phi^j_{\infty}}\right)_{T=0} \, ,
\end{equation}
getting around the potential issue of not knowing finite temperature solutions to the theory.

The thermodynamic prescription (\ref{originalprescription}) is often much more convenient than the asymptotic expansion prescription (\ref{scalarsasympt}) for calculating four-derivative corrections to the scalar charge. The latter requires perturbatively solving the four-derivative equations of motion, which in some cases is prohibitively difficult. 
% However 
On the other hand, the four-derivative corrections to the Helmholtz free-energy (and hence the mass) can be calculated by evaluating the four-derivative part of the Euclidean on-shell action on the two-derivative solution \cite{Reall:2019sah},
\begin{equation}
    F\left(T,Q_I,P^I,\phi_\infty^i\right) = F^{(2)}\left(T,Q_I,P^I,\phi_\infty^i\right)+T I^{(4)}_E\left(T,Q_I,P^I,\phi_\infty^i\right),
\end{equation}
where $F\equiv M-TS$. The key result of \cite{Reall:2019sah} was that while the four-derivative corrections to the solutions are generally non-zero, they give a vanishing contribution to the two-derivative part of the on-shell action; and hence the leading corrections to certain \textit{thermodynamic} quantities can be calculated without solving the corrected equations of motion. This is the approach we will take to evaluating the four-derivative corrections to the long-range, static force between extremal black holes.

For many of the theories we examine in this paper, the presence of \emph{shift symmetries} makes the computation of the scalar charges much simpler by reducing them to a calculation at the origin of moduli space. To be specific, we focus first on models of gravity coupled to a single dilatonic scalar $\phi$ and vector $A_\mu$, which will be relevant to several of the cases of interest to us here. For example, consider the two-derivative Lagrangian
\begin{equation}
   16\pi e^{-1} \mathcal{L}^{(2)} = R-\frac{1}{2}\left(\partial \phi\right)^2 - \frac{1}{4}e^{-a\phi}F^2,
\end{equation}
which has a shift symmetry
\begin{equation}
\label{amodelshift}
    g_{\mu\nu} \rightarrow g_{\mu\nu}, \hspace{10mm} A_\mu \rightarrow e^{\frac{a\phi_0}{2}} A_\mu, \hspace{10mm} \phi \rightarrow \phi +\phi_0,
\end{equation}
where $\phi_0$ is an arbitrary real constant. Four-derivative corrections to this model will be analyzed in detail in section \ref{SectionAModel}. In Schwarzschild-type coordinates, the electric and magnetic charges of the solution are, from (\ref{eq:PQdef}) 
\begin{equation}
    Q = \frac{1}{8\pi}\int_{S^2_\infty} \text{d}^2\sigma\left(e^{-a\phi}F^{tr}\right), \hspace{10mm}
    P = -\frac{1}{8\pi}\int_{S^2_\infty} \text{d}^2\sigma\left(\tilde{F}^{tr}\right),
\end{equation}
where $\tilde F_{\mu\nu}=\fft12\epsilon_{\mu\nu}{}^{\rho\sigma}F_{\rho\sigma}$. Under the action of the shift symmetry (\ref{amodelshift}), a solution with electric charge $Q$ and magnetic charge $P$ is transformed into a new solution with 
\begin{equation}
\label{amodelchargeshift}
  Q\rightarrow e^{-\frac{a\phi_0}{2}}Q, \hspace{10mm} P \rightarrow e^{\frac{a\phi_0}{2}}P.
\end{equation}
As a consequence, the two-derivative solution at an arbitrary point in moduli space is related to the solution at the origin by 
\begin{align}
    \label{twodersolshift}
    g^{(2)}_{\mu\nu}\left[T,Q,P,\phi_\infty\right] &= g_{\mu\nu}^{(2)}\left[T,e^{\frac{a\phi_\infty}{2}}Q, e^{-\frac{a\phi_\infty}{2}}P,0\right], \nonumber\\
    A^{(2)}_{\mu}\left[T,Q,P,\phi_\infty\right] &=  e^{\frac{a\phi_\infty}{2}} A_{\mu}^{(2)}\left[T,e^{\frac{a\phi_\infty}{2}}Q, e^{-\frac{a\phi_\infty}{2}}P,0\right], \nonumber\\
    \phi^{(2)}\left[T,Q,P,\phi_\infty\right] &= \phi^{(2)}\left[T,e^{\frac{a\phi_\infty}{2}}Q, e^{-\frac{a\phi_\infty}{2}}P,0\right] +\phi_\infty ,
\end{align}
where the superscript is used to emphasize that we are referring 
to 
the two-derivative solution. Since the metric is invariant under the symmetry, the temperature $T$ does not change. When the Lagrangian is deformed by adding four-derivative terms which respect the shift symmetry (\ref{amodelshift}), the transformation of the charges (\ref{amodelchargeshift}) remains valid and so the mass correction satisfies the relation
\begin{equation}
    \Delta M\left[T,Q,P,\phi_\infty\right] = \Delta M\left[T,e^{\frac{a\phi_\infty}{2}}Q, e^{-\frac{a\phi_\infty}{2}}P,0\right].
\end{equation}
If we apply the thermodynamic prescription for calculating the four-derivative correction to the extremal scalar charge (\ref{extscalarthermopres}), we derive the \textit{derivative formula}
\begin{equation}
\label{dcharge}
    \boxed{\Delta\mu_\phi\biggr\vert_{\genfrac{}{}{0pt}{1}{T=0,}{\phi_\infty = 0}} = -2aQ\left(\frac{\partial \Delta M}{\partial Q}\right)_{\genfrac{}{}{0pt}{1}{T=0,}{\phi_\infty = 0}}+2aP\left(\frac{\partial \Delta M}{\partial P}\right)_{\genfrac{}{}{0pt}{1}{T=0,}{\phi_\infty = 0}}.}
\end{equation}
Generalizations of this formula are used throughout this paper to simplify calculations of the extremal scalar charge, by reducing them to simple electric and magnetic charge derivatives of the extremal mass calculated at the origin of moduli space. In some cases this formula can be further generalized when the shift symmetry is broken in a controlled way; we will see an example of this in section \ref{subsec:a-model}.

A similar analysis can be done when the matter content also contains an axionic field, as will be the case in 
section \ref{SectionN2SUGRA}.
To illustrate this point, it suffices to work with a simple model of gravity coupled to a vector $A_\mu$ and axion $\chi$, of the form 
\begin{equation}
    16\pi e^{-1} \mathcal{L}^{(2)} = 
    R-\frac{1}{2}(\partial_\mu \chi)^2 -\frac{1}{4}F_{\mu\nu}F^{\mu\nu} -\lambda \chi F_{\mu\nu} \tilde{F}^{\mu\nu} \, ,
\end{equation}
which is invariant (up to a total derivative) under the shift symmetry
\begin{equation}
    \label{axionshiftsymmetry}
    g_{\mu\nu} \rightarrow g_{\mu\nu}, \hspace{10mm} A_\mu \rightarrow  A_\mu, \hspace{10mm} \chi \rightarrow \chi +\chi_0,
\end{equation}
where $\chi_0$ is an arbitrary real constant.
In Schwarzschild-type coordinates, the electric and magnetic charges of the solution are, from (\ref{eq:PQdef}) 
\begin{equation}
    Q = \frac{1}{8\pi}\int_{S^2_\infty} \text{d}^2\sigma\left(F^{tr}+4\lambda \chi \tilde{F}^{tr}\right), \hspace{10mm}
    P = -\frac{1}{8\pi}\int_{S^2_\infty} \text{d}^2\sigma\left(\tilde{F}^{tr}\right).
\end{equation}
Under the action of the shift symmetry (\ref{axionshiftsymmetry}), the two-derivative solution with electric charge $Q$ and magnetic charge $P$ is transformed into a new solution with
\begin{equation}
    Q \rightarrow Q - 4\lambda \chi_0 P, \hspace{10mm} P \rightarrow P.
\end{equation}
In this model, turning on a VEV for the axion is equivalent to adding a theta term to the Lagrangian, this is seen by the above to transform a magnetic monopole into a dyon via the Witten effect \cite{Witten:1979ey}. Following identical logic to the dilaton example, the two-derivative solution at an arbitrary point in moduli space is related to the solution at the origin by 
\begin{align}
    g^{(2)}_{\mu\nu}\left[T,Q,P,\chi_\infty\right] &= g_{\mu\nu}^{(2)}\left[T,Q+4\lambda \chi_\infty P, P,0\right], \nonumber\\
    A^{(2)}_{\mu}\left[T,Q,P,\chi_\infty\right] &= A_{\mu}^{(2)}\left[T,Q+4\lambda \chi_\infty P, P,0\right], \nonumber\\
    \chi^{(2)}\left[T,Q,P,\chi_\infty\right] &= \chi^{(2)}\left[T,Q+4\lambda \chi_\infty P, P,0\right] +\chi_\infty ,
\end{align}
and therefore the four-derivative correction to the mass satisfies
\begin{equation}
    \Delta M\left[T,Q,P,\chi_\infty\right] = \Delta M\left[T,Q+4\lambda \chi_\infty,P,0\right].
\end{equation}
Applying (\ref{extscalarthermopres}) we then derive an axionic derivative formula for the leading correction to the extremal scalar charge,
\begin{equation}
\label{acharge}
\boxed{
    \Delta \mu_\chi \biggr\vert_{\genfrac{}{}{0pt}{1}{T=0}{\chi_\infty = 0}} =  -16 \lambda P \left(\frac{\partial \Delta M}{\partial Q}\right)_{\genfrac{}{}{0pt}{1}{T=0}{\chi_\infty = 0}}.}
\end{equation}

Armed with the expressions for the scalar charges -- which can be computed using
(\ref{dcharge}) and (\ref{acharge}) -- we now have 
all the ingredients needed to obtain the long-range force. For all the cases we study, the force computed  at the origin of moduli space ($\phi^i_\infty=0$) will take the simple form
\begin{equation}
\label{force2}
    \lim_{r\rightarrow \infty} r^2f(r) = - M^2 -\fft14  \sum_i  \mu_i^2 + \sum_{I} (Q_I^2 + {P_I}^2)\, .
\end{equation}
In all of our examples, at the two-derivative level the force for extremal solutions is precisely zero, i.e. the repulsive gauge force exactly balances out the gravitational attraction and the interaction mediated by the scalars in the theory.
However, when we introduce higher-derivative corrections this is no longer the case.
Since we are holding the electric and magnetic charges fixed, the shift to the force due to the higher-derivative terms is
given by
\begin{align}
    \lim_{r\rightarrow \infty}r^2 \Delta f(r) = - 2M^{(2)} \Delta M
   -\frac{1}{2}(\mu_\phi^{(2)} \Delta \mu_\phi + \mu_\chi^{(2)} \Delta\mu_\chi) \, ,
\end{align}
where $M^{(2)}$, $\mu_\phi^{(2)}$, and $\mu_\chi^{(2)}$ denote the leading order (two-derivative) expressions for the mass and scalar charges, and $\Delta M$, $\Delta \mu_\phi$, and $\Delta \mu_\chi$ denote their higher-derivative corrections. 
In the next two sections we will compute this shift for a variety of examples. We will then show that the link between the WGC and the RFC ceases to be valid once higher-derivative operators are included in the theory. 

For every solution in this paper except for one, we have used used equation (\ref{originalprescription}) to compute the scalar charges, and found agreement where we could compare it to the result of the shift symmetry formulas (\ref{dcharge}) and (\ref{acharge}). The only exception is the 5d $\mathcal{N} = 2$ supergravity solution with $Q_1$ and $P_1$, where we do not know the solution away from $T = 0$.

%%%%%%%%%%%%%%%%%%%%%%%%%%%%%%%%%%%%%%%%%%%%%%%%%%%%%%%%%%%%%%

\section{Higher-Derivative Corrections to Non-Supersymmetric Black Holes}
\label{SectionAModel}

In this section we analyze the leading higher-derivative corrections to electrically charged black hole solutions with a non-trivial scalar profile. 
The models we consider in this section are generically non-supersymmetric (and cannot be embedded in a supersymmetric model) and serve to illustrate 
the main points we want to make in this paper.
%%%
In section \ref{subsec:a-model}, we consider a simple model with an exponential gauge kinetic function and consequent shift symmetry of the action; in this simple case the mild WGC and mild RFC are controlled by the same linear combination of Wilson coefficients. In section \ref{sec:moregeneral}, we consider a more generic model without such a symmetry and see, in the context of an explicit example, that the mild WGC and mild RFC are unrelated conjectures.

\subsection{Corrections to the \texorpdfstring{$a$}{a}-model}
\label{subsec:a-model}

Possibly the simplest and best-studied example of a black hole with scalar hair is given by the so-called $a$-model \cite{Horowitz:1991cd,Khuri:1995xk}, previously studied in the context of the mild WGC in \cite{Loges:2019jzs}. 
In $d=4$ this model is defined by the two-derivative Lagrangian
\begin{equation}
   16\pi e^{-1} \mathcal{L}^{(2)} = R-\frac{1}{2}\left(\partial \phi\right)^2 - \frac{1}{4}e^{-a\phi}F^2 \, ,
\end{equation}
and admits a well-known \cite{Garfinkle:1990qj} family of electrically charged black hole solutions with non-trivial scalar hair (and general values of $\phi_\infty$),
\begin{align}
    ds^2 &= -H^{-\frac{2}{1+a^2}}Wdt^2+H^{\frac{2}{1+a^2}}\left(W^{-1}dr^2+r^2d\theta^2 +r^2\sin^2(\theta)d\varphi^2\right),\nonumber\\
    A &= \alpha e^{\frac{a}{2}\phi_\infty}\left(H^{-1}-1\right)dt \, , \nonumber\\
    \phi &= \phi_\infty -\frac{2a}{1+a^2}\log\left(H\right),
\end{align}
where
\begin{equation}
    H = 1+\frac{h}{r},\hspace{10mm} W = 1+\frac{h}{r}\left(1-\left(1+a^2\right)\frac{\alpha^2}{4}\right).
\end{equation}
The (outer) horizon of this solution is located at
\begin{equation}
    r^{(2)}_+ = \frac{h}{4}\left(\left(a^2+1\right) \alpha ^2-4\right),
\end{equation}
with corresponding thermodynamic quantities
\begin{align}
    M^{(2)} &= \frac{h}{2}\left(\frac{1}{4} \left(a^2+1\right) \alpha ^2 +\frac{1-a^2}{ 1+a^2}\right) \, , \nonumber\\
    S^{(2)} &= \frac{\pi h^2}{16}  \left[\alpha^2\left(a^2+1\right)\right]^{\frac{2}{a^2+1}}\left[\left(a^2+1\right)\alpha^2-4\right]^{\frac{2a^2}{1+a^2}} \, , \nonumber\\
    T &= \frac{1}{\pi h}\left[\alpha^2\left(a^2+1\right)\right]^{-\frac{2}{a^2+1}}\left[\left(a^2+1\right)\alpha^2-4\right]^{\frac{1-a^2}{1+a^2}} \, , \nonumber\\
    F^{(2)} &= \frac{1}{16} \left(a^2+1\right) \alpha ^2 h-\frac{\left(a^2-3\right) h}{4 \left(a^2+1\right)} \, , \nonumber\\
    Q &= \frac{\alpha h}{2} e^{-\frac{a}{2}\phi_\infty} \, , \nonumber\\
    \Phi_e^{(2)} &= \frac{2}{\alpha(a^2+1)}e^{\frac{a}{2}\phi_\infty} \, , \nonumber\\
    \mu^{(2)} &= -\frac{2ah}{a^2+1}\, .
\end{align}
The superscript indicates quantities calculated in the two-derivative approximation that may shift when higher-derivative effective operators are included. The electric potential $\Phi_e$ and the scalar charge $\mu$ associated with $\phi$ are defined according to the conventions (\ref{electricmagneticpotentials}) and (\ref{scalarsasympt}). $Q$ and $T$ are held fixed and may be taken to be the canonical variables of our ensemble.

The extremal limit of this solution corresponds to $T\rightarrow 0$ only if $a^2<1$, so we will restrict ourselves to this region of parameter space\footnote{These extremal solutions have zero horizon area at the two-derivative level, so truncating the derivative expansion at first-order may not be a reliable approximation. However, this will not affect the purpose of this example, which is to highlight the role of shift symmetries in computing the scalar charges.}. In this range,
extremality corresponds to taking the limit
\begin{equation}
\alpha \xrightarrow[]{T=0} \frac{2}{\sqrt{a^2+1}}.
\end{equation}
From these expressions, we can verify that the long-range self-force vanishes at extremality, at all points in moduli space \cite{Heidenreich:2020upe}
\begin{equation}
    \lim_{r\rightarrow \infty} r^2f(r) = -(M^{(2)})^2 + e^{a\phi_\infty}Q^2 -\frac{1}{4}(\mu^{(2)})^2 \xrightarrow[]{T=0} 0.
\end{equation}
This is of the form (\ref{eq:total_force}) with gauge kinetic function $\mathcal{Z}(\phi_\infty)=e^{\frac{a\phi_\infty}{2}}$.
We now calculate the four-derivative corrections to the extremal mass and scalar charge. We will do this using two different bases of four-derivative operators. First we will use a basis which \textit{preserves} the shift symmetry (\ref{amodelshift})
\begin{align}
    16\pi e^{-1} \mathcal{L}^{(4)}_{\text{sym}} &= b_1 e^{-2 a\phi} (F^2)^2 + b_2 e^{-2 a\phi} \left(F\tilde{F}\right)^2 + b_3 e^{-a\phi} FFW + b_4 R_{\text{GB}} \nonumber\\
    &\hspace{10mm}+ b_5  \left((\partial \phi)^2\right)^2 +b_6 e^{- a\phi}(\partial\phi)^2 F^2 +b_7 e^{- a\phi} \partial\phi \partial\phi FF \,,
\end{align}
where
\begin{align}
    &F^2 = F_{\mu\nu}F^{\mu\nu}, \hspace{5mm} F\tilde{F} = \epsilon^{\mu\nu\rho\sigma}F_{\mu\nu}F_{\rho\sigma}, \hspace{5mm} FFW = F_{\mu\nu}F_{\rho\sigma}W^{\mu\nu\rho\sigma} \, , \nonumber\\
    &R_\text{GB} = R_{\mu\nu\rho\sigma}R^{\mu\nu\rho\sigma}-4R_{\mu\nu}R^{\mu\nu}+R^2 \, , \nonumber\\
    &W_{\mu\nu\rho\sigma} = R_{\mu\nu\rho\sigma}-\frac{1}{2}\left(g_{\mu[\rho}R_{\sigma]\nu}-g_{\nu[\rho}R_{\sigma]\mu}\right) + \frac{1}{6}g_{\mu[\rho}g_{\sigma]\nu}R \, , \nonumber\\
    &(\partial\phi)^2 = \partial_\mu \phi \partial^\mu \phi ,\hspace{5mm} \partial \phi \partial \phi FF = \partial_\mu \phi \partial_\nu \phi F^{\mu\rho}{F^\nu}_\rho.
\end{align}
Using the method of \cite{Reall:2019sah}, briefly reviewed in section \ref{sec:forces}, we calculate the leading corrections to the free-energy and find that they are finite only in the range $a^2 <1$. The four-derivative corrected extremal mass, including 
the leading two-derivative contribution,
is given by the compact formula 
\begin{align}
\label{amodelextmasssym}
M\biggr\vert_{\genfrac{}{}{0pt}{1}{\hspace{-2mm}T=0}{\phi_\infty = 0}} &= \frac{|Q|}{\sqrt{a^2+1}}+ \frac{4 \sqrt{a^2+1} }{\left(a^2+2\right) \left(a^2-1\right)
   \left(3a^2+5\right) \left(a^2+3\right) |Q|}\nonumber\\
   &\hspace{25mm}\times\left[24 b_1\left(a^2+1\right)^2 -b_3\left(a^2+1\right)
   \left(5a^2+3\right) \right.\nonumber\\
   &\hspace{35mm}\left.+6 b_5 a^4 -12 b_6 \left(a^2+1\right) a^2-6b_7(a^2+1)a^2\right] \,.
\end{align}
As a consistency check, when $a\rightarrow 0$ the total mass reduces to the expected expression \cite{Kats:2006xp},
\begin{equation}
    M\biggr\vert_{\genfrac{}{}{0pt}{1}{\hspace{-2mm}T=0}{\phi_\infty = 0}} \xrightarrow[]{a=0} |Q| - \frac{2}{5|Q|}\left(8b_1 -b_3\right).
\end{equation}
From the free-energy we can also calculate the corrected scalar charge $\Delta \mu$ at extremality. At the origin of moduli space we find the simple result
\begin{equation}
\label{amodelscalarchargecorrection}
    \Delta \mu\biggr\vert_{\genfrac{}{}{0pt}{1}{\hspace{-2mm}T=0}{\phi_\infty = 0}} = 2a\Delta M\biggr\vert_{\genfrac{}{}{0pt}{1}{\hspace{-2mm}T=0}{\phi_\infty = 0}},
\end{equation}
implying in turn that the  shift to the force is controlled entirely by the shift to the mass, 
\begin{equation}
\label{amodelforcecorrection}
    \lim_{r\rightarrow \infty} r^2\Delta f(r)\biggr\vert_{\genfrac{}{}{0pt}{1}{\hspace{-2mm}T=0}{\phi_\infty = 0}} =
    - 2(1-a^2)M^{(2)}\Delta M\biggr\vert_{\genfrac{}{}{0pt}{1}{\hspace{-2mm}T=0}{\phi_\infty = 0}} \, .
\end{equation}
This is consistent with the derivative formula (\ref{dcharge}). For $a^2<1$, the correction to the long-distance self-force (\ref{amodelforcecorrection}) is therefore positive (implying a repulsive interaction) if and only if the correction to the extremal mass is negative. In other words, in this case the condition on the Wilson coefficients $b_i$ for large extremal black holes to be the states required by the WGC is exactly the same as the condition required for large extremal black holes to be self-repulsive, and therefore to satisfy the RFC. 

It is interesting to repeat this analysis for a basis of four-derivative operators which \textit{violate} the shift symmetry. For this we use the same string-inspired basis of operators as \cite{Loges:2019jzs}, in the Einstein frame 
\begin{align}
    16\pi e^{-1} \mathcal{L}^{(4)}_{\text{string}} &= b_1 e^{-3a\phi} (F^2)^2 + b_2 e^{-3a\phi} \left(F\tilde{F}\right)^2 + b_3 e^{-2a\phi} FFW + b_4e^{-a\phi}R_{\text{GB}} \nonumber\\
    &\hspace{10mm}+ b_5 e^{-a\phi}\left((\partial \phi)^2\right)^2 +b_6 e^{-2a\phi}(\partial\phi)^2 F^2 +b_7 e^{-2a\phi} \partial\phi \partial\phi FF.
\end{align}
For $a^2 <\frac{1}{3}$, the four-derivative corrected extremal mass is given by 
\begin{align}
\label{amodelextmass}
M\biggr\vert_{\genfrac{}{}{0pt}{1}{\hspace{-2mm}T=0}{\phi_\infty = 0}} &= \frac{|Q|}{\sqrt{a^2+1}}+ \frac{2 \sqrt{a^2+1} }{\left(a^2-3\right) \left(a^2-1\right)
   \left(a^2+5\right) \left(3 a^2-1\right) |Q|}\nonumber\\
   &\hspace{25mm}\times\left[24 b_1\left(a^2+1\right)^2 +b_3\left(a^2+1\right)
   \left(a^2-3\right) -3b_4 \left(a^2+3\right) a^2 \right.\nonumber\\
   &\hspace{35mm}\left.+6 b_5 a^4 -12 b_6 \left(a^2+1\right) a^2-6b_7(a^2+1)a^2\right].
\end{align}
Interestingly, even though (\ref{amodelextmass}) was derived by performing an integral that only converges in the range $a^2<\frac{1}{3}$, the result can be analytically continued outside of this range except for isolated singularities at $a=\sqrt{3},1,\frac{1}{\sqrt{3}}$, corresponding to well-known values of $a$ for which this model arises as the consistent truncation of a model of supergravity \cite{Cremmer:1984hj,Duff:1995sm,Behrndt:1996hu} . 

In this case, since we have explicitly broken the shift symmetry at the four-derivative level, we cannot immediately apply the derivative formula (\ref{dcharge}). There is however a generalization that follows from the observation that the mass shift satisfies
\begin{equation}
    \Delta M\left[T,Q,P,\phi_\infty\right] = e^{-a\phi_\infty}\Delta M\left[T,e^{\frac{a\phi_\infty}{2}}Q, e^{-\frac{a\phi_\infty}{2}}P,0\right].
\end{equation}
This leads to a different relation between the extremal mass and scalar charge shifts
\begin{equation}
    \Delta \mu\biggr\vert_{\genfrac{}{}{0pt}{1}{\hspace{-2mm}T=0}{\phi_\infty = 0}} = 6a\Delta M\biggr\vert_{\genfrac{}{}{0pt}{1}{\hspace{-2mm}T=0}{\phi_\infty = 0}},
\end{equation}
as well as the shift to the force
\begin{equation}
    \lim_{r\rightarrow \infty} r^2\Delta f(r)\biggr\vert_{\genfrac{}{}{0pt}{1}{\hspace{-2mm}T=0}{\phi_\infty = 0}} =
    - 2(1-3a^2)M^{(2)}\Delta M\biggr\vert_{\genfrac{}{}{0pt}{1}{\hspace{-2mm}T=0}{\phi_\infty = 0}} \, .
\end{equation}
Again, if $a^2<\frac{1}{3}$, a negative mass shift implies a repulsive force, and so we see that the WGC and RFC imply equivalent conditions on the Wilson coefficients.

\subsection{More general gauge kinetic functions}
\label{sec:moregeneral}

In more complicated models, 
simple relations such as (\ref{amodelscalarchargecorrection}) do not hold. 
This can be illustrated, for example, by working with a model for which shifting the scalar by a constant is not a symmetry of the two-derivative action.
An interesting family of explicit black hole solutions was constructed \cite{Fan:2015oca} for models described by a Lagrangian
\begin{equation}
\label{actiona1a2}
    16\pi e^{-1} \mathcal{L}^{(2)} = R-\frac{1}{2}(\partial \phi)^2 - \frac{1}{4}Z(\phi,\omega) F^2,
\end{equation}
with a gauge kinetic function of the form
\begin{equation}
    Z(\phi,\omega) = \left(\cos^2(\omega)e^{a\phi} + \sin^2(\omega)e^{-\phi/a}\right)^{-1}.
\end{equation}
In the limit $\omega\rightarrow 0$, this reduces to the $a$-model studied in the previous subsection. At the origin of moduli space, $\phi_\infty=0$, a family of non-rotating, electrically charged black hole solutions was constructed in \cite{Fan:2015oca}
\begin{align}
\label{a1a2solorigin}
    ds^2 &= -\mathfrak{f} dt^2 + \frac{dr^2}{\mathfrak{f}} + r^{1+\nu} (r+h)^{1-\nu} \left(d\theta^2 + \sin^2(\theta)d\varphi^2\right) \, , \nonumber\\
    A &= \frac{q(r+h\cos^2(\omega))}{r(r+h)}dt \, , \nonumber\\
    \phi &= \sqrt{1-\nu^2} \log\left(1+\frac{h}{r}\right),
\end{align}
where 
\begin{equation}
    \mathfrak{f} = \left(1+\frac{h}{r}\right)^\nu \left(1+\frac{\cos^2(\omega)q^2}{2(1+\nu)hr}-\frac{\sin^2(\omega)q^2}{2(1-\nu)h(r+h)}\right), \hspace{5mm} \nu = \frac{1-a^2}{1+a^2}.
\end{equation}
Following the method described in the previous section, calculating the four-derivative corrections to the scalar charge requires knowing the two-derivative solution away from the origin of moduli space, $\phi_\infty \neq 0$. Unlike the $a$-model, if we shift the scalar by a constant $\phi \rightarrow \phi + \phi_0$ there is no compensating transformation law of the fields $g_{\mu\nu}$ and $A_\mu$ which is a symmetry of the action (\ref{actiona1a2}). 
To construct solutions at arbitrary points in moduli space we take the following approach;
we promote the parameter $\omega$ to be a background scalar field and allow it to transform under the shift symmetry as a \textit{spurion}. The action (\ref{actiona1a2}) is invariant under
\begin{equation}
    g_{\mu\nu} \rightarrow g_{\mu\nu}, \hspace{5mm} A_\mu \rightarrow Z(-\phi_{0},\omega)^{1/2} A_\mu, \hspace{5mm} \phi \rightarrow \phi + \phi_0, \hspace{5mm} \omega \rightarrow \tilde{\omega}(\omega) ,
\end{equation}
where
\begin{equation}
  \tilde{\omega}(\omega) = \tan^{-1}\left(e^{\frac{a^2-1}{2a}\phi_{0}}\tan(\omega)\right).  
\end{equation}
Using this spurionic symmetry of the action we can construct black hole solutions at arbitrary points in moduli space\footnote{This is an application of a general solution generating mechanism: let $I[X,g]$ be an action for degrees of freedom $X$, constant parameters (background scalar fields) $g$, and suppose that this model has a spurionic symmetry $I[\tilde{X}(X,g),\tilde{g}(g)] = I[X,g]$. Let $X_g$ denote a stationary point of the action $I[X,g]$. As a consequence of the spurionic symmetry, $\tilde{X}(X_g,g)$ is a stationary point of the action $I[X,\tilde{g}(g)]$. If $\tilde{g}(g)$ is invertible then it follows that $\tilde{X}(X_{\tilde{g}^{-1}(g)},\tilde{g}^{-1}(g))$ is a \textit{new} stationary point of the original action $I[X,g]$. }. Letting $g_{\mu\nu}(\omega,\phi_\infty)$, $A_\mu(\omega,\phi_\infty)$ and $\phi(\omega,\phi_\infty)$ denote the solution with modulus $\phi_\infty$, we find that this is given by
\begin{align}
    g_{\mu\nu}(\omega,\phi_\infty) &= g_{\mu\nu}(\tilde{\omega}^{-1}(\omega),0) \, , \nonumber\\
    A_\mu(\omega,\phi_\infty) &= Z(-\phi_\infty,\tilde{\omega}^{-1}(\omega))^{1/2} A_\mu (\tilde{\omega}^{-1}(\omega),0) \, , \nonumber\\
    \phi(\omega,\phi_\infty) &= \phi(\tilde{\omega}^{-1}(\omega),0)+\phi_\infty,
\end{align}
where $g_{\mu\nu}(\omega,0)$, $A_\mu(\omega,0)$, $\phi(\omega,0)$ corresponds to the solution (\ref{a1a2solorigin}).

To simplify the analysis we will focus on a specific case, $a=1$ and $\omega = \frac{\pi}{3}$, with gauge kinetic function
\begin{equation}
    Z\left(\phi\right) = \frac{4e^{-\phi}}{1+3e^{-2\phi}} = 4e^{-\phi} -12 e^{-3\phi} +  36 e^{-5\phi} + \mathcal{O}\left(e^{-7\phi}\right),
\end{equation}
which we can interpret as a kind of toy-model of the string genus expansion. The non-rotating, electrically charged black hole solution of this model, including general moduli, takes the form
\begin{align}
    \label{a1omegapi3model}
    ds^2 &= -\mathfrak{f} dt^2 + \frac{dr^2}{\mathfrak{f}} + r (r+h) \left(d\theta^2 + \sin^2(\theta)d\varphi^2\right) \, , \nonumber\\
    A &=\frac{q \left(\frac{h}{3 e^{-2 \phi_\infty}+1}+r\right)
   \sqrt{4 \cosh (\phi_\infty)-2 \sinh (\phi_\infty)}}{2 r
   (h+r)}\,dt \, , \nonumber\\
    \phi &= \phi_\infty+\log\left(1+\frac{h}{r}\right),
\end{align}
where 
\begin{equation}
    \label{moregenblackening}
    \mathfrak{f} = \frac{e^{2 \phi_\infty} (h+r) \left(2 h r+q^2\right)+6 h r (h+r)-3
   q^2 r}{2 h r \left(e^{2 \phi_\infty}+3\right) (h+r)}.
\end{equation}
For $h>0$ and $q>0$, this solution has a physical singularity located at $r=0$, which is hidden behind an event horizon only if
\begin{equation}
    0 < h < \frac{q}{\sqrt{2}} \sqrt{1+\frac{\sqrt{3}}{\sinh(\phi_\infty)-2\cosh(\phi_\infty)}}.
\end{equation}
General expressions for the two-derivative mass, entropy, temperature, electric potential, electric charge and scalar charge of the solution are found in appendix \ref{app:nonsusythermo}. For these black holes extremality is obtained when
\begin{equation}
    h \xrightarrow[]{T=0} q\left[\frac{2 \sinh (\phi_\infty)-4 \cosh (\phi_\infty)}{\sinh (\phi_\infty)-2 \cosh (\phi_\infty)+\sqrt{3}}\right]^{-1/2}.
\end{equation}
The extremal mass and scalar charge are given by
\begin{align}
    M^{(2)}\bigr\vert_{T=0} &=  -\frac{e^{-\phi_\infty/2} \left(e^{2 \phi_\infty}-3\right)}{\sqrt{2e^{2 \phi_\infty}-4 \sqrt{3} e^{\phi_\infty}+6}}|Q|\, , \nonumber\\ 
    \mu^{(2)}\bigr\vert_{T=0} &= 2 Q \sqrt{2 \cosh \phi_\infty)-\sinh (\phi_\infty)-\sqrt{3}} \; .
\end{align}
Using these expressions we can verify the cancellation of long-range static forces at extremality at all points of moduli space,
\begin{equation}
    \lim_{r\rightarrow \infty} r^2f(r) = -(M^{(2)})^2+Z\left(\phi_\infty,\frac{\pi}{3}\right)^{-1} Q^2 - \frac{1}{4}(\mu^{(2)})^2 \xrightarrow[]{T=0} 0.
\end{equation}
We now calculate the four-derivative corrections, which we parametrize as
\begin{align}
    16\pi e^{-1}\mathcal{L}^{(4)} &= b_1 (F^2)^2 + b_2 \left(F\tilde{F}\right)^2 + b_3 FFW + b_4 R_{\text{GB}} \nonumber\\
    &\hspace{10mm}+ b_5 \left((\partial \phi)^2\right)^2 +b_6 (\partial\phi)^2 F^2 +b_7 \partial\phi \partial\phi FF.
\end{align}
\begin{figure}[t]%
    \centering
    \includegraphics[width=10cm]{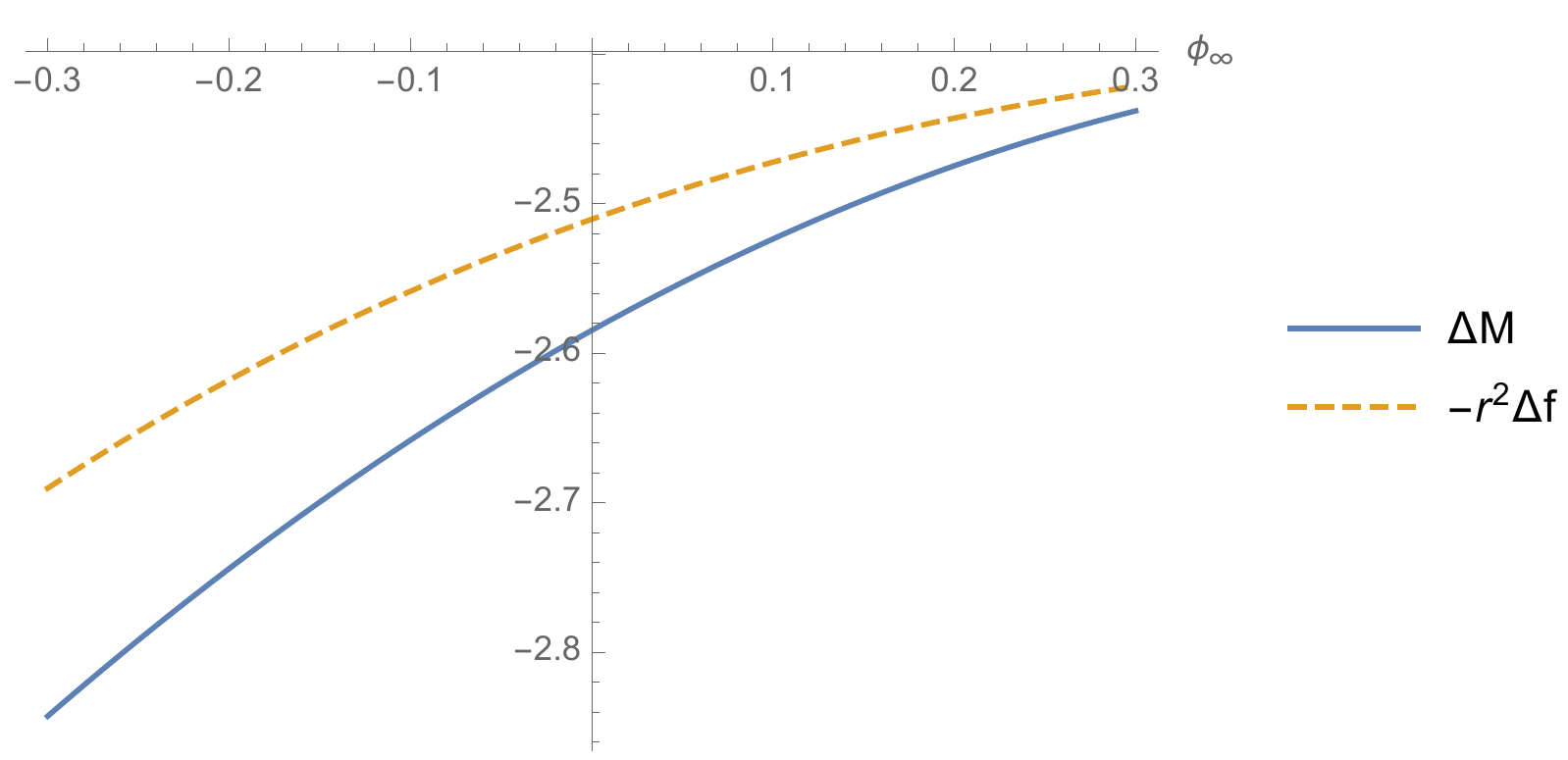}
    \caption{Corrections to the extremal mass and the (negative of the) static force for the model studied in section \ref{sec:moregeneral} on a neighbourhood of the origin of moduli space with $b_1> 0$ and $b_i=0, \hspace{2mm} i=2,...,7$. The numerical values on the vertical axis are the value of the respective quantity in Gaussian units, i.e. $G_\text{N}=1$ and $\epsilon_0=1/4\pi$. }%
    \label{fig:moregeneral}%
\end{figure}
From the four-derivative corrected free-energy we can calculate the correction to the extremal mass and scalar charge, as outlined earlier. The explicit expressions are very complicated functions of the modulus $\phi_\infty$ 
and the best we can do at this stage is to study them 
numerically point-by-point. At the origin of the moduli space we calculate 
\begin{align}
    \Delta M\Big|_{\genfrac{}{}{0pt}{1}{T=0}{\phi_\infty = 0}} &= \frac{1}{Q}\left[-2.585 b_1+0.3792 b_3-0.00003748 b_5\right.\nonumber\\
   &\hspace{15mm}\left.+0.005416 b_6+0.002708 b_7\right],
\end{align}
\begin{align}
     \Delta\mu\Big|_{\genfrac{}{}{0pt}{1}{T=0}{\phi_\infty = 0}} &=\frac{1}{Q}\left[-0.1065 b_1+0.1025 b_3-0.001072 b_5\right.\nonumber\\
   &\left.\hspace{15mm}+0.07988 b_6+0.03993 b_7\right],
\end{align}
\begin{align}
\lim_{r\rightarrow \infty} r^2 \Delta f(r)\Big|_{\genfrac{}{}{0pt}{1}{T=0}{\phi_\infty = 0}} &= 2.511 b_1-0.3796 b_3+0.0001750
   b_5\nonumber\\
   &\hspace{10mm}-0.01557 b_6-0.007785 b_7.
\end{align}
Unlike in the $a$-model calculation, for this purely electric black hole solution we find that the correction to the mass and static self-force are controlled by totally independent linear combinations of Wilson coefficients. Furthermore, as can be seen in figure \ref{fig:moregeneral}, for fixed values of the Wilson coefficients these quantities are controlled by unrelated functions on the moduli space. This example illustrates that, for generic gauge kinetic functions, the electric version of the mild WGC and mild RFC are distinct conjectures.

\section{Higher-Derivative Corrections in 5d \texorpdfstring{$\mathcal{N} = 2$}{N=2} Supergravity}
\label{SectionN2SUGRA}

The next system we analyze is the dimensional reduction of $\mathcal{N} = 2$ pure supergravity from five dimensions to four. We previously analyzed the higher-derivative corrections to the mass-to-charge ratio of a number of black hole solutions for this system  and the implications for the WGC in \cite{Cremonini:2020smy}. 
This construction has the advantage that the four-dimensional solutions allow for four different charges, as we  describe in more detail below.

The leading order Lagrangian for the bosonic fields of 5d $\mathcal N=2$ supergravity takes the form
\begin{equation}
    16\pi\hat{e}^{-1}\hat{\mathcal{L}}^{(2)} =  
    \hat{R} - \frac{1}{4} \hat{F}^2 + \frac{1}{12 \sqrt{3}}
    \epsilon^{\mu\nu\rho\sigma\lambda}\hat{A}_\mu\hat{F}_{\nu\rho}\hat{F}_{\sigma\lambda},
    \label{eq:2derLag}
\end{equation}
where hats are used to denote five-dimensional quantities.  The Chern-Simons coupling $\hat A\wedge\hat F\wedge\hat F$ is necessary for this action to be supersymmetric.  However, this term may vanish for certain black hole solutions.  Such black holes can also be viewed as solutions to pure 5D Einstein-Maxwell theory.

We are interested in four-derivative corrections to this theory. The most general set of such terms (up to reparametrizations) is given by
\begin{equation}
\label{finalL}
    16\pi\hat{e}^{-1}\hat{\mathcal{L}}^{(4)}  =
c_1 \hat{R}_{GB} +
c_2 \, \hat{W}_{\mu\nu\rho\lambda} \hat{F}^{\mu\nu} \hat{F}^{\rho\lambda}  + c_3 (\hat{F}^2)^2+ c_4 \hat{F}^4  +  c_5 \, \epsilon^{\mu\nu\rho\lambda\sigma} \hat{A}_\mu \hat{R}_{\nu\rho\delta\gamma} \hat{R}_{\lambda \sigma}{}^{\delta \gamma},
\end{equation}
where $\hat{R}_{GB}=\hat{R}_{\mu \nu \rho \sigma} \hat{R}^{\mu \nu \rho \sigma} - 4 \hat{R}_{\mu \nu} \hat{R}^{\mu \nu} + \hat{R}^2$ is the Gauss-Bonnet combination, $\hat{W}$ is the Weyl tensor, $ (\hat{F}^2)^2 = (\hat{F}_{\mu \nu} \hat{F}^{\mu \nu})^2$ and $\hat{F}^4 = \hat{F}_{\mu \nu} \hat{F}^{\nu \rho} \hat{F}_{\rho \sigma} \hat{F}^{\sigma \mu}$.  
The coefficients $c_i$ are assumed to be perturbatively small, and taken to be arbitrary.  In particular, we allow supersymmetry to be broken in the effective theory.
However, it is sometimes useful to consider unbroken SUSY. In this case, the coefficients take the values \cite{Cremonini:2008tw}
\begin{align}
    c \equiv c_1 = -2 c_2 = -6 c_3 = \frac{24}{11} c_4 = 2 \sqrt 3 c_5 \, .
\end{align}
In all the examples we examine in this paper, we find that the mass shifts due to the higher-derivative terms -- and hence the scalar charge shifts -- vanish\footnote{Actually, for non-zero values of $\chi_\infty$ we don't find a vanishing mass shift, which is still a puzzle for us. 
There may be additional boundary terms affecting the free-energy, and hence the mass, which only vanish when the moduli are turned off. This issue deserves further study but is beyond the scope of this paper.} when we choose these values and set $\phi^i_\infty \to 0$.

\subsection{Reduction Ansatz and Shift Symmetries}

Although our starting point is 5d $\mathcal N=2$ supergravity, we are mainly interested in four-dimensional black holes.  We thus compactify on a circle and reduce to 4d $\mathcal N=2$ supergravity coupled to one vector multiplet.  This reduction can be understood from the perspective of the 5d STU model compactified to the 4d STU model. Some pertinent information on the 4d STU model and its lift to five dimensions can be found in appendix~\ref{app:STU}.

The reduction of the five-dimensional action (\ref{eq:2derLag}) proceeds according to the 
ansatz
\begin{align}
d\hat s_5^2&=e^{\phi/ \sqrt{3}}g_{\mu\nu}dx^\mu dx^\nu+e^{-2 \phi / \sqrt{3} }(dz+\mathcal A)^2, \qquad 
\hat A=A_\mu dx^\mu+\chi(dz+\mathcal A) \, ,
\end{align}
where $g_{\mu\nu}$ is the 4d metric, the \textit{photon} $A_\mu$ and the \textit{graviphoton} $\mathcal{A}_\mu$ are 4d gauge fields, and the \textit{dilaton} $\phi$ and the \textit{axion} $\chi$ are a scalar and pseudoscalar, respectively. The result of the dimensional reduction is the 4d effective Lagrangian
\begin{align}
\begin{split}
        &  16\pi e^{-1} \mathcal{L}^{(2)} =  R \,   -\frac{1}{2} (\nabla \phi)^2  - \frac{1}{2} e^{2 \phi/ \sqrt{3}} (\nabla \chi)^2 - \frac{1}{4} e^{-\sqrt{3} \phi } \, G^2 -  \frac{1}{4} e^{- \phi / \sqrt{3}} (F + \chi G)^2 \\
        &\kern10em + \frac{1}{4 \sqrt 3} \epsilon^{\mu\nu\rho\sigma}\chi\left(F_{\mu\nu}F_{\rho\sigma}+\chi G_{\mu\nu}F_{\rho\sigma}+\frac{1}{3}\chi^2G_{\mu\nu}G_{\rho\sigma}\right) ,
        \label{eq:ReducedAction2der}
\end{split}
\end{align}
where $G \equiv d\mathcal A$ and $F \equiv dA$. The conserved electric and magnetic charges associated with $G$ will be labelled by $Q_0$ and $P_0$ respectively, and the charges associated with $F$ will be labelled by $Q_1$ and $P_1$.
In five-dimensions, $Q_0$ may be viewed as a boost parameter along the compactified direction, and $P_1$ is a topological charge equal to the NUT charge of a spacelike slice (see \cite{Cremonini:2020smy} for a more thorough discussion). This 4d action corresponds to the action for the bosonic fields of the 4d STU model with the three vector multiplets identified with each other.

At this point, it is worth noting that (\ref{eq:ReducedAction2der}) is invariant under shifts in $\phi$ and $\chi$, provided that the remaining fields are rescaled appropriately.  For the $\phi$ shift, we have
\begin{align}
    \phi \to \phi + \phi_0, \qquad \chi \to e^{- \phi_0 / \sqrt 3} \chi \, ,\qquad G \to e^{\sqrt 3 \phi_0 / 2} G \, , \qquad F \to e^{ \phi_0 / 2 \sqrt 3} F \, .
\label{eq:phishift}
\end{align}
From the five-dimensional point of view, this shift symmetry is realized by a general coordinate transformation
\begin{equation}
    x^\mu\to e^{-\phi_0/2\sqrt3}x^\mu,\qquad z\to e^{\phi_0/\sqrt3}z,
\end{equation}
along with the shifts (\ref{eq:phishift}).
The rescalings of the gauge fields are equivalent to a rescaling of the electric and magnetic charges,
\begin{equation}
\label{chargescalingsSec4}
Q_0 \rightarrow e^{-\sqrt 3 \phi_0 / 2} Q_0 \, , \quad 
P_0 \rightarrow e^{\sqrt 3 \phi_0 / 2} P_0\, , \quad 
    Q_1 \rightarrow e^{- \phi_0 / 2\sqrt 3} Q_1 \, , \quad P_1 \rightarrow  e^{ \phi_0 / 2\sqrt 3} P_1 \, .
\end{equation}
Invariance under the axion shift is given by
\begin{equation}
    \phi\to\phi,\qquad\chi\to\chi+\chi_0,\qquad G\to G,\qquad F\to F-\chi_0 G,
\label{eq:chishift}
\end{equation}
with corresponding charge shifts
\begin{align}
\label{axionchargescalingsSec4}
    Q_0&\to Q_0+\chi_0Q_1+\frac{1}{\sqrt3}\chi_0^2P_1-\fft1{3\sqrt3}\chi_0^3P_0,& P_0&\to P_0,\nn\\
    Q_1&\to Q_1+\fft2{\sqrt3}\chi_0P_1-\fft1{\sqrt3}\chi_0^2P_0,& P_1&\to P_1- \chi_0P_0.
\end{align}
In five dimensions, the axion shift is simply a gauge transformation
\begin{equation}
    \hat A\to\hat A+d(\chi_0z).
\end{equation}

Since the dilaton and axion shift symmetries correspond to symmetries of the five-dimensional action, so long as we work with covariant and gauge invariant higher derivative corrections, (\ref{finalL}), they will be preserved by the higher derivative four-dimensional action.  As a result, according to what we saw in section \ref{sec:forces}, these shift symmetries can be used for a simple calculation of the scalar charges in the theory.
In particular, for the four-charge solutions of this section, using (\ref{chargescalingsSec4}) with $\phi_0 = - \phi_\infty $ and denoting the black hole mass by $M$, the derivative formula (\ref{dcharge}) for the dilatonic charge becomes 
\begin{eqnarray}
\mu_\phi& =& -4 
\left[ \frac{\partial M}{\partial Q_0} \frac{\partial Q_0}{\partial \phi_\infty}+ \frac{\partial M}{\partial P_0} \frac{\partial P_0}{\partial \phi_\infty} +\frac{\partial M}{\partial Q_1} \frac{\partial Q_1}{\partial \phi_\infty} +\frac{\partial M}{\partial P_1} \frac{\partial P_1}{\partial \phi_\infty}  \right]_{\phi_\infty=0}  \nn \\
& =& 
 - 2 \sqrt{3} \, \frac{\partial M}{\partial Q_0} Q_0  +
2 \sqrt{3}  \,  \frac{\partial M}{\partial P_0} P_0 
- \frac{2}{\sqrt{3}}\frac{\partial M}{\partial Q_1} Q_1 
+ \frac{2}{\sqrt{3}} \frac{\partial M}{\partial P_1}P_1   \, .
\label{4Qcharge}
\end{eqnarray}
Similarly, using (\ref{axionchargescalingsSec4}) with $\chi_0=-\chi_\infty$, 
the derivative formula for the axionic charge is
\begin{eqnarray}
\mu_\chi& =& -4 
\left[ \frac{\partial M}{\partial Q_0} \frac{\partial Q_0}{\partial \chi_\infty}+ \frac{\partial M}{\partial P_0} \frac{\partial P_0}{\partial \chi_\infty} +\frac{\partial M}{\partial Q_1} \frac{\partial Q_1}{\partial \chi_\infty} +\frac{\partial M}{\partial P_1} \frac{\partial P_1}{\partial \chi_\infty}  \right]_{\chi_\infty=0} \nn \\
& =& 
  4  \frac{\partial M}{\partial Q_0} Q_1   
+ \frac{8}{\sqrt{3}} \frac{\partial M}{\partial Q_1} P_1  -4 \frac{\partial M}{\partial P_1}P_0   \, ,
\label{4Qaxioncharge}
\end{eqnarray}
which generalizes the simpler axion example we presented in section \ref{sec:forces} to the 
theory (\ref{eq:ReducedAction2der}).

Solutions to (\ref{eq:ReducedAction2der}) correspond to solutions of the five-dimensional theory that are locally of the form $\mathbb{R}^{3,1} \times S^1$.  In general, it is an assumption that these apparently four-dimensional solutions imply constraints on the five-dimensional solutions from which they arise. 
In our case, however, we do not really need this assumption -- sometimes called background independence -- as we are not trying to constrain the 5d theory so much as to simply understand the relation between the WGC and the RFC for different theories.

Solutions to this theory which include all four charges $(Q_0,Q_1,P_0,P_1)$ are known. However, there is a special class of two-charge solutions which are considerably simpler because they do not include sources for the axion $\chi$, which may then be set to zero. These include black hole solutions with $Q_0$ and $P_0$, solutions with $Q_0$ and $P_1$, and those with $Q_1$ and $P_0$.  The $(Q_0,P_0)$ black holes are always non-BPS, while the $(Q_0,P_1)$ and $(Q_1,P_0)$ solutions have BPS limits.  Here we compute the higher-derivative contributions to the scalar charge of such black holes using both of the methods outlined in section \ref{sec:forces}. We also analyze solutions with non-vanishing $Q_1$ and $P_1$ charges, which are interesting because they include a non-trivial axion profile in addition to the dilaton. However, we only compute their scalar charge using the derivative formula
(\ref{4Qaxioncharge}),
%of section \ref{sec:forces}, 
due to their increased complexity. For a more detailed account of this dimensional reduction, including the reduction of the four-derivative terms, see appendix B.

Recall that the
dilaton and axion shift symmetries (\ref{eq:phishift}) and (\ref{eq:chishift}) make it possible to relate our results at different locations at moduli space. 
Thus, for simplicity we consider the origin of moduli space, $\phi_\infty=0$ and $\chi_\infty=0$.  In this case, the general force between two black holes with mass $M$, charges $Q_0$, $P_0$, $Q_1$ and $P_1$, and scalar charges $\mu_\phi$ and $\mu_\chi$ given in (\ref{eq:total_force}), reduces to 
\begin{align}
    \lim_{r\rightarrow \infty} r^2f(r) = -M^2+(Q_0^2+Q_1^2+P_0^2+P_1^2)-\fft14(\mu_\phi^2+\mu_\chi^2) \, .
    \label{eq:Neq2force}
\end{align}
In the uncorrected, two-derivative theory, the force between extremal black holes always vanishes, but in theories with higher-derivative corrections this is not the case. The aim of this section is to analyze the corrections to the long-range force by computing the 
shifts
% corrections 
$\Delta M$, $\Delta \mu_\phi$, and $\Delta \mu_\chi$ for generic values of the four-derivative coefficients $c_i$.

The result of this analysis will be that there are black holes which are superextremal ($\Delta M < 0$) but self-attractive $(f <0)$. We will give  specific examples of values of EFT coefficients where this is the case. The black holes with $Q_0$ and $P_0$ charge
provide an especially striking example, because it is possible to choose the corrections such that the mass shift is always negative, but there is no way to choose them such that the force shift is sign definite.

\subsection{\texorpdfstring{$Q_0$, $P_0$}{Q0, P0} black hole}
\label{sec:q0p0}

The first solution we consider has non-zero values for $Q_0$ and $P_0$. This means setting the photon $A = 0$, so it corresponds to the Kaluza-Klein reduction of pure Einstein gravity in 5d. The solutions include the so-called Kaluza-Klein black holes and are always non-BPS. Static Kaluza-Klein black holes carrying $Q_0$ and $P_0$ charges were constructed in \cite{Gibbons:1985ac}, and rotating ones in \cite{Rasheed:1995zv,Matos:1996km,Larsen:1999pp} (see also \cite{Horowitz:2011cq}).  Following the notation of \cite{Larsen:1999pp}, the solution (with $\phi_\infty=0$) is 
\begin{align}
    & ds^2 \, = \, - \frac{\Delta}{\sqrt{H_1 H_2}} \, dt^2 + \frac{\sqrt{H_1 H_2}}{\Delta} \, dr^2 + \sqrt{H_1 H_2} \, d \Omega_{S^2}^2 \, ,\nn \\
    & \mathcal{A} = -  \, Q_0 \,(2r+ p -  2m) \, H^{-1}_2 \, dt - 2 P_0 \, \cos \theta \, d \varphi  \, ,\nn \\
    & e^{-2\phi/\sqrt{3}} = \frac{H_2}{H_1} \, ,
\end{align}
with the blackening function and harmonic functions given by
\begin{align}
    \begin{split}
        & \Delta = r^2 - 2 m r \, , \\
        & H_1 = r^2 + r(p - 2 m) + \frac{p (p - 2m)(q - 2 m)}{2 (p + q)}\, , \\
        & H_2 = r^2 + r(q - 2 m) + \frac{q (p - 2m)(q - 2 m)}{2 (p + q)}\, . \\
    \end{split}
\end{align}
These solutions are described by three parameters, $m$, $p$, and $q$. The physical charges are determined by these solution parameters and are given by:
\begin{align}
\begin{split}
    M^{(2)} \ &= \ \frac{p + q}{4} \, , \\
    Q_0^2 \ &= \frac{q(q^2 - 4 m^2)}{4(p + q)}\, , \\
    P_0^2 \ &= \frac{p (p^2 - 4 m^2)}{4(p + q)} \, ,\\
    \mu_\phi^{(2)} \ &= \ \frac{\sqrt{3}(p-q)}{2} \, .
\end{split}
\end{align}
The outer and inner horizons are located at $r=2m$ and $0$. Purely electric black holes are given by setting $p=2m$ and purely magnetic ones by $q=2m$. Extremality corresponds to taking the $m \to 0$ limit, where the temperature is zero. It can be easily verified that the force vanishes in this limit.

\begin{figure}[t]%
    \centering
    \subfloat[\centering]{{\includegraphics[width=7cm]{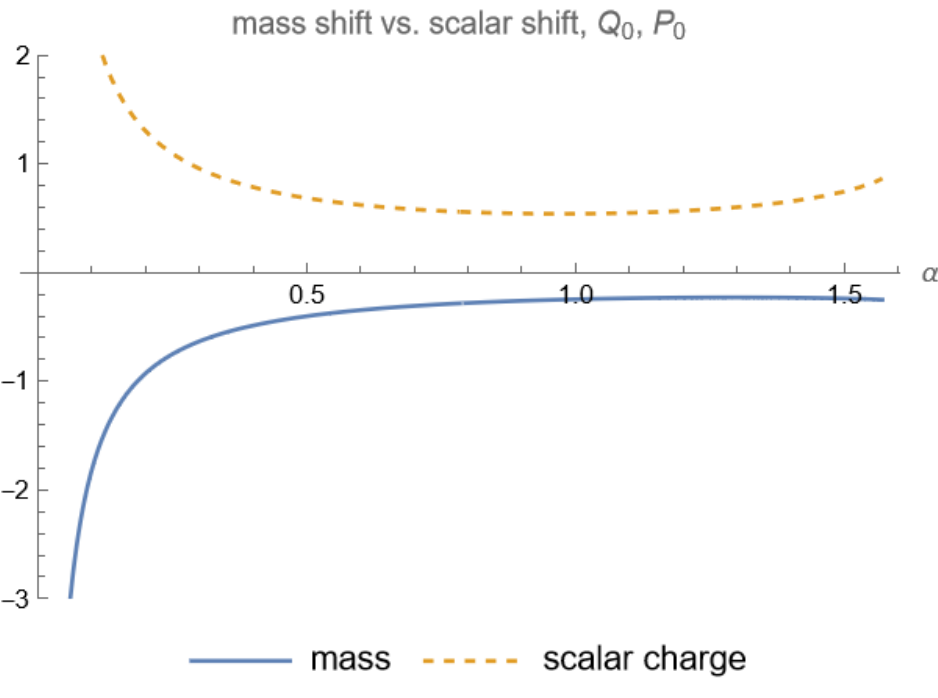} }}%
    \qquad
    \subfloat[\centering ]{{\includegraphics[width=7cm]{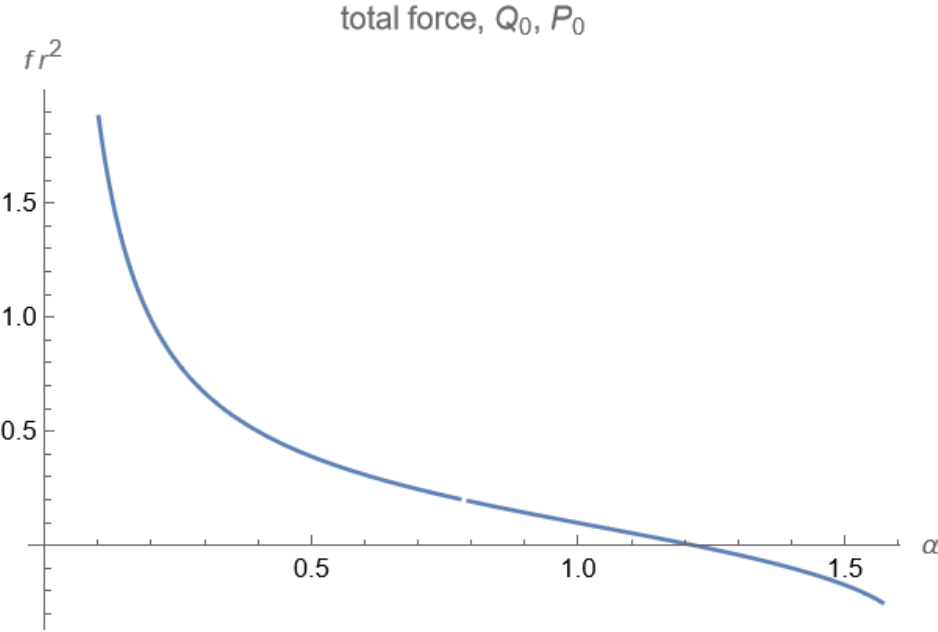} }}%
    \caption{Mass, scalar charge, and force shifts for the $Q_0$, $P_0$ black holes. We have set $p = \sin \alpha$, $q = \cos \alpha$, and chosen $c_1 = 1$. The force changes sign at $\alpha \simeq 1.21895$.}%
    \label{fig:Q0P0_shifts}%
\end{figure}

The shifts to the scalar charge may be computed using the methods outlined in section \ref{sec:forces}. For the explicit expressions, see appendix (\ref{app:Q0P0}). The result is that the mass shift and scalar charge shift are both sign-definite (figure \ref{fig:Q0P0_shifts}, a), but the force shift is not (figure \ref{fig:Q0P0_shifts}, b).
For $c_1 > 0$ the mass shift is always negative, implying that the mild form of the WGC is always satisfied. On the other hand, the scalar shift is always positive. The force shift has a more complicated behavior that originates from the fact that the leading-order scalar charge goes like $\mu_\phi \sim (p - q)$ and can be positive or negative. 
%%%%%%
It is convenient to parametrize $p$ and $q$ in terms of an angle $\alpha$, so that $\tan\alpha = p/q$, as shown in figure \ref{fig:Q0P0_shifts}. 
%%%%%%
For small $\alpha$, $\mu_\phi$ is negative and $M$ is positive. In this case, the force shift
\begin{equation}
    \lim_{r\rightarrow \infty} r^2 \Delta f(r) =  - 2 M^{(2)} \Delta M - \frac{1}{2} \mu_\phi^{(2)} \Delta \mu_\phi,
\end{equation}
is a sum of two positive terms. As $\alpha$ grows, we eventually have $p>q$. This flips the sign of the scalar contribution to the force, and eventually overwhelms the contribution from the mass. As a result, the force between the KK black holes may have either sign, depending on the charge. Thus, there are directions in charge space which don't support black hole solutions that are self-repulsive.

\subsection{\texorpdfstring{$Q_0$, $P_1$}{Q0, P1} Solutions}
\label{sec:p1q0}

We now turn to the $(Q_0,P_1)$ and $(Q_1,P_0)$ solutions which have BPS limits as three-equal-charge STU black holes, as reviewed in appendix~\ref{app:STU}.  We first consider the $(Q_0,P_1)$ case, corresponding to electric charge $Q_0$ for the graviphoton $\mathcal{A}$ and magnetic charge $P_1$ for the photon $A$. The general non-extremal solution (with $\phi_\infty=0$) takes the form
\begin{align}
    & ds^2 = - (H_0H_1^3)^{-1/2}f\, dt^2 + (H_0H_1^3)^{1/2}\left(\fft{dr^2}f+r^2d \Omega_{S^2}^2\right),\nn\\
    &\mathcal{A} = \left(1-\fft1{H_0}\right)\coth\beta_0dt,\quad
    e^{-2 \phi/\sqrt3} = \fft{H_0}{H_1}\,,\nn\\
    &A =\sqrt3m\sinh\beta_1\cosh\beta_1 \cos\theta d\varphi\,,
\label{eq:p1q0bh}
\end{align}
with
\begin{equation}
    H_0 =1+\fft{m\sinh^2\beta_0}r,\quad
    H_1 =1+\frac{m\sinh^2\beta_1}{r}, \quad
    f = 1-\frac{m}{r}\, .
\label{eq:H0H1f}
\end{equation}
The horizon is located at $r=m$. The charges and thermodynamic quantities of these solutions are given in terms of the solution parameters by 
\begin{align}
\begin{split}
    &M^{(2)} = \frac{m}{8}\,  \left( \cosh 2 \beta_0 + 3\cosh 2 \beta_1 \right) \, , \\
    &Q_0 = \frac{m}{4}  \sinh 2\beta_0  \, , \\
    &P_1 = \frac{\sqrt{3}m}{4}
    %\frac{m}{\sqrt 3} 
    \sinh 2 \beta_1 \,, \\
    &\mu_\phi^{(2)}= \frac{\sqrt{3}m }{4}\,  (\cosh 2 \beta_1 - \cosh 2 \beta_0) \, ,
\end{split}
\label{eq:MQP_P1Q0}
\end{align}
where the dilatonic scalar charge $\mu_\phi$ can be read off from the asymptotic expansion of the scalar.
The extremal limit is taken by letting $m\to0$ and $\beta_i\to\infty$ with $Q_0$ and $P_1$ fixed.  In this limit, $ T \to 0$ and we find 
\begin{equation}
    M^{(2)}\Big|_{T=0}=\fft12 |Q_0|+ \frac{\sqrt3}{2} |P_1|,\qquad \mu_\phi^{(2)}\Big|_{T=0} =  |P_1| - \sqrt 3 |Q_0|.
    \label{eq:MSTBPS_P1Q0}
\end{equation}
It is easy to verify that the leading order scalar charge at extremality is consistent with the derivative formula
(\ref{4Qcharge}).
% we introduced in Section 2.

The black hole is BPS when the signs of the charges are chosen appropriately; otherwise it is extremal but non-BPS. 
The higher-derivative terms introduce a mass shift \cite{Cremonini:2020smy} 
\begin{align}
    \Delta M \Big|_{T=0}  = -\frac{3 \sqrt3 (2 c_1 + 7 c_2 + 24 c_3 + 12 c_4)}{80 |P_1|} \, ,
\end{align}
and a scalar charge shift:
\begin{align}
    \Delta \mu_\phi \Big|_{T=0} & = \frac{3 (2 c_1 + 7 c_2 + 24 c_3 + 12 c_4)}{40 |P_1|}\,.
\end{align}
In this case, the scalar shift is proportional to the mass shift. 
This can be seen easily from
the derivative formula (\ref{4Qcharge}),
%of section \ref{sec:forces}, 
which here takes the simple form.
\begin{align}
    \Delta \mu_\phi =  \frac{2}{\sqrt3} |P_1| \frac{d \Delta M}{d P_1} \, ,
    % - 2 \sqrt{3} |Q_0| \frac{d \Delta M}{d Q_0} \, .
\end{align}
since the mass shift only depends on $P_1$, and not $Q_0$. This is in contrast to what we will find for the $Q_0$, $P_1$ black holes.
Finally, the corrected force is given by
\begin{equation}
\begin{split}
    \lim_{r\rightarrow \infty} r^2 \Delta f(r) \Big|_{T=0} &=  \frac{3 (|P_1| + \sqrt3 |Q_0|)}{40 |P_1|} (2 c_1 + 7 c_2 + 24 c_3 + 12 c_4) \\
    &= - \frac{2}{\sqrt3} (|P_1| + \sqrt3 |Q_0|) (\Delta M)_{T=0}\, .
\end{split}
\end{equation}
Since it is proportional to the shifted mass, we see
that any combination of the coefficients $c_i$ which makes the mass shift negative will also make the force shift positive. Thus, superextremal states are always self-repulsive.

\subsection{\texorpdfstring{$P_0$, $Q_1$}{P0, Q1} Solutions}
\label{sec:p0q1}

The solution with non-zero $Q_1$ and $P_0$ can be obtained by taking the overall electric/magnetic dual of the $(Q_0,P_1)$ solution given above.  To be explicit, the solution (with $\phi_\infty=0$) takes the form
\begin{align}
    & ds^2=-(H_0H_1^3)^{-1/2}f\,dt^2+(H_0H_1^3)^{1/2}\left(\fft{dr^2}f+r^2d\Omega_{S^2}^2\right),\nn\\
    & \mathcal A=m \sinh\beta_0\cosh\beta_0\cos\theta d\varphi,\qquad
    e^{-2 \phi/\sqrt3}=\frac{H_1}{H_0}\,,\nn\\
    & A=\sqrt3\left(1-\fft1{H_1}\right)\coth\beta_1dt\,,
\label{eq:p0q1bh}
\end{align}
with the functions $H_0$, $H_1$ and $f$ defined in (\ref{eq:H0H1f}).  Note that duality flips the sign of the dilaton, which
implies that the lifted 5d metrics of the $(Q_0,P_1)$ and $(P_0,Q_1)$ systems are not identical.

From the 4d point of view, the conserved charges and thermodynamic quantities are given by (\ref{eq:MQP_P1Q0}), (\ref{eq:TandS}) and (\ref{eq:MSTBPS_P1Q0}) with the duality replacement $Q_0\to P_0$ and $P_1\to Q_1$, and $\phi \to -\phi$. The result is
\begin{align}
\begin{split}
    &M^{(2)} = \frac{m}{8} \,  \left( \cosh 2 \beta_0 + 3\cosh 2 \beta_1 \right) \, , \\
    &P_0 = \frac{m}{4} \sinh 2 \beta_0  \, , \\
    &Q_1 = \frac{\sqrt3 m}{4}  \sinh 2 \beta_1\, ,\\
    &\mu_\phi^{(2)}= \frac{\sqrt{3}m}{4} \,  (\cosh 2 \beta_0 - \cosh 2 \beta_1) \, .
\end{split}
\label{eq:MQP_P0Q1}
\end{align}
The extremal limit is taken by letting $m\to0$ and $\beta_i\to\infty$ with $Q_1$ and $P_0$ fixed.  In this limit, $ T \to 0$ and we find
\begin{equation}
    M^{(2)}\Big|_{T=0}=\fft12 |P_0|+\frac{\sqrt3}{2}|Q_1|,\qquad \mu^{(2)}_\phi\Big|_{T=0} = \sqrt3 |P_0| - |Q_1|.
\label{eq:MSTBPS_P0Q1}
\end{equation}
In this limit, the solution reduces to the BPS case given in appendix~\ref{app:STU}.  The scalar shift can be obtained by a direct computation or using the derivative formula. 
In this case, the latter takes the form
\begin{align}
    \Delta \mu_\phi = -\frac{2}{\sqrt3} Q_1 \frac{d \Delta M}{d Q_1} + 2 \sqrt{3} P_0 \frac{d \Delta M}{d P_0} \, .
\end{align}
The result (see appendix \ref{app:Q1P0}) is a very non-trivial expression. It is interesting that the scalar shift for this solution takes a much more complicated form than that of the $Q_0$, $P_1$ solution, even though the two solutions are related by a duality transform. This traces back to the
complexity of the mass shift \cite{Cremonini:2020smy} for this solution. Ultimately, our set of higher-derivative corrections breaks duality invariance so the shifts to thermodynamic quantities need not be related by duality transformations. 

\begin{figure}[t]%
    \centering
    \subfloat[\centering]{{\includegraphics[width=7cm]{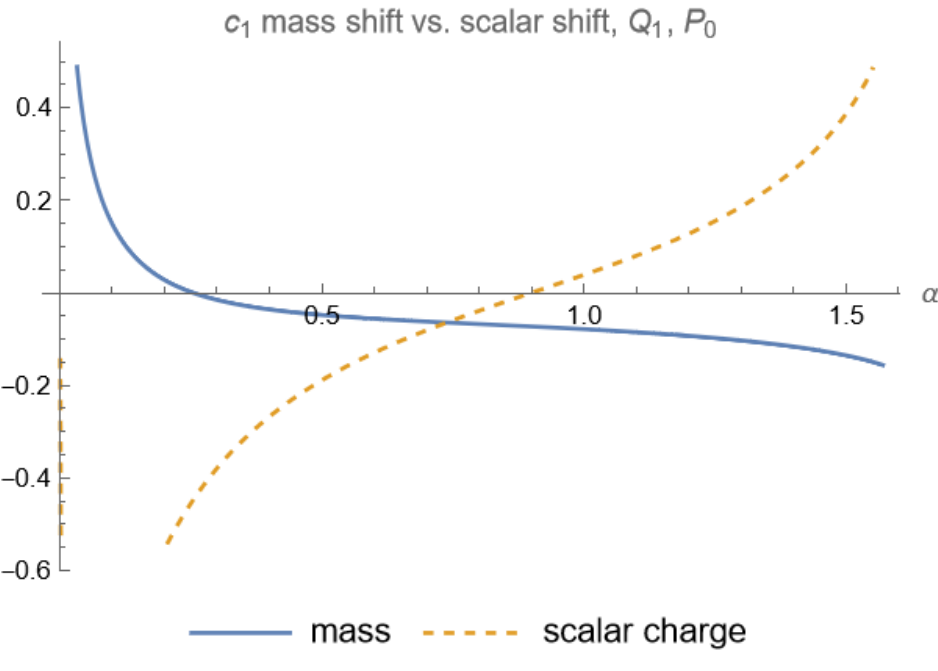} }}%
    \qquad
    \subfloat[\centering ]{{\includegraphics[width=7cm]{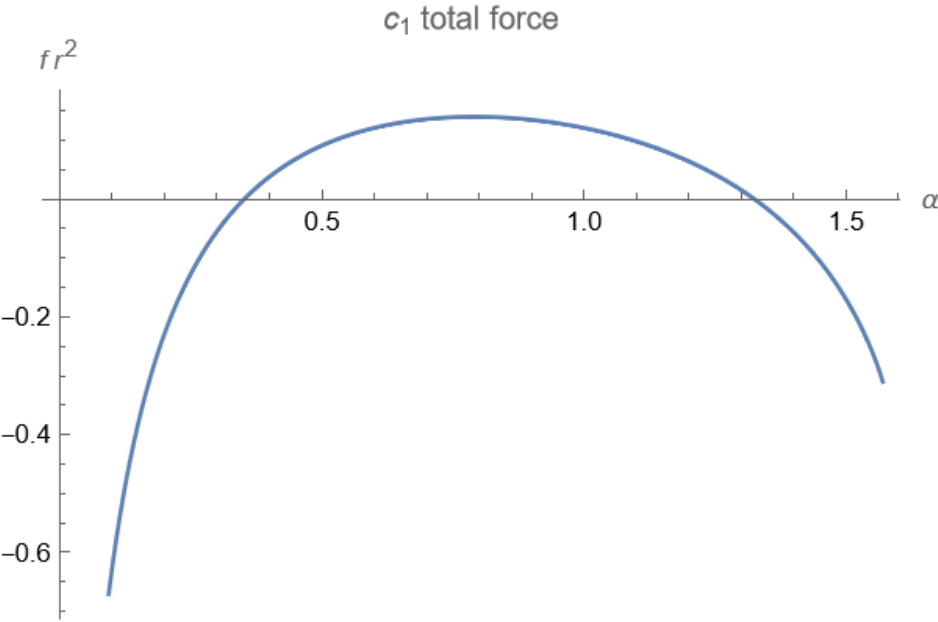} }}%
    \caption{Mass, scalar charge, and force shifts for the $Q_1$, $P_0$ black holes. We have set $P_0 = \sin \alpha$, $Q_0 = \cos \alpha$, and chosen $c_1 = 1$ and all other $c_i = 0$. We have used a ListPlot[] to reduce artifacts near where $P_0 = Q_1$.}%
    \label{fig:Q1P0_shiftsc1}%
\end{figure}

\begin{figure}[t]%
    \centering
    \subfloat[\centering]{{\includegraphics[width=7cm]{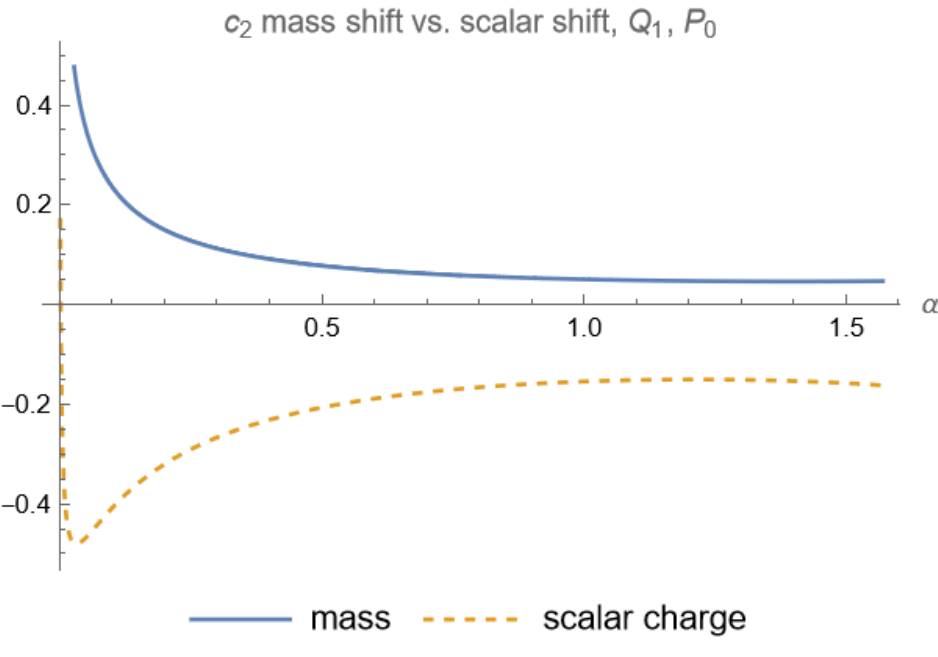} }}%
    \qquad
    \subfloat[\centering ]{{\includegraphics[width=7cm]{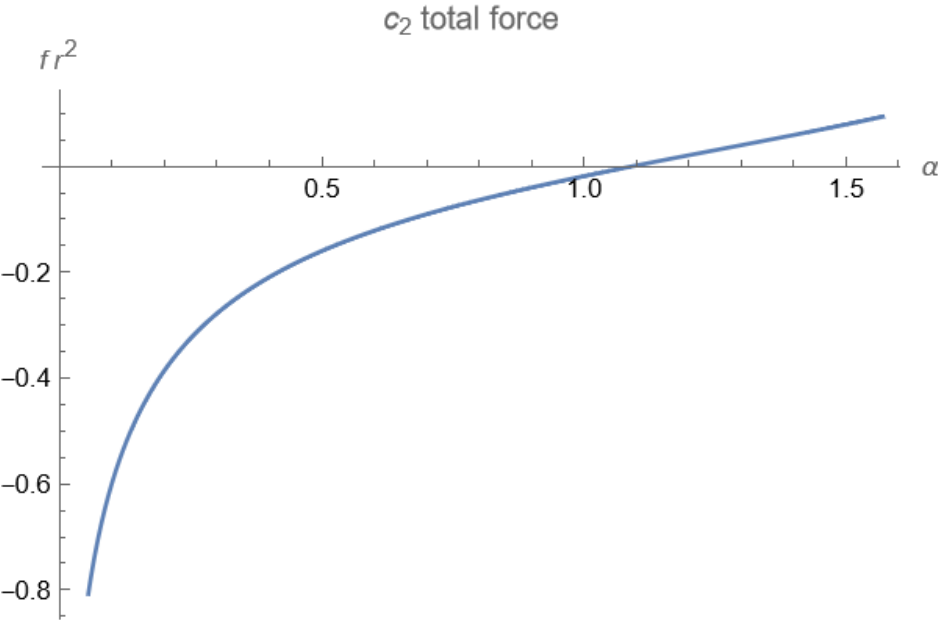} }}%
    \caption{Shifts for $Q_1$, $P_0$ black holes with $c_2 = 1$ and all other $c_i = 0$.}%
    \label{fig:Q1P0_shiftsc2}%
\end{figure}

\begin{figure}[t]%
    \centering
    \subfloat[\centering]{{\includegraphics[width=7cm]{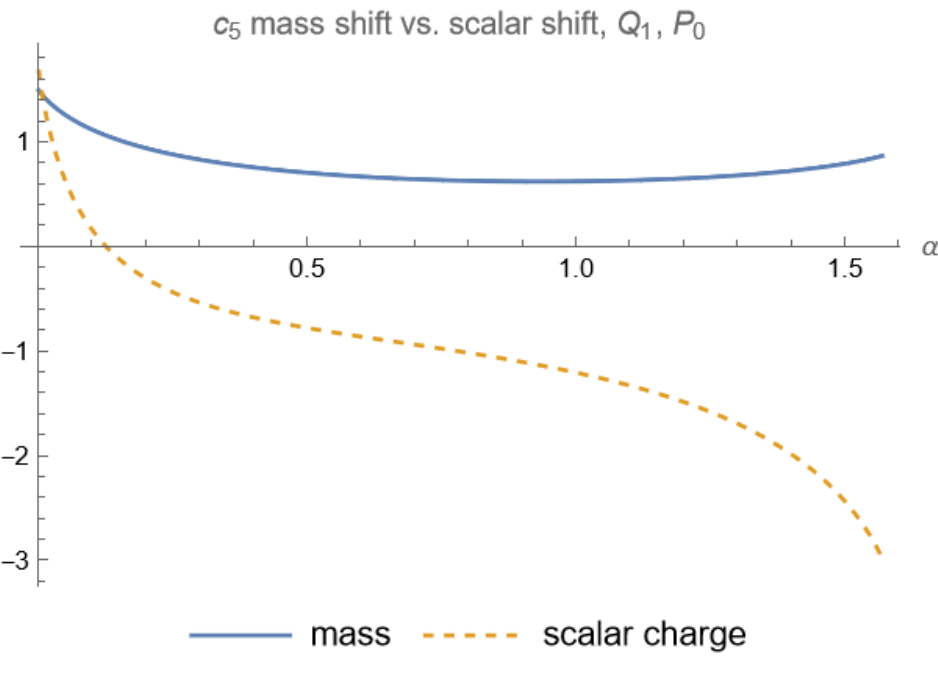} }}%
    \qquad
    \subfloat[\centering ]{{\includegraphics[width=7cm]{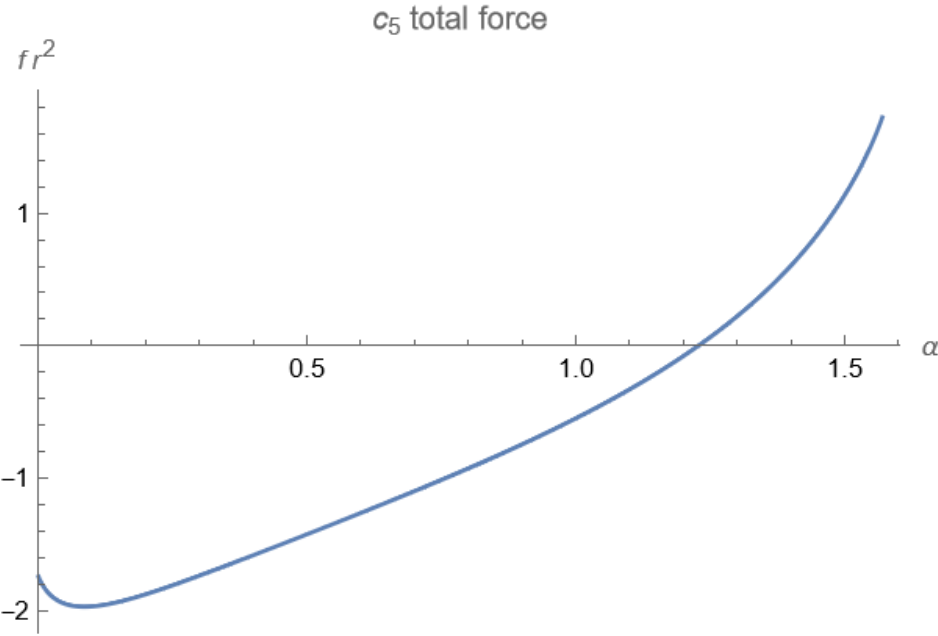} }}%
    \caption{Shifts for $Q_1$, $P_0$ black holes with $c_5 = 1$ and all other $c_i = 0$.}%
    \label{fig:Q1P0_shiftsc3}%
\end{figure}

We have plotted the contribution to the mass shift, scalar shift, and force for $c_1$, $c_2$, and $c_5$ ($c_3$ and $c_4$ give a contribution proportional to $c_2$). The results are shown in figures \ref{fig:Q1P0_shiftsc1}, \ref{fig:Q1P0_shiftsc2}, and \ref{fig:Q1P0_shiftsc3}. 
Note that all of the coefficients except for $c_1$ give sign-definite contributions to the mass shift. In every case, the force is not sign definite. 

\subsection{\texorpdfstring{$Q_1$, $P_1$}{Q1, P1} Solutions, and the axion shift}
\label{sec:p1q1}

The final two-charge solutions we consider are those with non-vanishing $Q_1$ and $P_1$ charges\footnote{As noted in (\ref{axionchargescalingsSec4}), the relevant shift symmetry to compute the axionic scalar charge for these solutions involves $Q_0$ as well, so the starting point will be black holes with non-vanishing $Q_0,Q_1,P_1$ charges.}. These are of special interest because they require a non-trivial axion profile $\chi$, which in turn introduces another possible scalar contribution to the long-range forces.
However, the presence of the axion makes the solutions significantly more complicated. As such, we will not consider the fully general finite temperature black holes, but rather work \textit{at extremality}, which greatly simplifies the analysis.

The precise form of the $Q_1, P_1$ solutions is discussed in appendix \ref{SingleVectorTruncation}, where we examine the truncation of the STU model to a single vector multiplet. 
Recall that in the presence of scalar VEVs,
the solutions are described in terms of the four parameters $(p^0_\infty,p^1_\infty,q_{\infty \, 0},q_{\infty \, 1})$ appearing in (\ref{allparameters}), of which only two are independent (due to the symplectic constraint and the requirement that asymptotically one recovers flat space). We use these constraints to eliminate $(q_{\infty \, 0},q_{\infty \, 1})$ 
and keep $(p^0_\infty,p^1_\infty)$ as our independent parameters.
In turn, these can be written explicitly in terms of the charges $Q_1, P_1$ and the scalar VEVs $\phi_\infty, \chi_\infty$ by making use of the 
expression for the central charge (\ref{eq:Zinfty4}).
In this analysis we work at the origin of moduli space for convenience, taking $\phi_\infty=\chi_\infty=0$. While the VEVs can be easily restored, taking them into account 
complicates the computations significantly.

At extremality and at the origin of moduli space, the leading expressions for the mass and scalar charges can be written in terms of the electric and magnetic charges as follows, 
\begin{align}
\label{Q1P1massandcharges}
    M^{(2)}\Big|_{T=0} =  \frac{\sqrt 3}{2} \sqrt{P_1^2 + Q_1^2}\, , \quad \mu_\phi^{(2)}\Big|_{T=0} =  \frac{P_1^2-Q_1^2}{\sqrt{P_1^2+Q_1^2}} \, ,
\quad \mu_\chi^{(2)}\Big|_{T=0} =  \frac{ 2 P_1 Q_1}{\sqrt{P_1^2+Q_1^2}} \, .
\end{align}
%
% Recall $\lambda=-\frac{1}{2\sqrt{3}}$
The scalar charges can be read off directly from the asymptotic expansion of the scalars.
The normalizations are 
 consistent with the convention adopted in the long-range force expression (\ref{eq:total_force}), i.e.
at the two-derivative level the quantities $M,\mu_\phi,\mu_\chi$ lead to a vanishing long-range force,
as expected,
\begin{equation}
    \lim_{r\rightarrow \infty} r^2f(r) = -(M^{(2)})^2 + Q_1^2 +P_1^2 -\frac{1}{4}(\mu^{(2)}_\phi)^2 -\frac{1}{4}(\mu^{(2)}_\chi)^2 \xrightarrow[]{T=0} 0.
\end{equation}
For this class of black holes, we are unable to directly evaluate the scalar shifts because we do not have the $Q_1$, $P_1$ solutions at general temperatures. Thus, we cannot compute the derivative of the mass at constant entropy-- such a derivative requires that the temperature is allowed to fluctuate. However, the shift symmetries of the action mean that we can adopt the derivative formulas 
(\ref{4Qcharge}) and (\ref{4Qaxioncharge}), which in this case yield
\begin{equation}
\Delta \mu_\phi =  - 4 \Bigl(   \frac{\partial \Delta M}{\partial Q_1} \frac{\partial Q_1}{\partial \phi_\infty} + \frac{\partial \Delta M}{\partial P_1} \frac{\partial P_1}{\partial \phi_\infty}  \Bigr) 
=   - \frac{2}{\sqrt 3} Q_1 \frac{\partial \Delta M}{\partial Q_1} +  \frac{2}{\sqrt 3}
P_1 \frac{\partial \Delta M}{\partial P_1}  \,,
\end{equation}
for the shift to the dilaton charge, and 
\begin{align}
    \Delta \mu_\chi = 4 Q_1  \frac{ \partial \Delta M}{\partial Q_0} + 
    \frac{8}{\sqrt{3}} P_1  \frac{ \partial \Delta M}{\partial Q_1} \, ,
\end{align}
for the shift to the axion. 

An important point to reiterate here is that the shift symmetry of the axion (\ref{axionchargescalingsSec4}) leads to a nontrivial mixing between the four charges. 
In particular, the shift in $Q_0$ generates $Q_1$ terms 
\begin{equation}
    Q_0 \rightarrow Q_0 - \chi_\infty Q_1 + \mathcal{O}(\chi_\infty^2)  
\end{equation}
which will contribute to the axionic scalar charge. Thus, even though we are interested in solutions with only $Q_1,P_1$ charges turned on, in order to extract the scalar charge $\Delta \mu_\chi$ using the derivative formula, the dependence of the mass 
on $Q_0$ cannot be neglected.
Indeed, it is easy to verify explicitly that this is the case by looking at the two-derivative expressions for the scalar charges. To properly reproduce $\mu_\chi$ in (\ref{Q1P1massandcharges}) (which can be independently verified by working out the asymptotic expansion of the scalar field) it is \emph{essential} that one takes into account derivatives of the mass $M^{(2)} = \frac{1}{2} \sqrt{(Q_0 - \sqrt{3}P_1)^2+(P_0+\sqrt{3}Q_1)^2}$ of the four-charge solutions with respect to the charge $Q_0$.

\begin{figure}[t]%
    \centering
    \subfloat[\centering]{{\includegraphics[width=7cm]{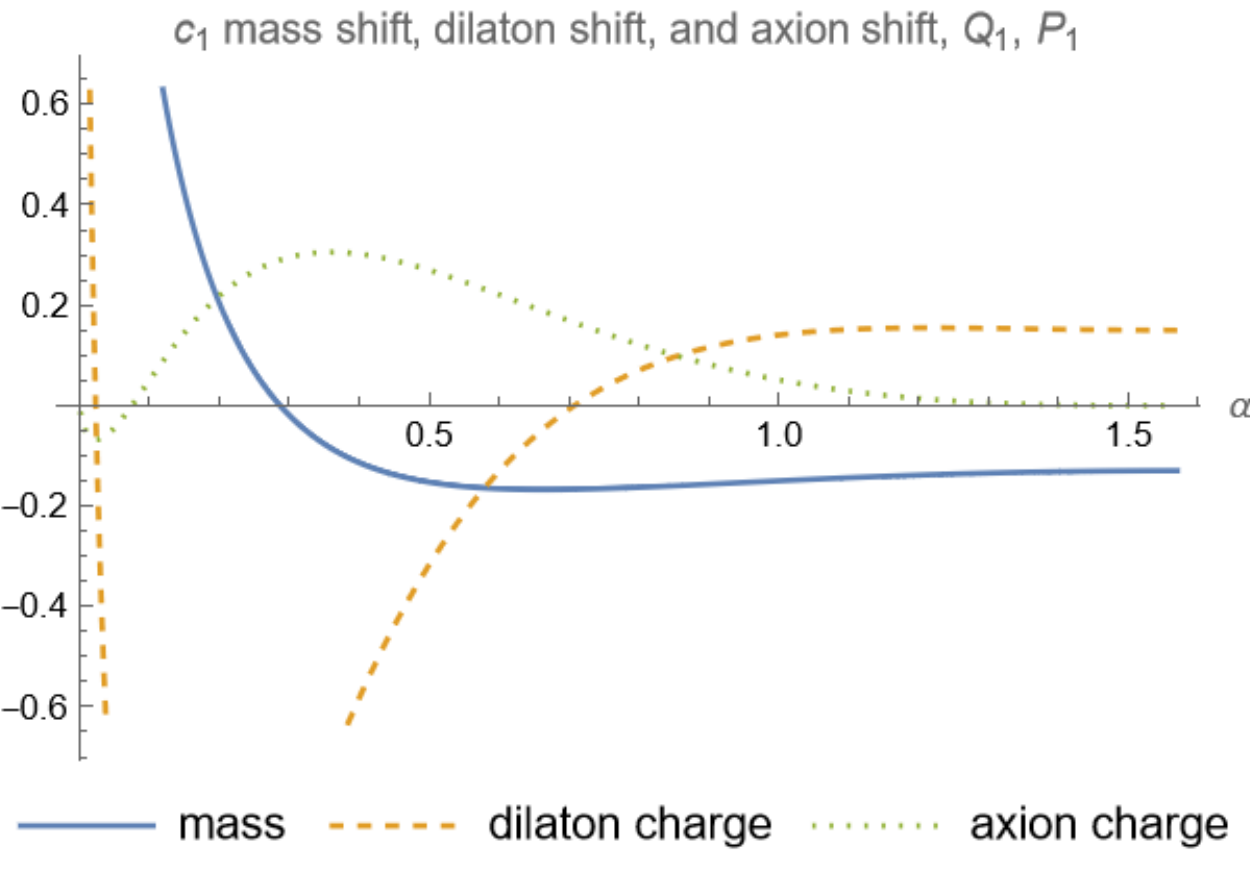} }}%
    \qquad
    \subfloat[\centering ]{{\includegraphics[width=7cm]{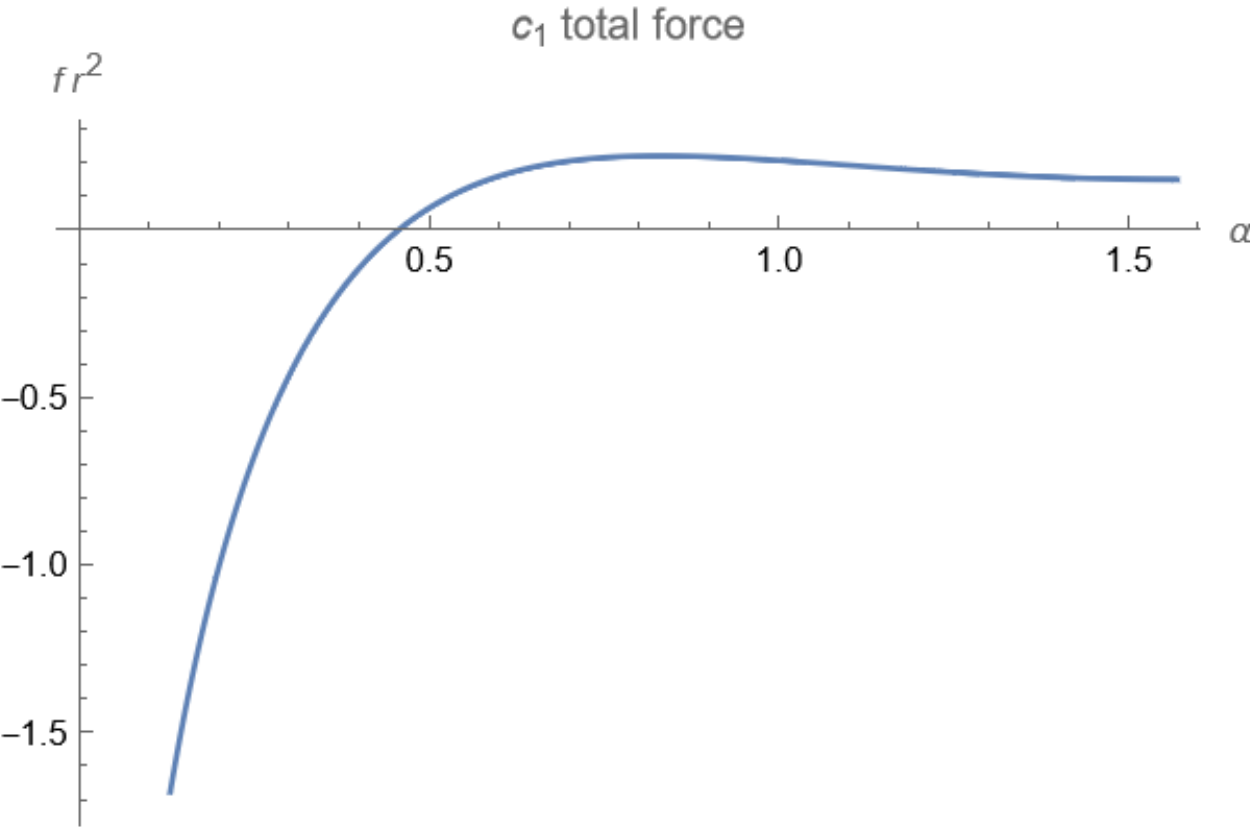} }}%
    \caption{Shifts for $Q_1$, $P_1$ black holes with $c_1 = 1$, all other $c_i = 0$.}%
    \label{fig:Q1P1_shiftsc1}%
\end{figure}

Adopting this logic with the higher-derivative contributions to the mass of the four-charge solutions, we can extract the shift to the axionic charge, along with the shift to the mass and dilatonic charge. The shifts and forces are plotted in figures (\ref{fig:Q1P1_shiftsc1}-\ref{fig:Q1P1_shiftsc4}).
The resulting final expressions are quite complicated, and contained in appendix (\ref{app:Q1P1}).

\begin{figure}[t]%
    \centering
    \subfloat[\centering]{{\includegraphics[width=7cm]{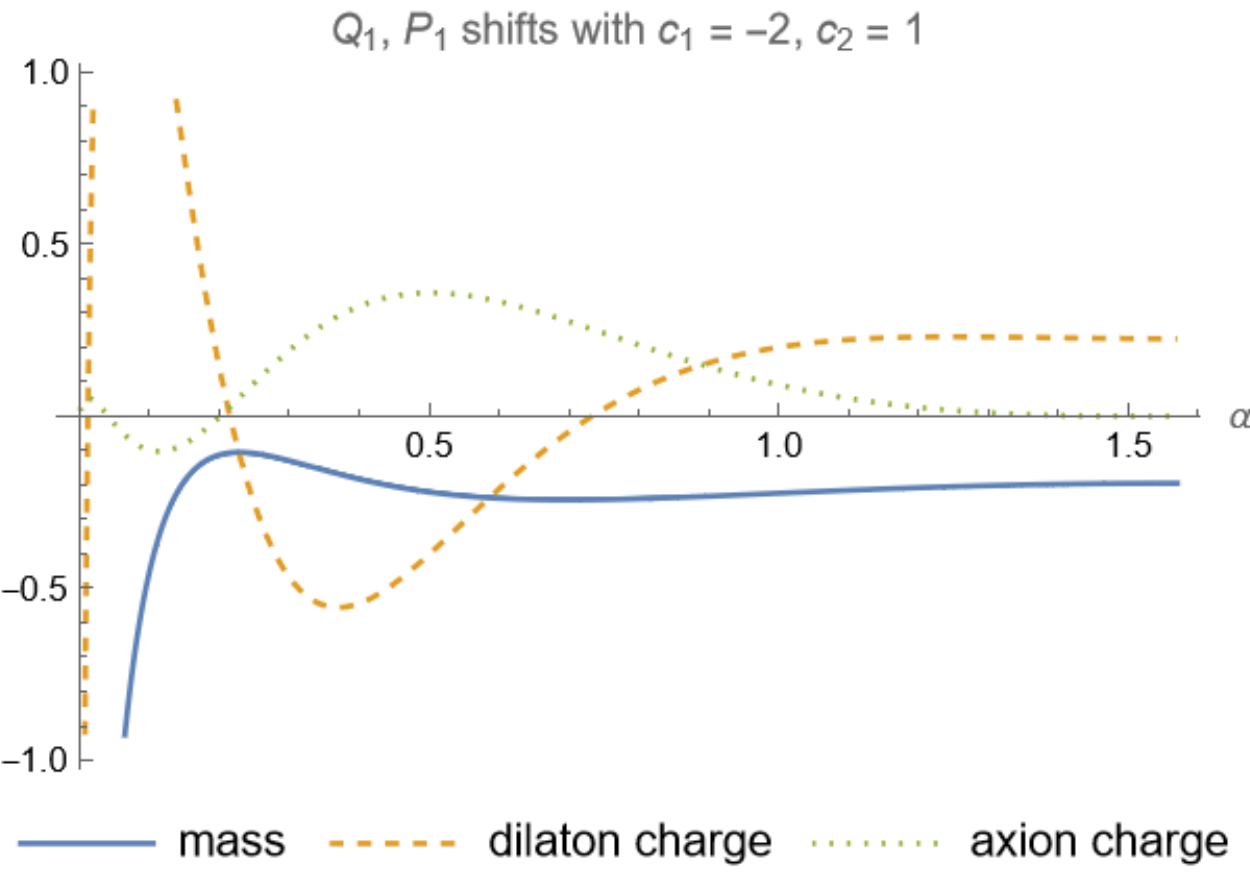} }}%
    \qquad
    \subfloat[\centering ]{{\includegraphics[width=7cm]{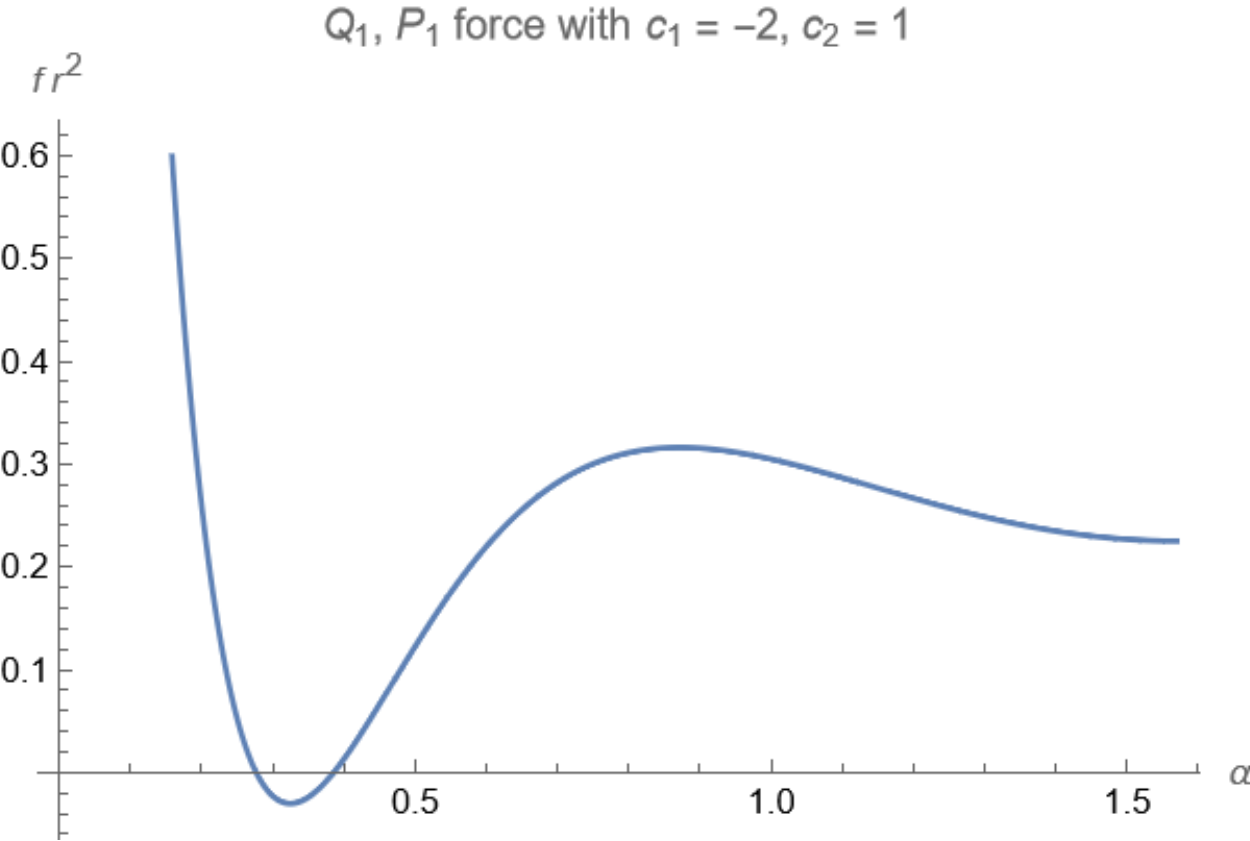} }}%
    \caption{Here $c_1 = -2$, $c_2 = 1$ and all other $c_i = 0$. This gives another example where the mass shift is always negative but the total force changes sign depending on the charge. }%
    \label{fig:Q1P1_shifts}%
\end{figure}

We emphasize the key point that the shifts to the force are controlled by combinations of the EFT coefficients and charges that are 
\emph{completely independent} of those controlling the mass shift. 
This indicates that in general there is no correlation between the WGC and the RFC in the presence of 
higher-derivative corrections. To be more explicit, we can identify examples for which $\Delta M<0$, as required by the WGC, and the force changes sign (figure \ref{fig:Q1P1_shifts}). For such examples, the weak form of the RFC could not be satisfied in every direction in charge space.

%%%%%%%%

%%%%%%%%

\begin{figure}[t]%
    \centering
    \subfloat[\centering]{{\includegraphics[width=7cm]{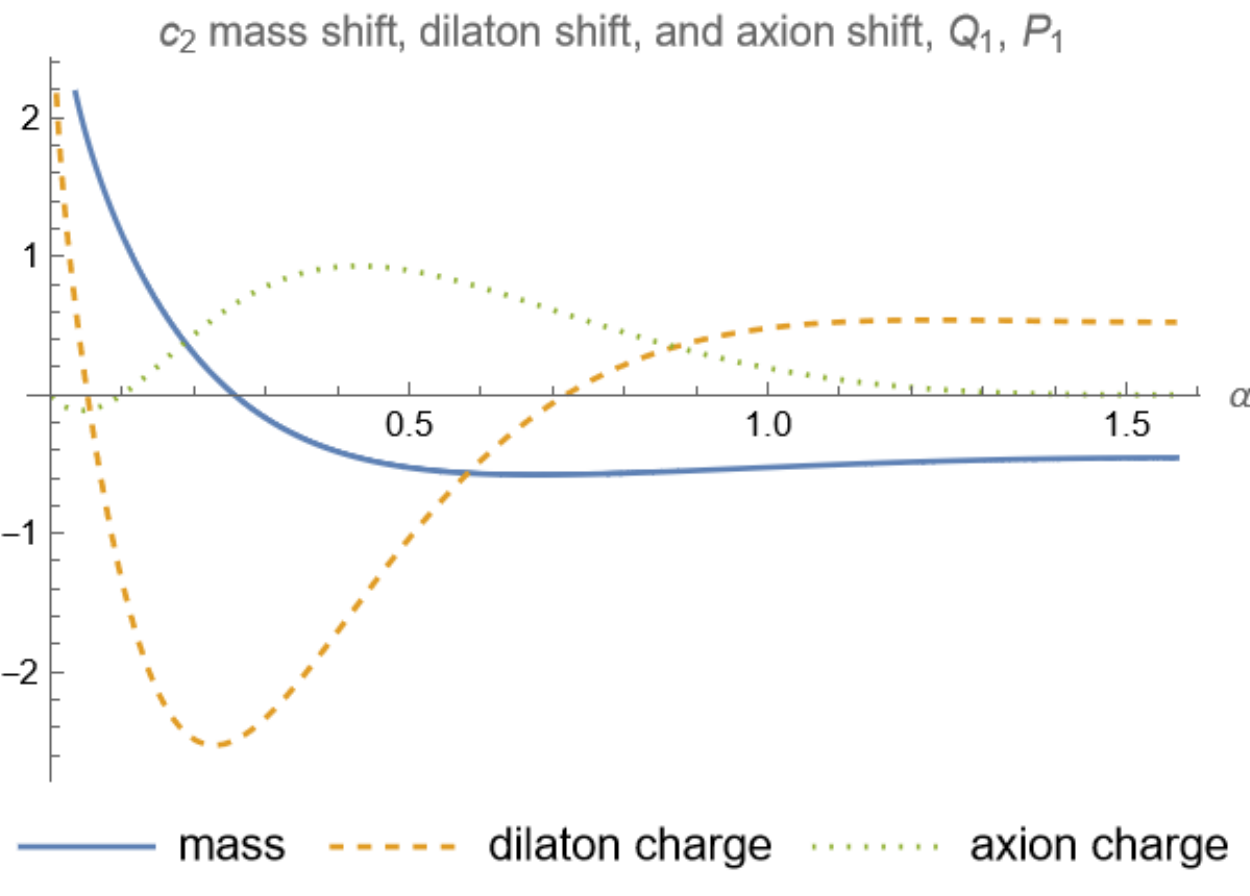} }}%
    \qquad
    \subfloat[\centering ]{{\includegraphics[width=7cm]{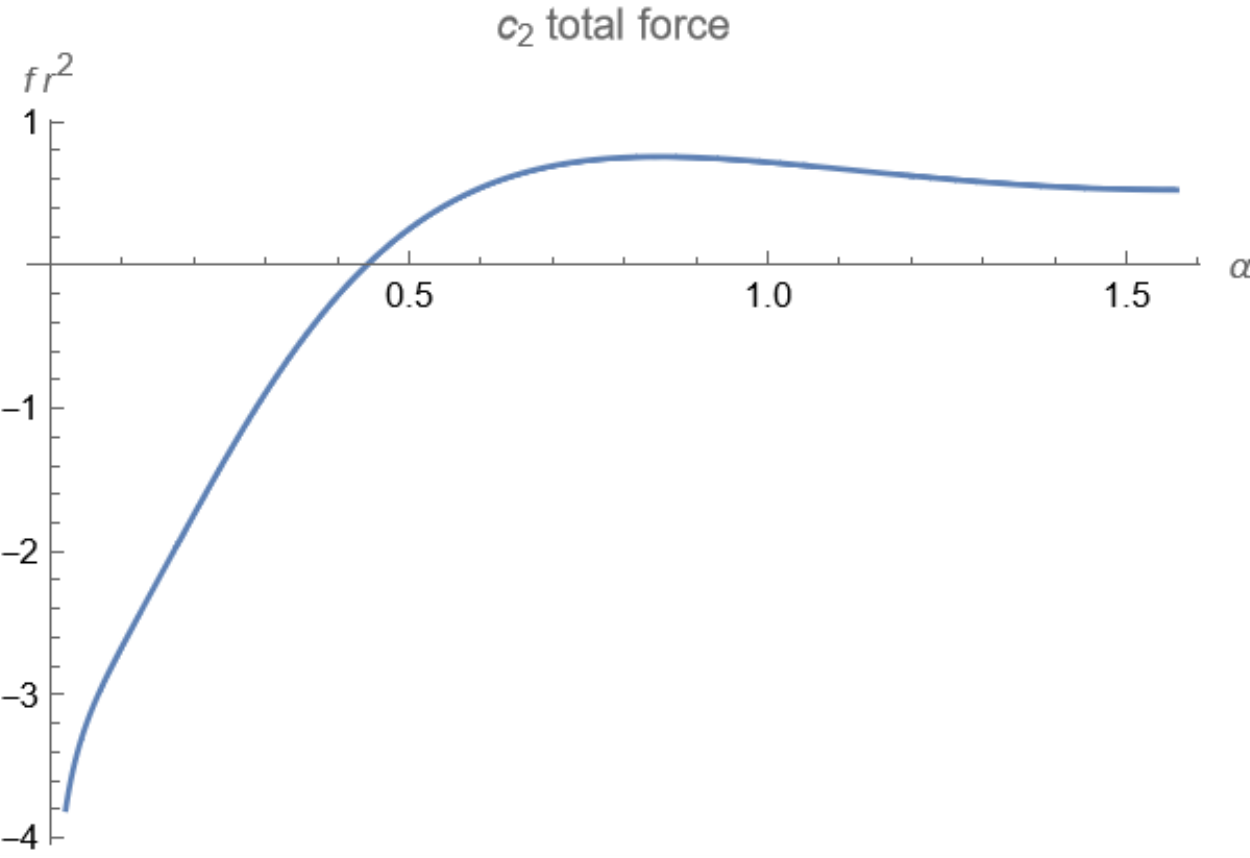} }}%
    \caption{Shifts for $Q_1$, $P_1$ black holes with $c_2 = 1$, all other $c_i = 0$.}%
    \label{fig:Q1P1_shiftsc2}%
\end{figure}

\begin{figure}[t]%
    \centering
    \subfloat[\centering]{{\includegraphics[width=7cm]{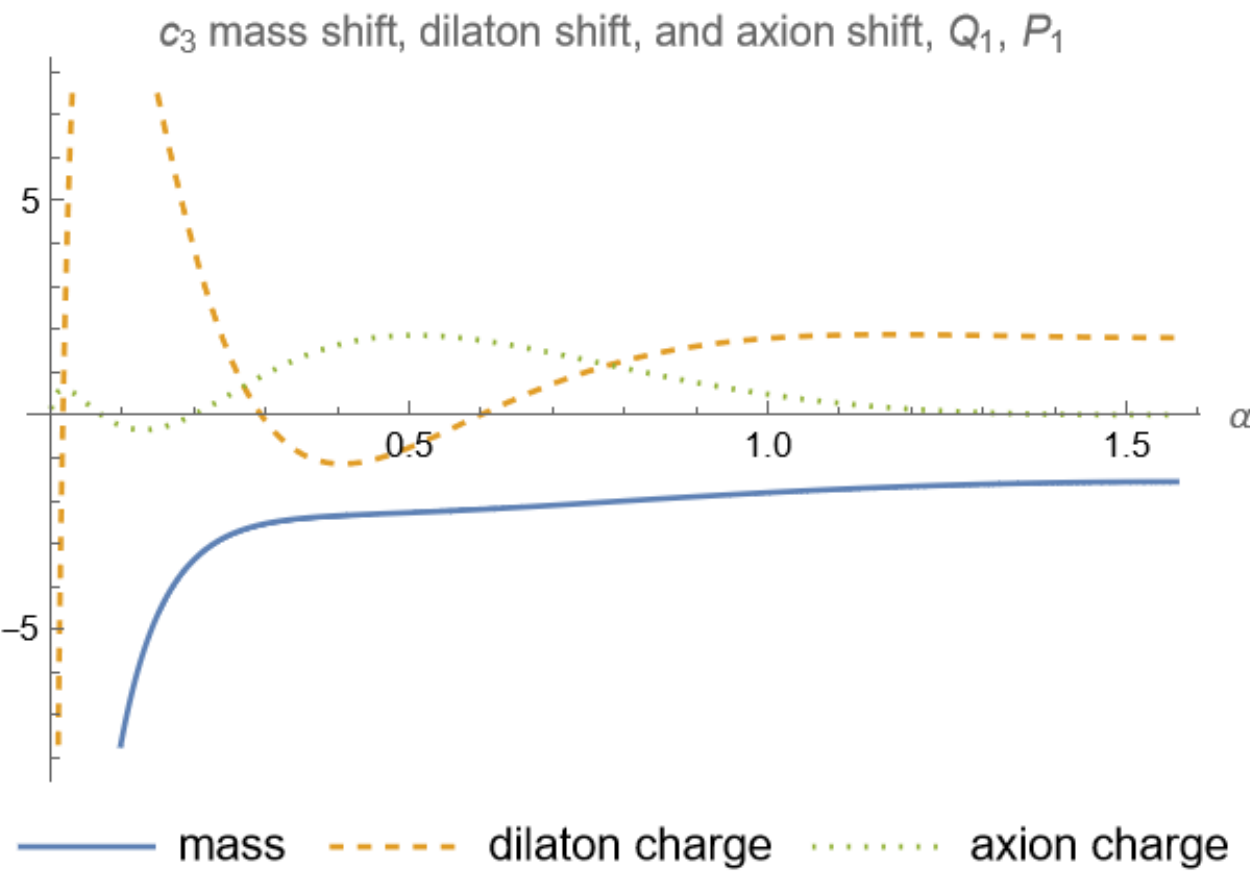} }}%
    \qquad
    \subfloat[\centering ]{{\includegraphics[width=7cm]{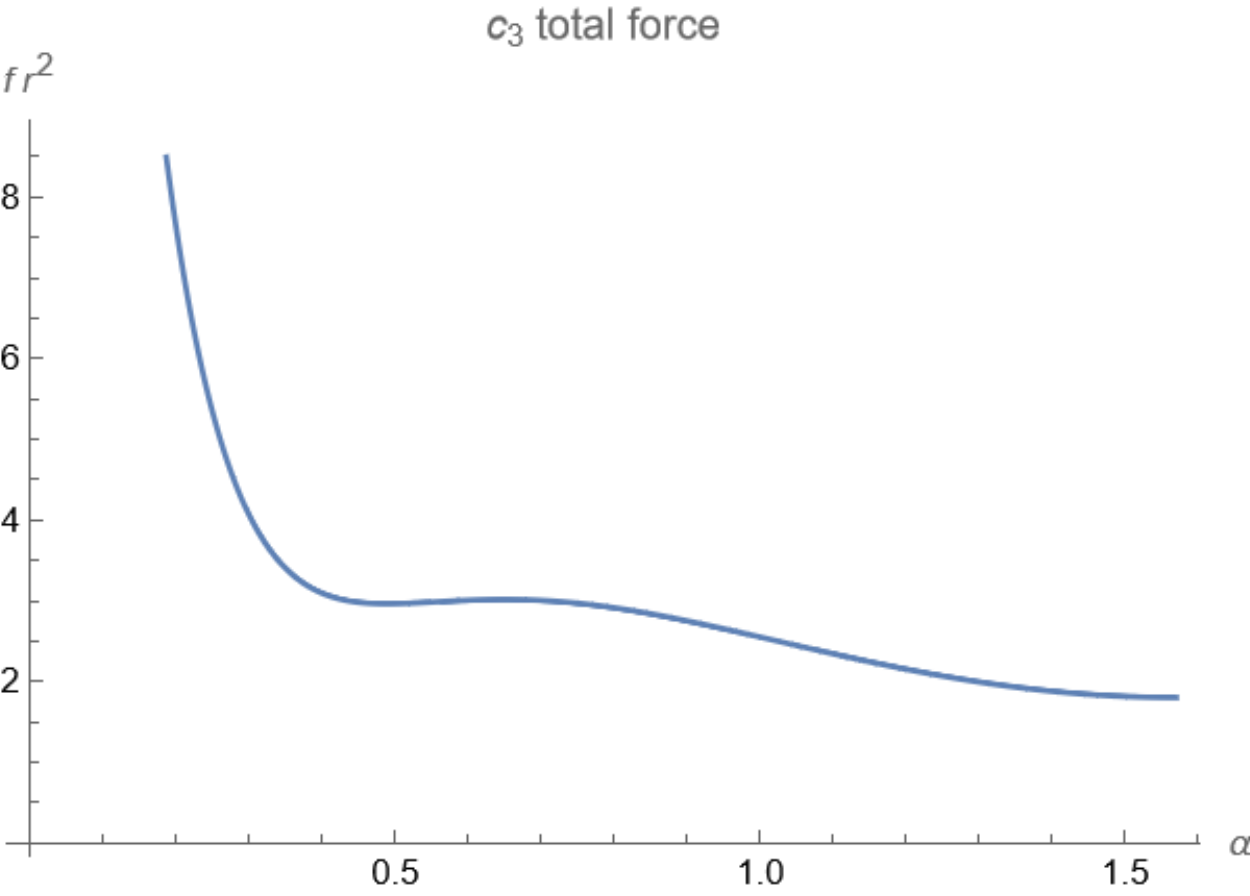} }}%
    \caption{Shifts for $Q_1$, $P_1$ black holes with $c_3 = 1$, all other $c_i = 0$.}%
    \label{fig:Q1P1_shiftsc3}%
\end{figure}

\begin{figure}[t]%
    \centering
    \subfloat[\centering]{{\includegraphics[width=7cm]{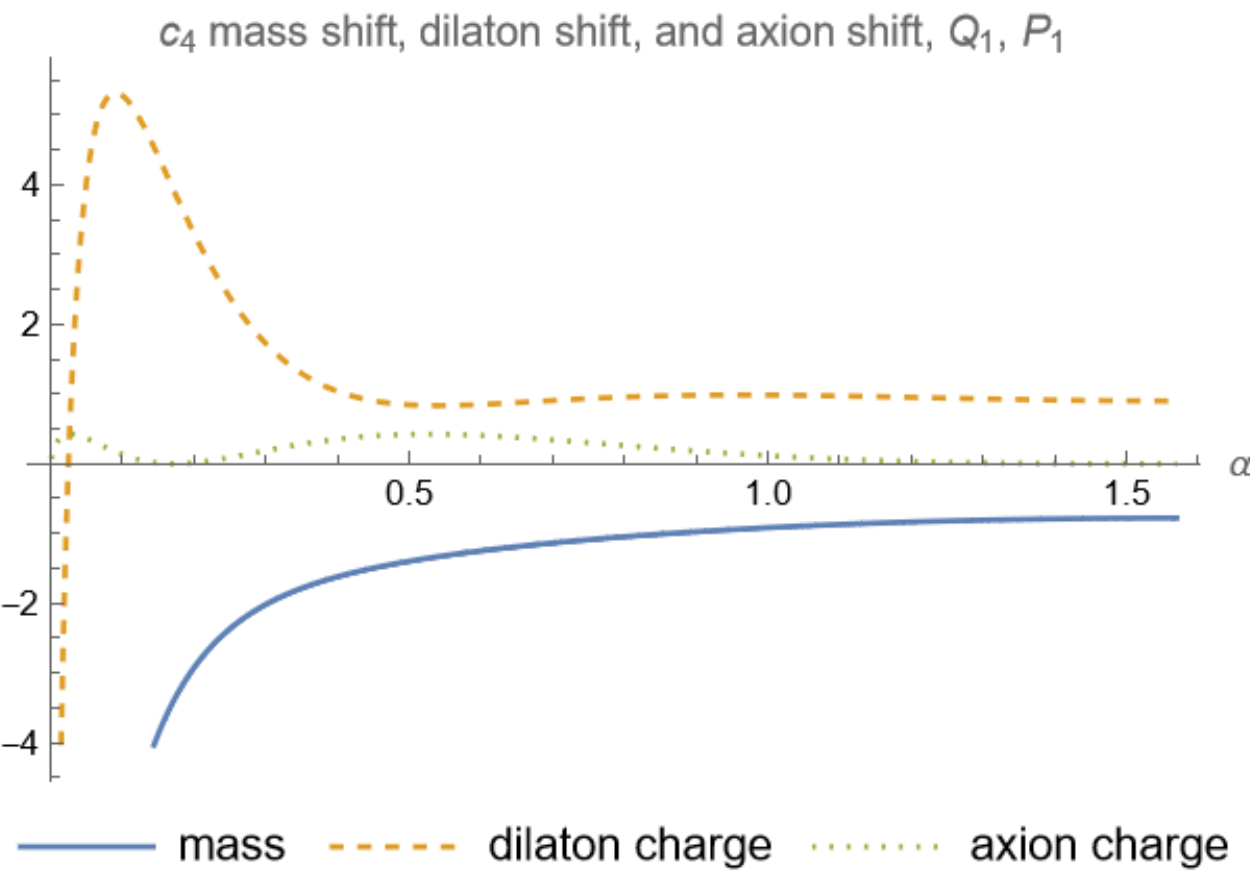} }}%
    \qquad
    \subfloat[\centering ]{{\includegraphics[width=7cm]{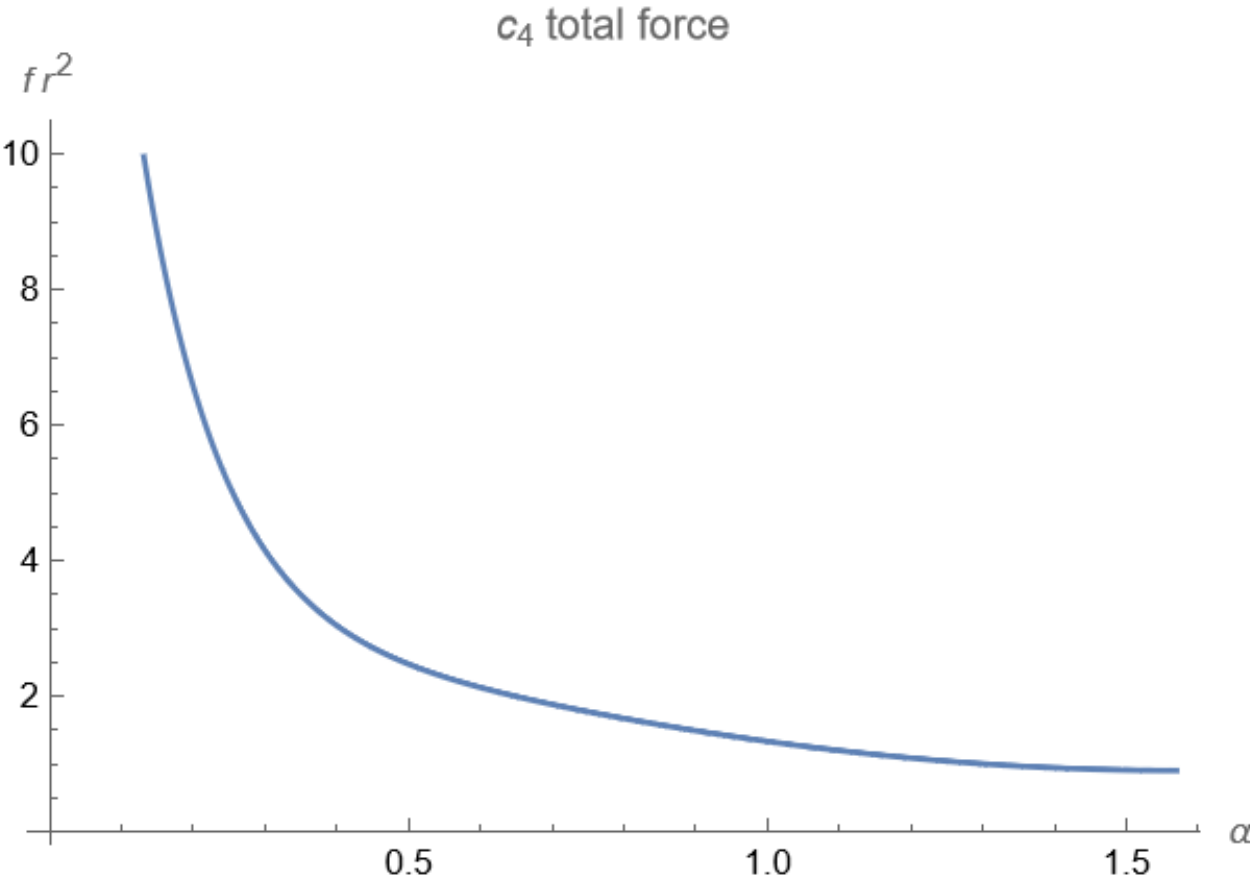} }}%
    \caption{Shifts for $Q_1$, $P_1$ black holes with $c_4 = 1$, all other $c_i = 0$.}%
    \label{fig:Q1P1_shiftsc4}%
\end{figure}

%%%%%%%

%%%%%%%

\section{Discussion}

Like the WGC, the RFC is a requirement on the spectrum of theories of quantum gravity. 
In its simplest form \cite{Palti:2017elp}, the RFC claims that 
EFTs consistent with quantum gravity must have a state which is self-repulsive. 
A number of refinements of the RFC were discussed in \cite{Heidenreich:2019zkl}, with the 
strongest among them 
stating that every direction in charge space should have a strongly self-repulsive multiparticle state.

The examples given in sections \ref{SectionAModel} and \ref{SectionN2SUGRA} demonstrate that higher-derivative corrected theories can have black hole solutions
for which ``gravity is the weakest force," in the sense that they are superextremal, but that are still attractive at large distances.
In particular, we have found explicit examples of 
directions in charge space along which 
there are no self-repulsive black holes, so the weak form of the RFC is not satisfied by the black hole spectrum. This is in contrast to the situation for the WGC, which may be satisfied by the black hole spectrum alone.
This has interesting implications for the relation between the two conjectures.

For a concrete example where the weak RFC requirement is not obeyed, consider the $Q_0$, $P_0$ black holes in $5d$ $\mathcal{N} = 2$ supergravity, with the higher-derivative action given in (\ref{finalL}). In this case we can ignore all the coefficients except for $c_1$. 
When $c_1 > 0$, the mass shift is always negative, ensuring that the WGC is satisfied at every direction in charge space, 
while the force can have either sign, depending on the values of the charges.
Moreover, it is easy to see from figure \ref{fig:Q0P0_shifts} that 
there is no choice for $c_1$ where the simple version of the RFC is obeyed: depending on the sign of $c_1$, either 
all electric extremal black holes will be attractive, or all magnetic ones will be. Thus no solution will be self-repulsive for every direction in charge space.

%%%

In fact, we find that the strong form of the conjecture is not satisfied for the $Q_0$, $P_0$ black holes either. To see this, we need to show the following fact: \textit{there is no pair of black holes which are simultaneously a) self-repulsive, b) mutually repulsive and c) oppositely charged (electrically)}. 

First, consider that non-extremal black holes are always attractive at the two-derivative level. So any black holes which satisfy the conjecture must be extremal and self-repulsive at the four-derivative level. Then we must look for extremal black holes with opposite $Q_0$ charges which are mutually repulsive and in the self-repulsive region of figure \ref{fig:Q0P0_shifts}. 
The force takes the form 
\begin{align}
    \lim_{r\rightarrow \infty} r^2f(r) = - M_1 M_2 - \frac{1}{4} |Q_1| |Q_2| + \frac{1}{4} |P_1| |P_2| - \frac{1}{4} \mu_1 \mu_2
\end{align}
where the minus sign before the $Q$ charges comes from the fact that we assume that the electric charges have the opposite sign. Using $p_i \to \sin \alpha_i, q_i \to \cos \alpha_i$ this implies
\begin{align}
\begin{split}
    \lim_{r\rightarrow \infty} r^2f(r) &= \frac{1}{8} \Bigg( -2 \cos(\alpha_1 - \alpha_2) + \sin(\alpha_1 + \alpha_2) \\
    & - 2 \sqrt{\frac{\cos^3 \alpha_1 \cos^3 \alpha_2}{(\cos \alpha_1 + \sin \alpha_1)(\cos \alpha_2 + \sin \alpha_2)}}+  2 \sqrt{\frac{\sin^3 \alpha_1 \sin^3 \alpha_2}{(\cos \alpha_1 + \sin \alpha_1)(\cos \alpha_2 + \sin \alpha_2)}} \Bigg)
    \label{mutual example}
\end{split}
\end{align}
Recall that $\alpha = 1.21895$ was where the \textit{self-force} changed sign from positive to negative. We have checked numerically that the \textit{mutual force} in equation (\ref{mutual example}) is always negative, and therefore attractive, if we require that $\alpha_i < 1.21895$. This means that the mutual force is always negative if we restrict ourselves to the region of parameter space where the self-force is positive. We draw two conclusions from this. The first is that there is no way to satisfy the strong form of the RFC using only the black hole solutions studied in this paper. The second is that this fact is highly sensitive to the four-derivative corrections. If we had found that the cross-over were at $\alpha = 1.3$ instead, then we would have a pair of black holes, $\alpha_1 = 1.29$, $\alpha_2 = 0$ which would be a) self-repulsive (at the four-derivative level), mutually repulsive (at the two-derivative level) and c) oppositely charged. Therefore we could take a collection of them which adds up to zero net electric charge to find a purely magnetic strongly self-repulsive multiparticle state.

The point of this exercise is to show that, while the WGC may be trivially satisfied by black holes with higher-derivative corrections, this does not appear to be true for the RFC. This means that the RFC, if true, requires that the spectrum include states beyond black hole solutions. It would be interesting to extend this simple analysis to other examples to see if there are any cases where the strong form of the RFC can be satisfied trivially by higher-derivative corrected black holes.

One might wonder about the ultimate fate of black holes which are attractive but superextremal. This paper concerns the long-range $\mathcal{O}(1/r^2)$ part of the force between extremal black holes, but at shorter distances this is expected to be modified by higher powers of $1/r$. These corrections are not accessible from ``thermodynamic" quantities like the free-energy and require a more detailed calculation. Nonetheless, we can still say something about their structure on general grounds. For black hole solutions which are both superextremal and attractive there are two logical possibilities. One option is that the corrections to the force become repulsive at short distances. This leads to the interesting possibility of forming stable bound states of non-BPS black holes. Another option is that these black holes simply coalesce to form a naked singularity, leading to a violation of weak cosmic censorship. It is natural to speculate that theories where this is possible are in the Swampland.

Another possible future direction would be to try to establish this conjecture in Anti-de Sitter space, where one might hope to put it on more rigorous footing using methods from the conformal bootstrap.  Naively, any conjecture about repulsive forces appears to require modification or clarification in this setting --- due to the confining potential of AdS space, no two massive objects can be truly self-repulsive at large distances. A very interesting recent idea in this direction is that theories of quantum gravity in AdS should have states with a non-negative binding energy \cite{Aharony:2021mpc}. Previous discussions of the WGC in AdS include \cite{Harlow:2015lma, Nakayama:2015hga,Montero:2016tif, Montero:2018fns, Cremonini:2019wdk, Agarwal:2019crm, Nakayama:2020dle, Liu:2020uaz}. Understanding these issues further is something we plan to pursue in future work.

%%%%%%%%%%%%
\section*{Acknowledgments}
We thank Matt Reece, Gary Shiu and Timm Wrase for insightful feedback on this paper. This work was supported in part by the U.S. Department of Energy under grant DE-SC0007859.
SC is supported  in  part  by  the  National  Science  Foundation grant  PHY-1915038.  YT is supported by Lehigh University’s Lee Fellowship. BM was supported by the European Research Council (ERC) under the European Union's Horizon 2020 research and innovation programme (grant agreement no.~758903). CRTJ acknowledges the support of the Mani L. Bhaumik Institute for Theoretical Physics.
%%%%%%%%%%%%%%%%%%%%%%%%%%%%%%%%%%%%%%%%%%%%%%
\appendix

%%%%%

\section{Further Details for Non-supersymmetric Black Holes}
\label{app:nonsusythermo}

In this appendix we present a detailed analysis of the thermodynamic properties of the non-supersymmetric black hole solution (\ref{a1omegapi3model}).
We begin by evaluating the Ricci scalar of this solution
\begin{equation}
    R = \frac{h \left(q^2 (h+2 r) \tanh (\phi_\infty)+h \left(4 r
   (h+r)+q^2\right)\right)}{8 r^3 (h+r)^3},
\end{equation}
which is singular at $r=0$ and $r=-h$. Since the Ricci scalar is a diffeomorphism invariant quantity, these must correspond to physical curvature singularities and, in a physical black hole solution, must be hidden behind an event horizon. The possible horizons are located at zeros of the blackening function (\ref{moregenblackening}), the largest of which we calculate to be
\begin{align}
    r_+ &= \frac{-2 h^2 \left(e^{2 \phi_\infty}+3\right)+\sqrt{4 h^4 \left(e^{2
   \phi_\infty}+3\right)^2-4 h^2 q^2 \left(e^{\phi_\infty}+3\right)^2+q^4 \left(e^{2 \phi_\infty}-3\right)^2}}{4 h \left(e^{\phi_\infty}+3\right)} \nonumber\\
   &\hspace{10mm}-\frac{q^2
   \left(e^{2 \phi_\infty}-3\right)}{4 h \left(e^{\phi_\infty}+3\right)}.
\end{align}
For the solution to be physically acceptable there are three conditions we must impose: $(i)$ the horizon must be real, $(ii)$ the singularity at $r_\text{sing} = \text{max}\{0,-h\}$ must be hidden behind a horizon, $r_\text{sing} \leq r_+$, $(iii)$ an extremal limit must exist whilst maintaining conditions (i) and (ii).

We find that condition (i) requires 
\begin{equation}
    4 h^4 \left(e^{2
   \phi_\infty}+3\right)^2-4 h^2 q^2 \left(e^{2 \phi
   _{\infty}}+3\right)^2+q^4 \left(e^{2 \phi_\infty}-3\right)^2 > 0,
\end{equation}
which, for positive $h$ and $q$  amounts to 
\begin{equation}
    0 < h < \frac{q}{\sqrt{2}} \sqrt{1+\frac{\sqrt{3}}{\sinh(\phi_\infty)-2\cosh(\phi_\infty)}}.
\end{equation}
In this same range and for small values of $\phi_\infty$ we can straightforwardly verify that the horizon area remains positive as we approach $T=0$. Thus, this range of parameters is physically acceptable. The mass, entropy, temperature, electric potential, electric charge and scalar charge of the solution are found to be 
\begin{align}
    M^{(2)} &= \frac{q^2 (\cosh (\phi_\infty)-2 \sinh (\phi_\infty))}{8 h \cosh (\phi_\infty)-4 h \sinh (\phi_\infty)}\nonumber\\
    S^{(2)} &= \frac{\pi  q^2 }{4 h^2 \left(e^{2 \phi_\infty}+3\right)^2}\left[2 h^4 \left(e^{2 \phi_\infty}+3\right)^2 \left(e^{4 \phi_\infty}+9\right)\right.\nonumber\\
    &\hspace{40mm}\left.+h^2 \left(e^{2
   \phi_\infty}-3\right) \left(e^{2 \phi_\infty}+3\right)^2 \left(g(h,q,\phi_\infty)-2 q^2 \left(e^{2 \phi_\infty}-3\right)\right)\right.\nonumber\\
   &\hspace{40mm}\left.+\frac{1}{2} q^2 \left(e^{2 \phi_\infty}-3\right)^3 \left(q^2 \left(e^{2 \phi_\infty}-3\right)-g(h,q,\phi_\infty)\right)\right]^{1/2}\nonumber\\
    T &=\frac{2 h q^2 e^{2 \phi_\infty}}{\pi } \left(e^{2 \phi_\infty}+3\right)\left[\frac{1}{\left(2 h^2 \left(e^{2
   \phi_\infty}+3\right)-g(h,q,\phi_\infty)+q^2 \left(e^{2 \phi_\infty}-3\right)\right)^2}\right.\nonumber\\
   &\hspace{50mm}\left.-\frac{3
   e^{-2 \phi_\infty}}{\left(2 h^2 \left(e^{2 \phi_\infty}+3\right)+g(h,q,\phi_\infty)-q^2 \left(e^{2
   \phi_\infty}-3\right)\right)^2}\right]^{1/2}\nonumber\\
    \Phi_e^{(2)} &= -\frac{e^{-\phi_\infty/2} \left(g(h,q,\phi_\infty)+q^2 \left(e^{2 \phi_\infty}-3\right)\right)}{4 h q
   \sqrt{e^{2 \phi_\infty}+3}}\nonumber\\
    Q &= \frac{q e^{\phi_\infty} \sqrt{4 \cosh (\phi_\infty)-2 \sinh (\phi_\infty)}}{e^{2 \phi_\infty}+3}\nonumber\\
    \mu^{(2)} &= h,
\end{align}
where 
\begin{equation}
    g(h,q,\phi_\infty) = \sqrt{4 h^4 \left(e^{2 \phi_\infty}+3\right)^2-4 h^2 q^2 \left(e^{2 \phi_\infty}+3\right)^2+q^4 \left(e^{2 \phi_\infty}-3\right)^2}.
\end{equation}
It is straightforward to verify that these expressions are consistent with the first law of black hole mechanics (\ref{firstlaw}).

\section{Two-Charge Solutions and the STU Model}
\label{app:STU}

The starting point for the dimensional reduction solutions that we consider is pure five-dimensional $\mathcal N=2$ supergravity, with boson fields $\hat g_{\mu\nu}$ and $\hat A_\mu$.  This can be viewed as a truncation of the five-dimensional STU model, which corresponds to $\mathcal N=2$ supergravity coupled to two vector multiplets.  When reduced on a circle, this yields the four-dimensional STU model, which corresponds to $\mathcal N=2$ supergravity coupled to three vector multiplets.  The truncation to pure supergravity in five dimensions corresponds to identifying the three four-dimensional vector multiplets.  The corresponding black holes can be thought of as three equal charge black holes of the four-dimensional STU model. Note that the conventions in this section may be translated directly into those of section \ref{SectionN2SUGRA} using equations (\ref{eq:normswitch})--(\ref{eq:chargemap}).

\subsection{The four-dimensional STU model}

The STU model \cite{Cremmer:1984hj,Duff:1995sm,Behrndt:1996hu} is a special case of $\mathcal N=2$ ungauged supergravity coupled to three vector multiplets. We start with a very minimal overview of $\mathcal N=2$ supergravity and special geometry \cite{deWit:1984wbb,deWit:1984rvr,Strominger:1990pd,Ceresole:1995jg,Andrianopoli:1996cm} before turning to the STU model itself.  To be specific, consider $\mathcal N=2$ supergravity coupled to $n_v$ vector multiplets.  The bosonic fields consist of the metric $g_{\mu\nu}$, $n_v+1$ gauge fields $A_\mu^I$ and $n_v$ unconstrained complex scalars $z^i$.  Here $I=0,1,\ldots,n_v$ and $i=1,2,\ldots,n_v$.  It is useful to introduce the constrained scalars $X^I$ with $z^i=X^i/X^0$. The bosonic Lagrangian can be written
\begin{equation}
    16\pi e^{-1}\mathcal L=R-2g_{i\bar j}\partial_\mu z^i\partial^\mu\bar z^{\bar j}+\fft12\Im\left(\mathcal N_{IJ}\mathcal F_{\mu\nu}^{+\,I}\mathcal F^{+\,\mu\nu\,J}\right),
\end{equation}
where $\mathcal F^I=dA^I$ and $\mathcal F^{+\,I}\equiv\fft12(\mathcal F^I+i\star\mathcal F^{I})$.  Expanding out $\mathcal F^+$ gives the equivalent expression
\begin{equation}
    16\pi e^{-1}\mathcal L=R-2g_{i\bar j}\partial_\mu z^i\partial^\mu\bar z^{\bar j}+\fft14\Im\mathcal N_{IJ}\mathcal F_{\mu\nu}^I\mathcal F^{\mu\nu\,J}+\fft18\Re\mathcal N_{IJ}\epsilon^{\mu\nu\rho\sigma}\mathcal F_{\mu\nu}^I\mathcal F_{\rho\sigma}^J.
\label{eq:N=2ReIm}
\end{equation}

The couplings $g_{i\bar j}$ and $\mathcal N_{IJ}$ in the Lagrangian are derived from a holomorphic prepotential $F(X^I)$.  In particular, the scalar metric is given by
\begin{equation}
    g_{i\bar j}=\partial_i\partial_{\bar j}K,
\end{equation}
where $K$ is the K\"ahler potential
\begin{equation}
    K=-\log\left(2\Im(\bar F_IX^I)\right).
\end{equation}
Here we have defined the derivatives
\begin{equation}
    F_I=\fft{\partial F}{\partial X^I},\qquad F_{IJ}=\fft{\partial^2 F}{\partial X^IX^J}.
\end{equation}
The gauge kinetic function has a somewhat more complicated form
\begin{equation}
    \mathcal N_{IJ}=\bar F_{IJ}+2i\fft{\Im F_{IK}\Im F_{JL}X^KX^L}{\Im F_{KL}X^KX^L},.
\label{eq:NIJ0}
\end{equation}

For the four-dimensional STU model \cite{Cremmer:1984hj,Duff:1995sm,Behrndt:1996hu}, we take $n_v=3$ and use the prepotential
\begin{equation}
    F=\fft{X^1X^2X^3}{X^0}.
\end{equation}
We also write the real and imaginary components of the complex scalars as
\begin{equation}
    z^i=x^i-iy^i
\end{equation}
(note the minus sign convention here).  Up to an overall scale fixed by working in the $X^0=1$ gauge, the K\"ahler potential is then
\begin{equation}
    K=-\log\left(-i(z^1-\bar z^1)(z^2-\bar z^2)(z^3-\bar z^3)\right)=-\log(8y^1y^2y^3),
\end{equation}
with corresponding K\"ahler metric $g_{i\bar j}=\delta_{i\bar j}/(2y^i)^2$.  We can also work out the components of $\mathcal N_{IJ}$
\begin{align}
    \mathcal N_{00}&=2x^1x^2x^3-iy^1y^2y^3-i\left((x^1)^2\fft{y^2y^3}{y^1}+(x^2)^2\fft{y^3y^1}{y^2}+(x^3)^2\fft{y^1y^2}{y^3}\right),\nn\\
    \mathcal N_{0i}&=\begin{pmatrix}-x^2x^3+ix^1\fft{y^2y^3}{y^1}~&-x^3x^1+ix^2\fft{y^3y^1}{y^2}~&-x^1x^2+ix^3\fft{y^1y^2}{y^3}\end{pmatrix},\nn\\
    \mathcal N_{ij}&=\begin{pmatrix}-i\fft{y^2y^3}{y^1}&x^3&x^2\\x^3&-i\fft{y^3y^1}{y^2}&x^1\\x^2&x^1&-i\fft{y^1y^2}{y^3}\end{pmatrix}.
\label{eq:NIJ}
\end{align}
The resulting Lagrangian can be obtained by substituting these expressions into (\ref{eq:N=2ReIm}).

To connect to explicit STU model Lagrangians in the literature, we can write the complex scalars in terms of their dilatonic and axionic components
\begin{equation}
    z^i=x^i-iy^i=\chi_i-ie^{-\varphi_i}.
\end{equation}
We then find
\begin{align}
    16\pi e^{-1}\mathcal L&=R-\fft12\sum_i((\partial_\mu\varphi_i)^2+e^{2\varphi_i}(\partial_\mu\chi_i)^2)\nn\\
    &\quad-\fft14e^{-\varphi_1-\varphi_2-\varphi_3}\mathcal (\mathcal F_{\mu\nu}^0)^2-\fft14\sum_ie^{2\varphi_i-\varphi_1-\varphi_2-\varphi_3}(\mathcal F_{\mu\nu}^i-\chi_i\mathcal F_{\mu\nu}^0)^2\nn\\
    &\quad+\chi_1\chi_2\chi_3\mathcal F^0\wedge\mathcal F^0-(\chi_2\chi_3\mathcal F^1+\chi_3\chi_1\mathcal F^2+\chi_1\chi_2\mathcal F^3)\wedge\mathcal F^0\nn\\
    &\quad+\chi_1\mathcal F^2\wedge\mathcal F^3+\chi_2\mathcal F^3\wedge\mathcal F^1+\chi_3\mathcal F^1\wedge\mathcal F^2.
\label{eq:STUchow}
\end{align}
This matches, \textit{e.g.}, (2.16) of \cite{Chow:2014cca}.

\subsection{The BPS black holes}

Black holes in the STU model have been extensively studied, as they have wide applicability to not just $\mathcal N=2$ supergravity but also to $\mathcal N=4$ and $\mathcal N=8$ theories and various brane configurations in string theory.  General non-extremal and rotating solutions have been constructed in \cite{Chow:2013tia,Chow:2014cca}, while extremal solutions are closely tied to the attractor mechanism \cite{Ferrara:1995ih,Strominger:1996kf,Ferrara:1996dd,Ferrara:1996um,Ferrara:1997tw}.  It is known that extremal STU black holes with non-degenerate horizons fall into three classes: $i)$ half-BPS black holes with $Z\ne0$, $ii)$ non-BPS black holes with $Z=0$ and $iii)$ non-BPS black holes with $Z\ne0$ \cite{Bellucci:2006ew,Bellucci:2006xz} (see also \cite{Bellucci:2008sv}).  Here
\begin{align}
    Z&=e^{K/2}W=e^{K/2}\fft{X^Iq_I-F_Ip^I}{\sqrt2}\nn\\
    &=\fft{q_0+q_1z^1+q_2z^2+q_3z^3+p^0z^1z^2z^3-p^1z^2z^3-p^2z^3z^1-p^3z^1z^2}{4\sqrt{y^1y^2y^3}},
\label{eq:Zccf}
\end{align}
is the central charge function, and $q_I$ and $p^I$ are the electric and magnetic charges given by
\begin{equation}
    p^I=\fft1{4\pi}\int_{S^2_\infty}\mathcal F^I,\qquad q_I=\fft1{4\pi}\int_{S^2_\infty}\tilde{\mathcal F}_I,
\label{eq:pIqI}
\end{equation}
where the dual field strength is
\begin{equation}
    \tilde{\mathcal F}_I=\Im\mathcal N_{IJ}*\mathcal F^J+\Re\mathcal N_{IJ}\mathcal F^J,
\end{equation}
with $\mathcal N_{IJ}$ given in (\ref{eq:NIJ0}), or explicitly for the STU model in (\ref{eq:NIJ}).

It is convenient to group the electric and magnetic charges together into a single charge vector
\begin{equation}
    \Gamma=\begin{pmatrix}p^I\\q_I\end{pmatrix},
\label{eq:Gvector}
\end{equation}
and introduce the symplectic inner product
\begin{equation}
    \langle A,B\rangle=A^IB_I-A_IB^I.
\end{equation}
Then, in addition to the central charge function, we also define the quartic invariant
\begin{equation}
    \mathcal I_4(\Gamma)=-\fft14(p^Iq_I)^2+\sum_{i<j}p^iq_ip^jq_j-p^0q_1q_2q_3+q_0p^1p^2p^3.
\label{eq:I4G}
\end{equation}
The BPS black holes have mass $M=|Z|$ and $\mathcal I_4(\Gamma)>0$, while the non-BPS black holes with $Z=0$ have $M>0$ and $\mathcal I_4(\Gamma)>0$.  The non-BPS black holes with $Z\ne0$ have $\mathcal I_4(\Gamma)<0$.  For the BPS mass, $M=|Z|$, the central charge function, (\ref{eq:Zccf}), is evaluated at asymptotic infinity.

The BPS black holes preserve four out of the eight real supersymmetries of the $\mathcal N=2$ theory and, in addition to $\mathcal I_4(\Gamma)>0$, satisfy the charge conditions
\begin{align}
    &p^0q_1-p^2p^3>0,\qquad p^0q_2-p^3p^1>0,\qquad p^0q_3-p^1p^2>0,\nn\\
    \mbox{or}\qquad&p^0q_1-p^2p^3<0,\qquad p^0q_2-p^3p^1<0,\qquad p^0q_3-p^1p^2<0.
\label{eq:BPSdomain}
\end{align}
In this case, the attractor mechanism fixes all scalars at the horizon.  The non-BPS black holes with $Z=0$ can be obtained from the BPS solution by flipping the signs of exactly two of $\{y^1,y^2,y^3\}$ in the BPS \textit{solution}.  Since physically $y^i=e^{-\varphi_i}$ is restricted to be positive, this means the sign flipped solution is physical in a different domain of charges than the BPS case, (\ref{eq:BPSdomain}).  There are three possible pair-wise sign flips that can be related by permuting $1,2,3$.  As one example, we can take
\begin{equation}
    y^1\to y^1,\qquad y^2\to-y^2,\qquad y^3\to-y^3,
\end{equation}
along with the non-BPS charge conditions
\begin{align}
    &p^0q_1-p^2p^3>0,\qquad p^0q_2-p^3p^1<0,\qquad p^0q_3-p^1p^2<0,\nn\\
    \mbox{or}\qquad&p^0q_1-p^2p^3<0,\qquad p^0q_2-p^3p^1>0,\qquad p^0q_3-p^1p^2>0.
\label{eq:nonBPSdomain}
\end{align}
The sign choices in (\ref{eq:BPSdomain}) along with (\ref{eq:nonBPSdomain}) and its permutations covers all possible charge assignments.  Note, also, that while the $Z=0$ solutions are non-BPS in $\mathcal N=2$ supergravity, the full set of $\mathcal I_4(\Gamma)>0$ solutions are all $1/8$ BPS in $\mathcal N=8$ supergravity, preserving four out of the 32 real supersymmetries.  The reason the $Z=0$ solutions are non-BPS from the $\mathcal N=2$ point of view is that the preserved supersymmetries lie outside of the $\mathcal N=2$ truncation.

In contrast to the $Z\ne0$ BPS and $Z=0$ non-BPS black holes, which both have $\mathcal I_4(\Gamma)>0$, the extremal $Z\ne0$ non-BPS black holes have $\mathcal I_4(\Gamma)<0$.  As a result, they necessarily lie on a disconnected branch of solutions.  Furthermore, in this case, the attractor mechanism no longer fixes all the scalars, leaving two real scalars as moduli.  From now on, we will disregard these $Z\ne0$ non-BPS black holes, and focus on the BPS branch (noting that the $Z=0$ non-BPS solutions can be obtained from the BPS solutions by appropriate sign flips).

The general asymptotically Minkowski BPS black holes are given in terms of 14 parameters: eight charges, collected in the charge vector (\ref{eq:Gvector}), and six real scalar VEVs.  The solution can be built out of eight harmonic functions, grouped as the symplectic vector
\begin{equation}
    \mathcal H=\begin{pmatrix}H^I\\H_I\end{pmatrix},\qquad H^I=p_\infty^I+\fft{p^I}r,\qquad H_I=q_{\infty\,I}+\fft{q_I}r.
\end{equation}
To avoid a naked singularity along the flow, the charges and corresponding asymptotic constants must have the same sign within each harmonic function.  While there appears to be eight real parameters $(p_\infty^I,q_{\infty\,I})$ related to the scalar VEVs, they have to satisfy two constraints
\begin{equation}
    \langle\Gamma,\Gamma_\infty\rangle=0,\qquad\mathcal I_4(\Gamma_\infty)=1,
\label{eq:twocons}
\end{equation}
leaving the appropriate six independent VEVs.  Here, in analogy to the charge vector $\Gamma$ defined in (\ref{eq:Gvector}), we have also defined the asymptotic `charge' vector
\begin{equation}
   \Gamma_\infty=\begin{pmatrix}p_\infty^I\\q_{\infty\,I}\end{pmatrix}.
\end{equation}
The first constraint is a requirement to solve the BPS equations, while the second ensures the metric is asyptotically Minkowski with $\eta_{\mu\nu}=(-1,1,1,1)$ at infinity.

The explicit BPS black hole solution is given as follows.  The metric takes the form
\begin{equation}
    ds^2=-e^{2U}dt^2+e^{-2U}[dr^2+r^2(d\theta^2+\sin^2\theta d\phi^2)],
\end{equation}
where
\begin{equation}
    e^{-4U}=I_4(\mathcal H).
\label{eq:4UBPS}
\end{equation}
The complex scalars are
\begin{equation}
    z^i=\fft{H^i+2i\partial_{q_{\infty\,i}}\sqrt{\mathcal I_4(\mathcal H)}}{H^0+2i\partial_{q_{\infty\,0}}\sqrt{\mathcal I_4(\mathcal H)}}.
\label{eq:ziBPS}
\end{equation}
The gauge fields carry conserved (Noether) electric charges $q_I$ and topological magnetic charges $p_I$ as defined in (\ref{eq:pIqI}).  In particular, we can write
\begin{equation}
    A^I=-f^I(r)dt-p^I\cos\theta\,d\phi,
\end{equation}
or
\begin{equation}
    \mathcal F^I=\partial_rf^I(r)dt\wedge dr+p^I\sin\theta\,d\theta\wedge d\phi.
\end{equation}
Here $f^I(r)$ satisfies
\begin{equation}
    \Im\mathcal N_{IJ}\partial_rf^I(r)=\fft{e^{2U}}{r^2}(q_I-\Re\mathcal N_{IJ}p^J).
\end{equation}

At the two-derivative level, the mass of the BPS black hole is $M=|Z|$ where $Z$ is given in (\ref{eq:Zccf}), and is evaluated at infinity.  The Bekenstein-Hawking entropy is
\begin{equation}
    S=\fft{A_{\mathrm{horizon}}}4=\pi\sqrt{\mathcal I_4(\Gamma)},
\end{equation}
where $\mathcal I_4(\Gamma)$ is given in (\ref{eq:I4G}).  The entropy depends only on the electric and magnetic charges, while the mass also depends on the scalar VEVs at infinity.  Note also that the scalar charges can be obtained by differentiating the mass with respect to the scalar VEVs.

\subsection{Axion-free solutions}

For general eight charge BPS solutions, the expression for the scalars, (\ref{eq:ziBPS}), can be somewhat complicated.  However, there are three distinct cases where the axions can be turned off throughout the attractor flow.  These are $i$) charges $(p^0,q_0)$ only, $ii$) charges $(p^0,q_1,q_2,q_3)$ only, and $iii$) charges $(p^1,p^2,p^3,q_0)$ only.  The first case, with charges $(p^0,q_0)$, gives rise to a negative quartic invariant
\begin{equation}
    \mathcal I_4(\Gamma)~\to~-\fft14(p^0q_0)^2,
\end{equation}
and is thus non-BPS.  Actually, for this reason, it does not fall into the class of solutions given by (\ref{eq:4UBPS}) and (\ref{eq:ziBPS}).  From a five-dimensional point of view, this corresponds to the Kaluza-Klein black hole with electric and magnetic charges turned on.  Setting $p^i=0$ and $q_i=0$ removes the five-dimensional vector multiplet charges, leaving pure geometry in five dimensions.

The other two axion-free cases, $(p^0,q_1,q_2,q_3)$ and $(p^1,p^2,p^3,q_0)$, are indeed BPS, and are often referred to as `electric' and `magnetic', respectively.  The electric black holes have
\begin{equation}
    \mathcal I_4(\Gamma)~\to~-p^0q_1q_2q_3.
\end{equation}
Since we demand $\mathcal I_4(\Gamma)>0$, this indicates that either one or three of the charges must be negative.  Compatibility with (\ref{eq:BPSdomain}) then demands either $p^0>0$ and $q_i<0$ or $p^0<0$ and $q_i>0$.  The electric BPS solution takes the form
\begin{align}
    ds^2&=-(-H^0H_1H_2H_3)^{-1/2}dt^2+(-H^0H_1H_2H_3)^{1/2}(dr^2+r^2(d\theta^2+\sin^2\theta d\phi^2)),\nn\\
    e^{-\varphi_i}=y^i&=\fft1{H_i}\sqrt{\fft{H_1H_2H_3}{-H^0}}.
\label{eq:eBPS}
\end{align}
The mass is
\begin{equation}
    M=|Z|_{\infty}=\left.\fft{|p^0y^1y^2y^3-q_1y^1-q_2y^2-q_3y^3|}{4\sqrt{y^1y^2y^3}}\right|_\infty,
\end{equation}
and the entropy is
\begin{equation}
    S=\pi\sqrt{-p^0q_1q_2q_3}.
\end{equation}

Imposing the asymptotic Minkowski condition and specializing to the $q_i>0$ case, the harmonic functions take the explicit form
\begin{align}
    H^0&=-e^{\fft12(\varphi_{1,\infty}+\varphi_{2,\infty}+\varphi_{3,\infty})}+\fft{p^0}r,\nn\\
    H_1&=e^{\fft12(\varphi_{1,\infty}-\varphi_{2,\infty}-\varphi_{3,\infty})}+\fft{q_1}r,\qquad\mbox{etc.},
\end{align}
and the expression for the mass becomes
\begin{equation}
    M=\fft14\left[q_1e^{\fft12(-\varphi_1+\varphi_2+\varphi_3)}+q_2e^{\fft12(\varphi_1-\varphi_2+\varphi_3)}+q_3e^{\fft12(\varphi_1+\varphi_2-\varphi_3)}-p^0e^{\fft12(-\varphi_1-\varphi_2-\varphi_3)}\right]_\infty.
\label{eq:eBHmass}
\end{equation}
(Recall that $p^0<0$ in this case.)

The magnetic black holes have
\begin{equation}
    \mathcal I_4(\Gamma)~\to~p^1p^2p^3q_0,
\end{equation}
and compatibility with (\ref{eq:BPSdomain}) then demands that the four charges are all positive or all negative.  These black holes have the explicit form
\begin{align}
    ds^2&=-(H_0H^1H^2H^3)^{-1/2}dt^2+(H_0H^1H^2H^3)^{1/2}(dr^2+r^2(d\theta^2+\sin^2\theta d\phi^2)),\nn\\
    e^{-\varphi_i}&=H^i\sqrt{\fft{H_0}{H^1H^2H^3}}
\label{eq:mBPS}
\end{align}
The mass is
\begin{equation}
    M=|Z|_{\infty}=\left.\fft{|q_0+p^1y^2y^3+p^2y^3y^1+p^3y^1y^2|}{4\sqrt{y^1y^2y^3}}\right|_\infty,
\end{equation}
and the entropy is
\begin{equation}
    S=\pi\sqrt{q_0p^1p^2p^3}.
\end{equation}

When all four charges are positive and the asymptotic Minkowski condition is satisfied, the harmonic functions take the explicit form
\begin{align}
    H_0&=e^{-\fft12(\varphi_{1,\infty}+\varphi_{2,\infty}+\varphi_{3,\infty})}+\fft{q_0}r,\nn\\
    H^1&=e^{\fft12(-\varphi_{1,\infty}+\varphi_{2,\infty}+\varphi_{3,\infty})}+\fft{p^1}r,\qquad\mbox{etc.},
\end{align}
and the expression for the mass becomes
\begin{equation}
    M=\fft14\left[p^1e^{\fft12(\varphi_1-\varphi_2-\varphi_3)}+p^2e^{\fft12(-\varphi_1+\varphi_2-\varphi_3)}+p^3e^{\fft12(-\varphi_1-\varphi_2+\varphi_3)}+q_0e^{\fft12(\varphi_1+\varphi_2+\varphi_3)}\right]_\infty.
\end{equation}
This expression is essentially the overall electric/magnetic dual of (\ref{eq:eBHmass}).

\subsection{Truncating to a single vector multiplet}
\label{SingleVectorTruncation}

We can truncate the STU model to a single vector multiplet by identifying the three vector multiplets with each other. This truncation can be lifted to ungauged $\mathcal N=2$ supergravity in five dimensions coupled to a single vector multiplet.  In particular, by setting
\begin{equation}
    z^1=z^2=z^3,\qquad A^1=A^2=A^3,
\label{eq:zAtrunc}
\end{equation}
the STU model Lagrangian (\ref{eq:STUchow}) reduces to
\begin{align}
    16\pi e^{-1}\mathcal L&=R-\fft32\left((\partial_\mu\varphi)^2+e^{2\varphi}(\partial_\mu\chi)^2\right)-\fft14e^{-3\varphi}(\mathcal F_{\mu\nu}^0)^2-\fft34e^{-\varphi}(\mathcal F^1_{\mu\nu}-\chi\mathcal F^0_{\mu\nu})^2\nn\\
    &\quad+\chi^3\mathcal F^0\wedge\mathcal F^0-3\chi^2\mathcal F^1\wedge\mathcal F^0+3\chi\mathcal F^1\wedge\mathcal F^1.
    \label{eq:truncSTU}
\end{align}
The lifting to five-dimensional $\mathcal N=2$ supergravity
\begin{equation}
    16\pi\hat e^{-1}\hat{\mathcal L}=\hat R-\fft14\hat F_{\mu\nu}^2+\fft1{3\sqrt3}\hat A\wedge\hat F\wedge\hat F,
\end{equation}
then proceeds according to
\begin{align}
    d\hat s^2&=e^{\varphi}g_{\mu\nu}dx^\mu dx^\nu+e^{-2\varphi}\eta^2,\nn\\
    \hat A&=\sqrt 3(A^1_\mu dx^\mu-\chi\eta),\qquad \eta=dz+A^0_\mu dx^\mu.
\end{align}
The truncated STU model Lagrangian, (\ref{eq:truncSTU}), is equivalent to the dimensionally reduced action, (\ref{eq:ReducedAction2der}), considered in section~\ref{SectionN2SUGRA}, however with the following scaling/redefinitions of the fields
\begin{equation}
\varphi=\fft1{\sqrt3}\phi^{(IV)},\quad\chi=-\fft1{\sqrt3}\chi^{(IV)},\qquad
A^0=\mathcal A^{(IV)},\qquad A^1=\fft1{\sqrt3}A^{(IV)},
\label{eq:normswitch}
\end{equation}
along with a reversal in orientation in the reduction of $\epsilon_5$ to $\epsilon_4$ in the CP-odd sector.

For the black hole solution, the truncation (\ref{eq:zAtrunc}) amounts to taking the charges
\begin{equation}
    p^1=p^2=p^3,\qquad q_1=q_2=q_3,
\end{equation}
while leaving $p^0$ and $q_0$ free.  The solution is given in terms of four harmonic functions (\textit{e.g.} $H^0$, $H^1$, $H_0$ and $H_1$) and two constraints, (\ref{eq:twocons}), leaving six independent parameters which can be taken as four real charges and the complex scalar VEV, $\langle z^1\rangle$, at infinity.  Note that, compared to the conventions of section~\ref{SectionN2SUGRA}, the dilaton and axion VEVs are given by
\begin{equation}
    \varphi_\infty=\fft1{\sqrt3}\phi^{(IV)}_\infty,\qquad\chi_\infty=-\fft1{\sqrt3}\chi^{(IV)}_\infty,
\end{equation}
while the charges are mapped according to
\begin{equation}
    q_0=2Q_0,\qquad q_1=\fft2{\sqrt3}Q_1,\qquad p^0=-2P_0,\qquad p^1=-\fft2{\sqrt3}P_1.
\label{eq:chargemap}
\end{equation}
In particular, the magnetic charges flip sign because of the orientation reversal when switching between (\ref{eq:truncSTU}) and (\ref{eq:ReducedAction2der}).

After truncation, the four-dimensional solution takes the form
\begin{align}
    ds^2&=-e^{2U}dt^2+e^{-2U}(dr^2+r^2(d\theta^2+\sin^2\theta d\phi^2)),\nn\\
    A^0&=-f_0(r)dt-p^0\cos\theta d\phi,\nn\\
    A^1=A^2=A^3&=-f_1(r)dt-p^1\cos\theta d\phi,\nn\\
    z^1=z^2=z^3&=\chi-ie^{-\varphi},
\end{align}
where the functions $f_0(r)$ and $f_1(r)$ are related to the electric charges
\begin{align}
    \partial_rf_0(r)&=-e^{3\varphi(r)}\fft{e^{2U}}{r^2}\left(-q_0-3q_1\chi(r)+3p^1\chi(r)^2-p^0\chi(r)^3\right),\nn\\
    \partial_rf_1(r)&=-e^{\varphi(r)}\fft{e^{2U}}{r^2}\bigl(-q_1+2p^1\chi(r)-p^0\chi(r)^2\nn\\
    &\kern6.5em+\chi(r)e^{2\varphi(r)}\left(-q_0-3q_1\chi(r)+3p^1\chi(r)^2-p^0\chi(r)^3\right)\bigr).
\end{align}
The metric function is
\begin{equation}
    e^{2U}=\fft1{\sqrt{\mathcal I_4(\mathcal H)}},
\label{eq:e2Ut}
\end{equation}
where
\begin{equation}
    \mathcal I_4(\mathcal H)=((H_1)^2+H_0H^1)((H^1)^2-H^0H_1)-\fft14(H^0H_0+H^1H_1)^2,
\label{eq:I44}
\end{equation}
and the axion and dilaton are
\begin{equation}
    \chi=\fft12\fft{H^0H_0+H^1H_1}{(H^1)^2-H^0H_1},\qquad e^{-\varphi}=\fft{e^{-2U}}{(H^1)^2-H^0H_1}.
\label{eq:cpt}
\end{equation}

The asymptotic parameters $(p^0_\infty,p^1_\infty,q_{\infty0},q_{\infty1})$ must be chosen to satisfy the constraints (\ref{eq:twocons}).  Written in terms of the scalar VEVs $\chi_\infty=\langle\chi\rangle$ and $\varphi_\infty=\langle\varphi\rangle$, we find explicitly
\begin{align}
\label{allparameters}
    p^0_\infty&=\fft{e^{\fft32\varphi_\infty}\Im Z_\infty}{|Z_\infty|},\nn\\
    p^1_\infty&=\fft{e^{\fft12\varphi_\infty}\Re Z_\infty+\chi_\infty e^{\fft32\varphi_\infty}\Im Z_\infty}{|Z_\infty|},\nn\\
    q_{\infty1}&=-\fft{(e^{-\fft12\varphi_\infty}-\chi_\infty^2e^{\fft32\varphi_\infty})\Im Z_\infty-2\chi_\infty e^{\fft12\varphi_\infty}\Re Z_\infty}{|Z_\infty|},\nn\\
    q_{\infty0}&=\fft{(e^{-\fft32\varphi_\infty}-3\chi_\infty^2e^{\fft12\varphi_\infty})\Re Z_\infty+\chi_\infty(3e^{-\fft12\varphi_\infty}-\chi_\infty^2e^{\fft32\varphi_\infty})\Im Z_\infty}{|Z_\infty|},
\end{align}
where the central charge is
\begin{align}
    Z_\infty&=\fft14\Bigl[\left(e^{\fft32\varphi_\infty}(q_0+3\chi_\infty q_1-3\chi_\infty^2p^1+\chi_\infty^3p^0)+3e^{-\fft12\varphi_\infty}(p^1-\chi_\infty p^0)\right)\nn\\
    &\qquad+i\left(e^{-\fft32\varphi_\infty}p^0-3e^{\fft12\varphi_\infty}(q_1-2\chi_\infty p^1+\chi_\infty^2p^0)\right)\Bigr].
\label{eq:Zinfty4}
\end{align}
These expressions simplify considerably if we set $\chi_\infty=0$, yielding
\begin{equation}
    e^{-\fft32\varphi_\infty}p^0_\infty=-e^{\fft12\varphi_\infty}q_{\infty1}=\fft{\Im Z_\infty}{|Z_\infty|},\qquad e^{-\fft12\varphi_\infty}p^1_\infty=e^{\fft32\varphi_\infty}q_{\infty0}=\fft{\Re Z_\infty}{|Z_\infty|},
\end{equation}
where
\begin{equation}
    Z_\infty=\fft14\left[\left(e^{\fft32\varphi_\infty}q_0+3e^{-\fft12\varphi_\infty}p^1\right)+i\left(e^{-\fft32\varphi_\infty}p^0-3e^{\fft12\varphi_\infty}q_1\right)\right].
\end{equation}
Finally, the mass and entropy of the BPS solution can be read off from (\ref{eq:Zinfty4}) and (\ref{eq:I44}), respectively.

\subsection{Two-charge solutions to the truncated STU model}

In addition to the non-BPS Kaluza-Klein black hole with charges $(p^0,q_0)$ considered in section~\ref{sec:q0p0}, there are two axion-free BPS solutions with corresponding charges $(p^0,q_1)$ and $(p^1,q_0)$.  The BPS $(p^0,q_1)$ solution can be obtained from (\ref{eq:eBPS}) by setting $H_1=H_2=H_3$, and takes the form
\begin{align}
    ds^2&=-(H^0(H_1)^3)^{-1/2}dt^2+(H^0(H_1)^3)^{1/2}(dr^2+r^2(d\theta^2+\sin^2\theta d\phi^2)),\nn\\
    e^{-\varphi}&=\sqrt{\fft{H_1}{H^0}},\qquad\chi=0,\nn\\
    A^0&=p^0\cos\theta d\phi,\qquad A^1=\left(e^{\fft12\varphi_\infty}-\fft1{H_1}\right)dt,
\end{align}
where
\begin{equation}
    H^0=e^{\fft32\varphi_\infty}+\fft{p^0}r,\qquad H_1=e^{-\fft12\varphi_\infty}+\fft{q_1}r.
\end{equation}
Note that we have flipped the sign of the magnetic charge $p^0$ and the corresponding harmonic function $H^0$ compared with (\ref{eq:eBPS}).  The mass and entropy of this BPS black hole are
\begin{equation}
    M=\fft14(e^{-\fft32\varphi_\infty}p^0+3e^{\fft12\varphi_\infty}q_1),\qquad S=\pi\sqrt{p^0(q_1)^3}.
\end{equation}
The non-extremal generalization of this solution is considered in section~\ref{sec:p0q1} and is given in (\ref{eq:p0q1bh}), keeping in mind the different normalization conventions, (\ref{eq:normswitch}), and the mapping of charges, (\ref{eq:chargemap}), which for this solution is given by
\begin{equation}
    q_1=\fft2{\sqrt3}Q_1,\qquad p^0=2P_0.
\end{equation}
Note that there is no minus sign in the relation between $p^0$ and $P_0$ since we have already flipped the sign of $p^0$.

The BPS $(p^1,q_0)$ solution to the truncated STU model is obtained from (\ref{eq:mBPS}), and takes the form
\begin{align}
    ds^2&=-(H_0(H^1)^3)^{-1/2}dt^2+(H_0(H^1)^3)^{1/2}(dr^2+r^2(d\theta^2+\sin^2\theta d\phi^2)),\nn\\
    e^{-\varphi}&=\sqrt{\fft{H_0}{H^1}},\qquad\chi=0,\nn\\
    A^0&=\left(e^{\fft32\varphi_\infty}-\fft1{H_0}\right)dt,\qquad A^1=-p^1\cos\theta d\phi,
\end{align}
where
\begin{equation}
    H_0=e^{-\fft32\varphi_\infty}+\fft{q_0}r,\qquad H_1=e^{\fft12\varphi_\infty}+\fft{p^1}r.
\end{equation}
The mass and entropy of this BPS black hole are
\begin{equation}
    M=\fft14(e^{\fft32\varphi_\infty}q_0+3e^{-\fft12\varphi_\infty}p^1),\qquad S=\pi\sqrt{q_0(p^1)^3}.
\end{equation}
Note that this solution can be obtained from the $(p^0,q_1)$ one by electric/magnetic duality.  The non-extremal generalization of this solution is considered in section~\ref{sec:p1q0} and is given in (\ref{eq:p1q0bh}), again noting the change in normalization.

We also consider the BPS two-charge $(p^1,q_1)$ solution in section~\ref{sec:p1q1}.  Here the axion is turned on, and the solution is given by setting $q_0=0$ and $p^0=0$ in (\ref{eq:e2Ut}), (\ref{eq:I44}) and (\ref{eq:cpt}).  In this case, the quartic invariant is
\begin{equation}
    \mathcal I_4(\Gamma)~\to~\fft34(p^1q_1)^2,
\end{equation}
and both charges can take either sign.

%%%%%

\section{Details of Dimensional Reduction}

Below we present details of the dimensional reduction of Gauss-Bonnet gravity on a circle. 
We start by considering Einstein gravity coupled to Gauss-Bonnet curvature corrections in five dimensions, 
\begin{align}
    S = -\frac{1}{16 \pi} \int d^5 \hat{x} \, \sqrt{-\hat{g}} \, \hat{R} \, + c_1 \left( \hat{R}_{\mu \nu \rho \sigma} \hat{R}^{\mu \nu \rho \sigma}- 4 \hat{R}_{\mu \nu } \hat{R}^{\mu \nu } + \hat{R}^2 \right) ,
\end{align}
where the parameter $c_1$ is assumed to be small, 
so that the higher-derivative terms can be treated perturbatively.
The dimensionally reduced theory is obtained from the 
reduction ansatz 
\begin{align}
    \hat{g}_{\hat{\mu} \hat{\nu}} = 
    \begin{pmatrix}
    e^{2 \alpha \phi} g_{\mu \nu} + e^{2 \beta \phi}  A_\mu A_\nu  &  e^{2 \beta \phi}  A_\mu\\
    e^{2 \beta \phi}  A_\nu & e^{2 \beta \phi}
    \end{pmatrix}  ,
\end{align}
where $g_{\mu\nu}$ is the four-dimensional metric, $A_\mu$ is the graviphoton and the scalar field $\phi$ describes the radion or dilaton.
We use hats to denote five-dimensional quantities, while 
the unhatted indices $\mu,\nu$ run over $0, \ldots, 3$ and the index $5$ labels the compactified dimension. 
Finally, the two parameters $\alpha$ and $\beta$ are left arbitrary for now.
The inverse metric is then given by
\begin{align}
    \hat{g}^{\hat{\mu} \hat{\nu}} =  e^{-2 (\alpha+\beta) \phi}
    \begin{pmatrix}
    e^{2 \beta \phi} g^{\mu \nu}  &  -A^\mu e^{2 \beta \phi}\\
    -A^\nu e^{2 \beta \phi} &  e^{2 \alpha \phi} + e^{2 \beta \phi} A^2
    \end{pmatrix}  .
\end{align}
Using this ansatz and taking $\mathcal{F}=dA$, the five-dimensional Christoffel symbols can be related to four-dimensional quantities in the following way,
\begin{align}
\hat{\Gamma}^\mu_{\nu\lambda}
= \Gamma^\mu_{\nu\lambda} + \alpha(\partial_\nu \phi \, \delta^\mu _\lambda + \partial_\lambda \phi \, \delta^\mu _\nu - \partial^\mu \phi \, g_{\nu_\lambda})-\beta e^{2(\beta - \alpha)\phi} \partial^\mu \phi A_\nu A_\lambda + \frac{1}{2} e^{2(\beta - \alpha)\phi}(A_\nu \mathcal{F}_\lambda {} ^\mu+ A_\lambda \mathcal{F}_\nu {} ^\mu)\, , 
\end{align} 
\begin{align}
\hat{\Gamma}^5_{\nu\lambda}
=(\beta-\alpha)(\partial_\nu \phi \, A_\lambda + \partial_\lambda \phi \, A_\nu)+\alpha (A^\mu \partial_\mu \phi) \, g_{\nu \lambda}+\beta e^{2(\beta - \alpha)\phi} (A^\mu \partial_\mu \phi)A_\nu A_\lambda \nonumber\\
+\frac{1}{2}\Bigl( \partial_\nu A^\mu \, g_{\mu \lambda} + 
\partial_\lambda A^\mu \, g_{\mu \nu}+ A^\mu \partial_\mu g_{\nu \lambda}\Bigr)-\frac{1}{2}e^{2(\beta - \alpha)\phi}(A_\mu A_\nu \mathcal{F}_\lambda {} ^\mu+ A_\mu A_\lambda \mathcal{F}_\nu {} ^\mu) \, , 
\end{align}
\begin{align}
\hat{\Gamma}^\mu_{5 \lambda}=-\beta e^{2(\beta - \alpha)\phi}\partial^\mu \phi A_\lambda+\frac{1}{2}e^{2(\beta - \alpha)\phi}\mathcal{F}_\lambda {} ^\mu \, ,
\end{align}
\begin{align}
\hat{\Gamma}^5_{5 \mu}=\beta \, e^{2(\beta - \alpha)\phi} A_\mu A^\nu \partial_\nu \phi -\frac{1}{2}e^{2(\beta - \alpha)\phi} A^\nu \mathcal{F}_{\mu\nu}+\beta \partial_\mu \phi \, ,
\end{align}
\begin{align}
\hat{\Gamma}^\mu_{55}=-\beta e^{2(\beta - \alpha)\phi}\partial^\mu \phi \, ,
\end{align}
\begin{align}
\hat{\Gamma}^5_{55}=\beta e^{2(\beta - \alpha)\phi} \,  A^\nu \partial_\nu \phi \, .
\end{align}
The dimensional reduction of the components of the Ricci tensor then becomes
\begin{align}
\label{RicciTensorCoord1}
 \hat{R}_{\mu \nu}
& =R_{\mu \nu}-\alpha g_{\mu \nu}\nabla^2\phi-\beta(\beta-\alpha)\partial_\mu \phi\partial_\nu \phi -\frac{1}{2}e^{2(\beta - \alpha)\phi}\mathcal{F}_\mu{}^\lambda \mathcal{F}_{\nu \lambda}\nonumber\\
& -(2\alpha+\beta)(\nabla_\mu \nabla_\nu \phi
+\alpha g_{\mu \nu}(\partial\phi)^2-\alpha\partial_\mu \phi\partial_\nu\phi)\nonumber\\
& +e^{2(\beta - \alpha)\phi}A_\mu (-\frac{1}{2}\nabla_\lambda \mathcal{F}^\lambda{}_\nu-\frac{3}{2}\beta\partial_\lambda \phi \mathcal{F}^\lambda{}_\nu)+e^{2(\beta - \alpha)\phi}A_\nu (-\frac{1}{2}\nabla_\lambda \mathcal{F}^\lambda{}_\mu-\frac{3}{2}\beta\partial_\lambda \phi \mathcal{F}^\lambda{}_\mu)\nonumber\\
& -e^{2(\beta - \alpha)\phi} \beta A_\mu A_\nu  \nabla^2\phi
+\frac{1}{4}e^{4(\beta-\alpha)\phi}A_\mu A_\nu \mathcal{F}^2
- e^{2(\beta - \alpha)\phi}\beta(2\alpha+\beta)A_\mu A_\nu(\partial \phi)^2 \, ,
\end{align}
\begin{align}
\label{RicciTensorCoord2}
\hat{R}_{5\mu} & =
-\beta e^{2(\beta-\alpha)\phi}\nabla^2 \phi A_\mu
+\frac{3}{2} \beta e^{2(\beta-\alpha)\phi}
\partial^\rho \phi \mathcal{F}_{\mu \rho} 
+ \frac{1}{2} e^{2(\beta-\alpha)\phi}\nabla^\rho \mathcal{F}_{\mu \rho}\nonumber\\
& +\frac{1}{4} e^{4(\beta - \alpha)\phi}A_\mu \mathcal{F}^2-
\beta(2\alpha+\beta)e^{2(\beta-\alpha)\phi} (\partial \phi)^2 A_\mu \, ,
\end{align}
\begin{align}
\label{RicciTensorCoord3}
\hat{R}_{55}=-\beta e^{2(\beta - \alpha)\phi}\nabla^2 \phi+\frac{1}{4} e^{4(\beta - \alpha)\phi}\mathcal{F}^2-\beta(2\alpha+\beta)e^{2(\beta - \alpha)\phi}(\partial \phi)^2 \, .
\end{align}
Finally, the dimensionally reduced Ricci scalar is given by
    \begin{align}
\hat{R}=R-2(3\alpha+\beta)\nabla^2 \phi-2(\beta^2+3\alpha^2+2\alpha\beta)(\partial \phi)^2 -\frac{1}{4}e^{2(\alpha-\beta)\phi}\mathcal{F}^2 .
\end{align}

Calculating the dimensional reduction of the Riemann tensor in the coordinate basis is extremely cumbersome. 
Thus, to compute it we switch to working with vielbeins and using the 
Cartan structure equations. 
The vielbeins take the form
\begin{align}
 &\hat{e}^a = e^{\alpha \phi} e^a, \qquad \qquad  
 \hat{e}^y = e^{\beta \phi} (d y + A) \, ,
\end{align}
where Latin letters $a,b,$ etc denote tangent space indices 
and $y$ denotes the fifth dimension.
In what follows, for convenience we will absorb 
$e^{(\beta - \alpha)\phi}$ into the definition of the flux, and denote the resulting quantity by $F$, 
%$\mathcal{F}$ for simplicity:
\begin{align}
    e^{(\beta - \alpha)\phi} \mathcal{F}\longrightarrow F \, .
\end{align}
The first structure equation tells us that five-dimensional spin-connection takes the form, 
\begin{align}
    & \hat{\omega}^{ab} \ = \ \omega^{ab} + \alpha \,  \left( \partial^b \phi  \ e^{a} - \partial^a  \phi \ e^b \right) - \frac{1}{2} \, e^{-2 \alpha \phi} \, F^{ab}  \, e^y \, , \\
    & \hat{\omega}^{a y} \ = \ - e^{( \beta - \alpha)\phi} \left( \beta  \ \partial^a \phi \ e^y  + \frac{1}{2} \,F^a_b \, e^b\right) \, .
 \end{align}
The second structure equation gives us the Riemann tensor, which in the vielbein basis is 
\begin{align}
    \begin{split}
        & e^{2 \alpha \phi} \hat{R}^{ab}{}_{cd} =  R^{a b}{}_{cd} - \frac{1}{2}  \, \left(  F^{ab}  \, F_{cd} +  \, F^a{}_{[c} \, F^{b}{}_{d]}  \right) \\
        & \qquad \qquad    - 4\, \alpha \ \delta^{[a}_{[c}  \  \nabla_{d]} \nabla^{b]} \phi   + 4\, \alpha^2 \  \delta^{[a}_{[c} \nabla_{d]}  \phi \ \nabla^{b]}  \phi   -2 \, \alpha^2 \ \delta^{[a}_{[c} \ \delta^{b]}_{d]} \  (\nabla \phi)^2  ,\\
        \\
        & e^{2 \alpha \phi} \hat{R}^{y b}{}_{cd} = \  ( \beta - \alpha) \   \left( \nabla_{[c} \phi  \ F^b{}_{d]}  + \nabla^b \phi \ F_{cd} \right)    + \frac{1}{2} \, \nabla^b F_{cd}   + \alpha \ \delta^{b}_{[c} \, \nabla^a  \phi \ F_{a d] } \, , \\
        \\
        & e^{2 \alpha \phi} \hat{R}^{y b}{}_{yd} = - \beta \, ( \beta - 2 \alpha) \ \nabla^b \phi \ \nabla_d \phi  - \beta  \  \nabla^b \, \nabla_d \phi - \alpha \,  \beta  \ (\nabla \phi)^2 \ \delta^b_d  + \frac{1}{4}  \,F_{cd} \, F^{cb} ,
    \end{split}
\end{align}
where the pairs $[a, \, b]$ and $[c,\, d]$ are anti-symmetrized over, with weight 1. Now we can compute the Ricci tensor and Ricci scalar, 
%with $D=4$
%
\begin{align}
\label{RicciVielbein}
    \begin{split}
        & e^{2 \alpha \phi} \hat{R}_{ab} =   R_{ab} - \frac{1}{2}  \,  F_{ac}  \, F_{bc}    - \, \alpha  \ \eta_{ab}  \  \nabla^2 \phi \, - \left( 2 \alpha + \beta \right) \nabla_a \nabla_b \phi  \\
        & \qquad \qquad \qquad - \left( 2 \alpha + \beta \right) \alpha \eta_{a b} (\nabla \phi)^2 + \left( 2 \alpha^2 -  \beta \  ( \beta - 2 \alpha) \right)  \ \nabla_a  \phi \ \nabla_b  \phi  \, ,\\
        \\
        & e^{2 \alpha \phi} \hat{R}_{yb} = \ \frac{3 \beta}{2} \, \nabla_d  \phi \, F_{bd} + \frac{1}{2} \   \nabla^d F_{bd} \, , \\
        \\
        & e^{2 \alpha \phi} \hat{R}_{yy} =  - \beta ( \beta + 2 \alpha ) (\nabla \phi)^2 - \beta  \, \nabla^2 \phi + \frac{1}{4}  \,F^2 \, ,
    \end{split}
\end{align}
and
\begin{align}
    \begin{split}
        &e^{2 \alpha \phi} \hat{R} =  R - \frac{1}{4}  \,  F^2   -2 \, \left( 3 \alpha + \beta \right) \  \nabla^2 \phi - \left( 6 \alpha^2 + 4 \alpha \beta + 2 \beta^2  \right) \  (\nabla  \phi)^2  \, .
    \end{split}
\end{align}

The curvature tensors in the coordinate basis can be easily related to 
those in the vielbein basis.  \\
For the components of the Ricci tensor, we have
\begin{align}
\hat{R}_{\mu\nu}=e^a{}_\mu \, e^b{}_\nu \; e^{2\alpha \phi}\hat{R}_{ab}+2 e^{(\beta + \alpha)\phi}A_\mu \; e^a{}_\nu \; \hat{R}_{ay}+e^{2\beta \phi} A_\mu A_\nu \hat{R}_{yy} \, ,
\end{align}
\begin{align}
\hat{R}_{5\mu}=e^a{}_\mu \; e^{(\beta+\alpha)\phi}\hat{R}_{ya}+e^{2\beta\phi}A_\mu \hat{R}_{yy} ,
\end{align}
\begin{align}
    \hat{R}_{55}= e^{2\beta \phi}\hat{R}_{yy} .
\end{align}
These relations can be used to check that the Ricci tensor expressions (\ref{RicciTensorCoord1}), (\ref{RicciTensorCoord2}) and (\ref{RicciTensorCoord3})  obtained in the coordinate basis are indeed consistent with (\ref{RicciVielbein}). Similarly, one could relate the Riemann tensor in the coordinate basis to that in the vielbein basis by a similar transformation. Including such a transformation is beyond the scope of this paper.

We now have all the ingredients needed to compute the reduction of the
five-dimensional Gauss-Bonnet term $\hat{R}_{\text{GB}} = \hat{R}_{\mu \nu \rho \sigma} \hat{R}^{\mu \nu \rho \sigma}- 4 \hat{R}_{\mu \nu } \hat{R}^{\mu \nu } + \hat{R}^2 $. 
One obtains the following expression,
\begin{align}
   \hat{R}_{\text{GB}}=R_{\text{GB}} -\frac{3}{8}F^4+\frac{3}{16}(F^2)^2 % F_{ab}F^{ab}F_{cd}F^{cd}
   +F_a{}^c F^{ab} R_{bc}-\frac{1}{3}F^2 R
   -\frac{1}{2}F^{ab}F^{cd}W_{abcd}\nonumber\\
   + (2\alpha^2 +3\alpha\beta + 2\beta^2)F^2  (\partial \phi)^2  +(-8\alpha^2 -8\alpha\beta -4 \beta^2)(\partial\phi)^2R\nonumber\\
   +(24\alpha^3 + 48\alpha^2\beta +24\alpha\beta^2)(\partial\phi)^2(\nabla^2\phi)+(-6\alpha^2 -8\alpha\beta - 6 \beta^2)F_a{}^cF_{bc}(\partial^a\phi)(\partial^b\phi)\nonumber\\
   +(16\alpha^2 + 16\alpha\beta +8\beta^2)R_{ab}(\partial^a\phi)(\partial^b\phi)+(16\alpha^3\beta+8\alpha^2\beta^2)\left((\partial\phi)^2\right)^2\nonumber\\
   +(2\alpha+2\beta)F_a{}^b(\partial^a\phi)(\nabla_c F_b{}^c)+(-\frac{\alpha}{2}-\frac{\beta}{2})F^2(\nabla^2 \phi) \, ,
\end{align}
where 
$R_{\text{GB}}$ denotes the four-dimensional Gauss-Bonnet combination, and we have used the Mathematica package xTensor to perform tensor manipulations. 
The expression can be simplified further using the equations of motion, 
giving
\begin{align}
\label{finalGB}
    \hat{R}_{\text{GB}}=R_{\text{GB}} +\frac{1}{8}F^4 + \frac{2\alpha+\beta}{32\alpha}(F^2)^2-\frac{1}{2}F^{ab}F^{cd}W_{abcd}\nonumber\\
    +3\alpha(\alpha + \beta)F^2(\partial \phi)^2+(8\alpha^2 +6\alpha\beta - 2 \beta^2)F_a{}^cF_{bc}(\partial^a\phi)(\partial^b\phi)\nonumber\\
    -8\alpha(12\alpha^3+25\alpha^2\beta+20\alpha\beta^2+6\beta^3)\left((\partial\phi)^2\right)^2 \, .
\end{align}
Thus far we have left the parameters $\alpha$ and $\beta$ unspecified. Note that for 
general choices of $\alpha$ and $\beta$ we would have a non-canonical Einstein term 
$\mathcal{L}= e^{(\beta+(D-2)\alpha)\phi}\sqrt{-g} \, R$ in D dimensions. 
Thus, in four dimensions in order to work with the standard Einstein-Hilbert action we choose $\beta=-2\alpha$, which in turn gives 
\begin{align}
     \hat{R}_{\text{GB}}=R_{\text{GB}} +\frac{1}{8}F^4-\frac{1}{2}F^{ab}F^{cd}W_{abcd}
    -3\alpha^2 F^2(\partial \phi)^2\nonumber\\
    -12\alpha^2 F_a{}^cF_{bc}(\partial^a\phi)(\partial^b\phi) +48\alpha^4\left((\partial\phi)^2\right)^2 .
\end{align}

The dimensional reduction of Riemann squared, Ricci tensor squared and Ricci scalar squared 
respectively take the form
\begin{align}
    \hat{R}_{abcd}\hat{R}^{abcd} & =R^{abcd}R_{abcd}-\frac{3}{2}F^{ab}F^{cd}W_{abcd}-\frac{7}{8}F^4 +\frac{(48\alpha^2+92\alpha\beta+45\beta^2)}{64(\alpha+\beta)^2}(F^2)^2\nonumber\\
  & +(-6\alpha+6\beta)F^{bc}(\nabla_aF_{bc})(\partial^a\phi)+\frac{(6\alpha^3-30\alpha^2\beta-23\alpha\beta^2+11\beta^3)}{2(\alpha+\beta)}F^2(\partial\phi)^2\nonumber\\
  &  -4(7\alpha^4+6\alpha^3\beta-\alpha^2\beta^2+4\alpha\beta^3-\beta^4)\left((\partial\phi)^2\right)^2 \nonumber \\ & +8(4\alpha^3+6\alpha^2\beta-2\alpha\beta^2+\beta^3)(\partial^a\phi)(\partial^b\phi)(\nabla_a\nabla_b\phi)\nonumber\\
 &  +4(2\alpha^2+\beta^2)(\nabla_b\nabla_a\phi)(\nabla^b\nabla^a\phi)
    -2(4\alpha^2-11\alpha\beta-2\beta^2)F_a{}^cF_{bc}(\partial^a\phi)(\partial^b\phi)\nonumber\\
  &  +2(\alpha-\beta)F_a{}^cF^{ab}(\nabla_c\nabla_b\phi)+(\nabla_cF_{ab})(\nabla^cF^{ab}) ,
\end{align}
\begin{align}
    -4\hat{R}_{ab}\hat{R}^{ab} & =-4R_{ab}R^{ab}+F^4-\frac{(48\alpha^4+72\alpha^3\beta+36\alpha^2\beta^2+6\alpha\beta^3+\beta^4)}{64\alpha^2(\alpha+\beta)^2}(F^2)^2\nonumber\\
&   6\alpha(\alpha+\beta)F^2(\partial\phi)^2 -6(16\alpha^2-8\alpha\beta+3\beta^2)F_a{}^cF_{bc}(\partial^a\phi)(\partial^b\phi)\nonumber\\
&    +4(2\alpha^2+2\alpha\beta-\beta^2)(10\alpha^2+10\alpha\beta+\beta^2)\left((\partial\phi)^2\right)^2\nonumber\\
&    -8(2\alpha+\beta)^3(\nabla_b\nabla_a\phi)(\partial^a\phi)(\partial^b\phi)-4(2\alpha+\beta)^2(\nabla_b\nabla_a\phi)(\nabla^b\nabla^a\phi) ,
\end{align}
\begin{align}
\hat{R}^2=R^2+\frac{\left((\alpha-\beta)(2\alpha+\beta)\right)^2}{64\alpha^2(\alpha+\beta)^2} (F^2)^2-36\alpha^2(\alpha+\beta)^2\left((\partial\phi)^2\right)^2 ,
\end{align}
Adding these terms to form the Gauss Bonnet combination $\hat{R}_{\text{GB}}$ will yield (\ref{finalGB}), 
up to terms that involve total derivatives, which we show below:
%\subsection{Total Derivatives}
\begin{align}
   e^{2\alpha\phi}\nabla^a\left(e^{-2\alpha\phi}(\nabla_b\nabla_a\phi)(\partial^b\phi)-e^{-2\alpha\phi}(\partial_a\phi)\nabla^2\phi\right)=\nonumber\\
    -(\nabla^2 \phi)^2+2\alpha(\partial\phi)^2\nabla^2\phi+R_{ab} (\partial^a\phi)(\partial^b\phi)-2\alpha(\nabla_b\nabla_a \phi)(\partial^a\phi)(\partial^b\phi)+(\nabla_b\nabla_a\phi)(\nabla^b\nabla^a\phi)\, ,
\end{align}
\begin{align}
    e^{2\alpha\phi} \nabla_b \left(e^{-2\alpha\phi}(\partial\phi)^2\partial^b\phi\right)=\nonumber\\
    2(\partial^a\phi) (\partial^b\phi)(\nabla_b\nabla_a\phi)+ (\partial\phi)^2(\nabla^2\phi-2\alpha(\partial\phi)^2) \, ,
\end{align}
\begin{align}
    2 e^{-2(\beta-2\alpha)\phi}\nabla_a \left(e^{2(\beta-2\alpha)\phi}F^{ab} \nabla_c F^c{}_b\right)
    -e^{-2(\beta-2\alpha)\phi}\nabla_c
    \left(e^{2(\beta-2\alpha)\phi}F_{ab} \nabla^c F^{ab}\right)=\nonumber\\
    -2F_a{}^cF^{ab}R_{bc}+2F^{ab}F^{cd}R_{abcd}
    +(4\alpha-2\beta) F^{bc}(\partial^a\phi\nabla_a F_{bc})
    -2(\nabla_a F^{ab})(\nabla_c F_b{}^c)\nonumber\\
    +(8\alpha-4\beta) F_a{}^b(\nabla_c F_b{}^c)-(\nabla_c F_{ab})(\nabla^c F^{ab}) \, ,
\end{align}
\begin{align}
 e^{-2(\beta-2\alpha)\phi}\nabla_c(e^{2(\beta-2\alpha)\phi}\partial_b\phi F_a{}^c F^{ab})=\nonumber\\
-\partial^a\phi\left(F^{bc}(\nabla_c F_{ab})+F_a{}^b(\nabla_cF_b{}^c)\right)
+F_a{}^c\left(2(-2\alpha+\beta)F_{bc}(\partial^a\phi)(\partial^b\phi)+F^{ab}(\nabla_c\nabla_b\phi)\right)\, ,
\end{align}
\begin{align}
e^{-2(\beta-2\alpha)\phi}\nabla_a\left(e^{2(\beta-2\alpha)\phi}F^2\partial^a\phi \right)=\nonumber\\
2F^{bc}\left(\partial^a\phi \nabla_a F_{bc}+(\beta-2\alpha)(\partial\phi)^2F_{bc}\right)+F^2\nabla^2\phi \, ,
\end{align}
\begin{align}
e^{2\alpha\phi}\nabla_b(e^{-2\alpha\phi}R^{ab}\partial_a\phi)=\nonumber\\
\frac{1}{2}(\partial_a\phi)(\partial^a R)+R^{ab}(\nabla_b\nabla_a\phi)-2\alpha R_{ab}(\partial^a\phi)(\partial^b\phi) \, ,
\end{align}
\begin{align}
e^{2\alpha\phi}\nabla^a\left( e^{-2\alpha\phi}R \partial_a\phi \right)=
%\nonumber\\
R \nabla^2\phi + \partial_a\phi\partial^a R-2\alpha R (\partial\phi)^2 \, .
\end{align}

\section{Mass and Scalar Shifts in 5d SUGRA Solutions}
We relegate to this appendix some of the technical aspects of the solutions presented in section \ref{SectionN2SUGRA}, and the corresponding expressions for the higher-derivative corrections to the mass and scalar charges.

\subsection{\texorpdfstring{$Q_0$, $P_0$}{Q0, P0} Solutions}
\label{app:Q0P0}

The temperature and entropy are given by
\begin{align}
    F \ &= \ \frac{p + q - 2 m}{4} \, , \\
    T \ &= \ \frac{m (p + q)}{\pi\sqrt{p q} (2 m + p)(2 m + q)} \, , \\
    S \ &= \ \pi  \frac{\sqrt{p q} (q + 2m)(p +2 m )}{2 (p + q)} \, .
\end{align}
It is clear that $T \to 0$ in the extremal $m \to 0$ limit%
\footnote{There is a subtlety in 
% how to take 
taking the extremal limit when either one of the charges vanish.  Consider \emph{e.g.} a purely electric black hole with vanishing magnetic charge $P_0$ obtained by setting $p=2m$.  In this case $T = 1/({4 \pi \sqrt{2 q m}})$, which diverges in the extremal limit when $m \to 0$.  In order to obtain zero temperature at extremality, we have to instead take $m\to0$ first before sending $p\to0$.  A similar procedure can be used for the purely magnetic solutions.  Dyonic black holes always have zero temperature in the extremal limit.}.

The four-derivative corrections to the free-energy are computed from the four-derivative on-shell action evaluated on the two-derivative solution \cite{Reall:2019sah}, and are given by 
\begin{align}
\begin{split}
    \Delta F &= -\frac{c_1}{4(p-q)^{5/2}}\Bigg( 3 i \pi q^2 \sqrt{\frac{p+q}{p^2 - 4 m^2}} + \frac{\sqrt{p-q}(p+q)(p^2q^2 (p-4q) + 2 m p q (8p^2 - 19 p q + 5 q^2))}{p^2  (p + 2m)(q+2m)^2} \\
    & \qquad  + 4 m^2(8 p^3 - 16 p^2 q + 4 p q^2 + q^3) + \frac{6 q^2 (p + q)\arctanh \left[ \frac{(p + 2m) \sqrt{p + q}}{\sqrt{(p-q)(p^2 - 4 m^2}} \right] }{\sqrt{(p + q)(p^2 - 4 m^2)}}\Bigg)\, .
\end{split}
\end{align}
The free energy allows us to compute the corrections to all other quantities. 
For the correction to the scalar charge we find
\begin{align}
\begin{split}
    \Delta \mu_\phi &= c_1 \Bigg( \frac{\sqrt{3} q^2(11p- q ) \sqrt{p + q} \arctanh \left[\frac{\sqrt{p + q}}{\sqrt{p - q}} \right]}{p (p-q)^{7/2}} \\
    & +  \frac{1}{2 \sqrt{3} p (p - q)^3} \left( 3p^3 - 18 p^2 q - 49 p q^2 + 4 q^3 + 33 p^2 \pi q^2 \sqrt{\frac{p + q}{p^2(q-p)}} - 3 p \pi q^3 \sqrt{\frac{p + q}{p^2(q-p)}} \right) \, \Bigg)  .
\end{split}
\end{align}
Finally, we can use this to compute the long-range force. Recall that the forces cancel at the two-derivative level. Therefore the total force between two extremal black holes with charges $P_0$ and $Q_0$, will be, to leading order in the EFT coefficients, 
\begin{align}
\begin{split}
    r^2 \Delta f = & - 2  M \left(\Delta M \right)_{Q, P, T = 0} - 1/2 \mu_\phi \left(\Delta \mu_\phi \right)_{Q, P, T = 0}  \\
    & = \frac{c_1}{4 p (p - q)^{2}} \Bigg( -p^3 - 4 q^3 + 21 p q^2 + 8 p^2 q - 3 \pi  (5p-q) q^2 \sqrt{\frac{p+q}{ (q - p)}} \\
    & \qquad \qquad  - 6 (5p - q)q^2 \frac{ \sqrt{p + q}}{\sqrt{p - q}} \arctanh \frac{\sqrt{p + q}}{\sqrt{p-q}} \Bigg) \,.
\end{split}
\end{align}

\subsection{\texorpdfstring{$Q_0$, $P_1$}{Q0, P1} Solutions}
\label{app:Q0P1}

The temperature and entropy for these solutions are
\begin{align}
\begin{split}
    T&=\fft1{4\pi m\cosh\beta_0\cosh^3\beta_1}\,,\\
    S&=\pi m^2\cosh\beta_0\cosh^3\beta_1\,.
\end{split}
\label{eq:TandS}
\end{align}
At the two-derivative level, the free energy and mass are given by
\begin{align}
\begin{split}
F&= \frac{m}{8} ( \cosh 2 \beta_0 + 3 \cosh 2 \beta_1 - 2) \, , \\
    &M = \frac{m}{8}\,  \left( \cosh 2 \beta_0 + 3\cosh 2 \beta_1 \right) \, . \\
\end{split}
\end{align}
The four-derivative corrections to the mass and dilatonic scalar charge
take the simple form
\begin{align}
\begin{split}
    \Delta M  &= -\frac{3 \sqrt3 (2 c_1 + 7 c_2 + 24 c_3 + 12 c_4)}{80 |P_1|} \, , \\
    \Delta \mu_\phi &= \frac{3 (2 c_1 + 7 c_2 + 24 c_3 + 12 c_4)}{40 |P_1|}\, ,
\end{split}
\end{align}
and the corrected force is 
\begin{align}
\begin{split}
    r^2 \Delta f &=  \frac{3 (|P_1| + \sqrt3 |Q_0|)}{40 |P_1|} (2 c_1 + 7 c_2 + 24 c_3 + 12 c_4)  \, .
\end{split}
\end{align}

\subsection{\texorpdfstring{$Q_1$, $P_0$}{Q1, P0} Solutions}
\label{app:Q1P0}

The corrections for this solution are far more complicated than for its dual, the $Q_0$, $P_1$ black hole solution. The mass shift is given by 
\begin{align}
\begin{split}
     (\Delta M)_{T = 0} & = \frac{1}{54 \left(2 P_0-\frac{2 Q_1}{\sqrt{3}}\right){}^5} \Bigg( 36 \log \left(\frac{ Q_1}{\sqrt{3} P_0}\right) \left(48 c_1 P_0^2 Q_1^2-\left(2 c_1+c_2-12 \left(2 c_3+c_4\right)\right) Q_1^4\right)\\
     &-96 \sqrt{3} c_5 \left(24 \sqrt{3} P_0^3 Q_1-8 \sqrt{3} P_0 Q_1^3-36 P_0^2 Q_1^2 \log \left(\frac{\sqrt{3} P_0}{Q_1}\right)-9
   P_0^4+Q_1^4\right) \\
   & + \left(3 P_0-\sqrt{3} Q_1\right) \Big(\left(c_2-12 \left(2 c_3+c_4\right)\right) \left(-39 \sqrt{3} P_0^2 Q_1+69 P_0 Q_1^2+27 P_0^3-25
   \sqrt{3} Q_1^3\right) \\
   & \qquad \qquad +6 c_1 \left(43 \sqrt{3} P_0^2 Q_1+79 P_0 Q_1^2-15 P_0^3-11 \sqrt{3} Q_1^3\right)\Big)\Bigg)\, ,
\end{split}
\end{align}
while the scalar charge correction is  
\begin{align}
\begin{split}
    (\Delta \mu_\phi)_{T = 0} &=  \frac{1}{72 \sqrt{3} \left(2 P_0-\frac{2 Q_1}{\sqrt{3}}\right){}^6} \Bigg( 64 Q_1^2 \log \left(2 P_0\right) \Big(2 c_1 \left(216 \sqrt{3} P_0^2 Q_1-57 P_0 Q_1^2+792 P_0^3-\sqrt{3}
   Q_1^3\right) \\
   & -\left(c_2-12 \left(2 c_3+c_4\right)\right) Q_1^2 \left(57 P_0+\sqrt{3} Q_1\right)\Big) \\
   &+ 64 Q_1^2 \log \left(\frac{2 Q_1}{\sqrt{3}}\right) \Big(\left(c_2-12 \left(2 c_3+c_4\right)\right) Q_1^2 \left(57
   P_0+\sqrt{3} Q_1\right) \\
   & +2 c_1 \left(-216 \sqrt{3} P_0^2 Q_1+57 P_0 Q_1^2-792 P_0^3+\sqrt{3} Q_1^3\right)\Big) \\
   & + \frac{16}{3} \left(3 P_0-\sqrt{3} Q_1\right) \Big(2 c_1 \left(-708 \sqrt{3} P_0^3 Q_1-4920 P_0^2 Q_1^2+100 \sqrt{3}
   P_0 Q_1^3+135 P_0^4+81 Q_1^4\right) \\
   & -\left(c_2-12 \left(2 c_3+c_4\right)\right) \left(-156 \sqrt{3} P_0^3 Q_1+408
   P_0^2 Q_1^2-260 \sqrt{3} P_0 Q_1^3+81 P_0^4-73 Q_1^4\right) \Big) \\
   &  -512 c_5 \Big(-351 P_0^4 Q_1+21 \sqrt{3} P_0 Q_1^4+  6 \sqrt{3} P_0^3 Q_1^2 \left(66 \log
   \left(\frac{\sqrt 3 P_0}{Q_1}\right)-76\right) \\
    & +18 P_0^2 Q_1^3 \left(18 \log \left(\frac{P_0}{Q_1}\right)+28+9
   \log (3)\right)+27 \sqrt{3} P_0^5-Q_1^5\Big) \, .
    \Bigg)
\end{split}
\end{align}
Finally, the long-range force at zero temperature is given by
\begin{align}
\begin{split}
        r^2 \Delta f 
    & = \frac{1}{27 \left(2 P_0-\frac{2 Q_1}{\sqrt{3}}\right){}^5} \Bigg( 
    \left(3 P_0-\sqrt{3} Q_1\right) \Big(-24 \sqrt{3} \left(3 c_2-72 c_3-36 c_4+64 \sqrt{3} c_5\right) P_0^3 Q_1 \\
    & +12 \left(19 c_2-4 \left(114 c_3+57 c_4+160 \sqrt{3} c_5\right)\right) P_0^2 Q_1^2-152 \sqrt{3} \left(c_2-12 \left(2 c_3+c_4\right)\right) P_0 Q_1^3 \\
    & +2 c_1 \left(312 \sqrt{3} P_0^3 Q_1+2148 P_0^2 Q_1^2-152 \sqrt{3} P_0 Q_1^3-45 P_0^4+9 Q_1^4\right)\\
    & +9 \left(3 c_2-72 c_3-36 c_4+32 \sqrt{3} c_5\right) P_0^4+\left(c_2-4 \left(6 c_3+3 c_4+8 \sqrt{3} c_5\right)\right) Q_1^4\Big) \\
    & -6 Q_1^2 \Big(-288 c_5 P_0^2 \left(5 \sqrt{3} P_0+3 Q_1\right)+\left(c_2-12 \left(2 c_3+c_4\right)\right) Q_1^2
   \left(\sqrt{3} Q_1-27 P_0\right) \\ 
   & +2 c_1 \left(72 \sqrt{3} P_0^2 Q_1-27 P_0 Q_1^2+360 P_0^3+\sqrt{3}
   Q_1^3\right)\Big) \log \left(\frac{3 P_0^2}{Q_1^2}\right) 
   \Bigg) \, .
\end{split}
\end{align}
We can examine this expression in the single charge limits, where we find 
\begin{align}
    \lim_{P_0 \to 0} r^2 \Delta f &= \frac{3}{8} (2 c_1 + c_2 - 24 c_3 - 12 c_4) \log \left(P_0/Q_1\right) \, , \\
    \lim_{Q_1 \to 0} r^2 \Delta f &= \frac{1}{32} (-10 c_1 + 3 c_2 - 72 c_3 - 36 c_4 + 32 \sqrt{3} c_5) \, .
\end{align}
The force diverges in the $P_0 \to 0$ limit, just like the mass shift. 

\subsection{\texorpdfstring{$Q_1$, $P_1$}{Q1, P1} Solutions}
\label{app:Q1P1}

For black holes with non-vanishing $Q_1$, $P_1$ charges, 
the mass shift at extremality in the presence of higher-derivative terms is quite non-trivial, and takes the form
\begin{align}
\begin{split}
        (\Delta M)_{T= 0 } &= -\frac{\sqrt3 \sqrt{Q_1^2 + P_1^2}}{16 Q_1^{10}}
        \Bigl[ -Q_1^2 (Q_1^2+ 2P_1^2) \Bigl( c_2 (678 P_1^4 + 678 P_1^2 Q_1^2 + Q_1^4) -2 c_1 (6 P_1^4 + 6 P_1^2 Q_1^2 + 11 Q_1^4) \\
    &  + 12 \left(348 \, c_3 \, P_1^4 + 114 \, c_4 \, P_1^4 + 348 \, c_3 \, P_1^2 Q_1^2 + 114 \, c_4 \, P_1^2 Q_1^2 + 
  26 \, c_3 \, Q_1^4 + 11 c_4 \, Q_1^4 \right) \Bigr)\\
    &  - 3 \Bigl(c_2 (452 P_1^8 + 904 P_1^6 Q_1^2 + 528 P_1^4 Q_1^4 + 76 P_1^2 Q_1^6 - 
      Q1^8) \\
    &- 2 c_1 (4 P_1^8 + 8 P_1^6 Q_1^2 + 12 P_1^4 Q_1^4 + 
      8 P_1^2 Q_1^6 + Q_1^8)\\
    & +12 c_4 (76 P_1^8 + 152 P_1^6 Q_1^2 + 96 P_1^4 Q_1^4 + 20 P_1^2 Q_1^6 + 
     Q_1^8) + 24 c_3 (116 P_1^8 + 232 P_1^6 Q_1^2 + \\
     & 144 P_1^4 Q_1^4 + 
     28 P_1^2 Q_1^6 + Q_1^8)) \Bigr) \log\frac{P_1^2}{Q_1^2 + P_1^2}\Bigr] \, .
\end{split}
\end{align}
%Note that it is expressed entirely in terms of the charges $Q_1, P_1$ and the Wilson coefficients $c_i$.
Recall that these solutions are supported by both a dilatonic scalar and an axion.
The dilatonic scalar charge shift is given by
\begin{align}
\begin{split}
        \Delta \mu_\phi &= -\frac{1}{8 Q_1^{10}\sqrt{Q_1^2 + P_1^2}}
        \Bigl[ Q_1^2 \Bigl( -c_2 (25764 P_1^8 + 50850 P_1^6 Q_1^2 + 29402 P_1^4 Q_1^4 + 
    4315 P_1^2 Q_1^6 - 11 Q_1^8) + \\
    &
 2 c_1 (228 P_1^8 + 450 P_1^6 Q_1^2 + 502 P_1^4 Q_1^4 + 269 P_1^2 Q_1^6 + 
    23 Q_1^8) \\
&    - 
 12 (c_4 (4332 P_1^8 + 8550 P_1^6 Q_1^2 + 5182 P_1^4 Q_1^4 + 
       953 P_1^2 Q_1^6 + 23 Q_1^8) + \\
&    2 c_3 (6612 P_1^8 + 13050 P_1^6 Q_1^2 + 7826 P_1^4 Q_1^4 + 
       1375 P_1^2 Q_1^6 + 25 Q_1^8)) \Bigr)\\
    &  - 3 \Bigl( c_2 (8588 P_1^{10} + 21244 P_1^8 Q_1^2 + 17560 P_1^6 Q_1^4 + 5284 P_1^4 Q_1^6 + 
    377 P_1^2 Q_1^8 - Q_1^{10})  \\
& -2 c_1 (76 P_1^{10} + 188 P_1^8 Q_1^2 + 236 P_1^6 Q_1^4 + 164 P_1^4 Q_1^6 + 
    43 P_1^2 Q_1^8 + Q_1^{10}) \\
&    + 
 12 c_4 (1444 P_1^{10} + 3572 P_1^8 Q_1^2 + 3032 P_1^6 Q_1^4 + 
    1004 P_1^4 Q_1^6 + 103 P_1^2 Q_1^8 + Q_1^{10}) + \\
& 24 c_3 (2204 P_1^10 + 5452 P_1^8 Q_1^2 + 4600 P_1^6 Q_1^4 + 
    1492 P_1^4 Q_1^6 + 143 P_1^2 Q_1^8 + Q_1^{10}) \Bigr) \log\frac{P_1^2}{Q_1^2 + P_1^2}\Bigr] \, .
\end{split}
\end{align}
For the axionic charge we obtain instead 
\begin{align}
\begin{split}
        \Delta \mu_\chi &= - \frac{ P_1}{4 Q_1^{11}\sqrt{Q_1^2 + P_1^2}}
        \Bigl[ Q_1^2 (c_2 (13560 P_1^8 + 27120 P_1^6 Q_1^2 + 16058 P_1^4 Q_1^4 + 
     2498 P_1^2 Q_1^6 - 5 Q_1^8) \\ 
     & - 
  2 c_1 (120 P_1^8 + 240 P_1^6 Q_1^2 + 274 P_1^4 Q_1^4 + 154 P_1^2 Q_1^6 + 
     17 Q_1^8) \\ 
     & + 
  12 c_4 (2280 P_1^8 + 4560 P_1^6 Q_1^2 + 2830 P_1^4 Q_1^4 + 
     550 P_1^2 Q_1^6 + 17 Q_1^8) + \\
     &
  24 c_3 (3480 P_1^8 + 6960 P_1^6 Q_1^2 + 4274 P_1^4 Q_1^4 + 
     794 P_1^2 Q_1^6 + 19 Q_1^8)) 
        \\
        & + 3 (2 P_1^2 + Q_1^2) 
        (c_2 (2260 P_1^8 + 4520 P_1^6 Q_1^2 + 2488 P_1^4 Q_1^4 + 228 P_1^2 Q_1^6 - 
     Q_1^8) \\
     &- 2 c_1 (20 P_1^8 + 40 P_1^6 Q_1^2 + 44 P_1^4 Q_1^4 + 
     24 P_1^2 Q_1^6 + Q_1^8) \\
     & + 
  12 c_4 (380 P_1^8 + 760 P_1^6 Q_1^2 + 440 P_1^4 Q_1^4 + 60 P_1^2 Q_1^6 + 
     Q_1^8) \\ 
     &+ 24 c_3 (580 P_1^8 + 1160 P_1^6 Q_1^2 + 664 P_1^4 Q_1^4 + 
     84 P_1^2 Q_1^6 + Q_1^8))
        \log\frac{P_1^2}{Q_1^2 + P_1^2}\Bigr] \, .
        \end{split}
\end{align}
We emphasize that the shifts to the two scalar charges are controlled by combinations of the Wilson coefficients and charges that are 
completely independent of those controlling the mass shift, indicating that in general there is no correlation between the WGC and the RFC in the presence of 
higher-derivative corrections.

%%%%%
\bibliographystyle{JHEP}
\bibliography{cite.bib}
%%%%%
\end{document}